\documentclass[iop,apj]{emulateapj}

\usepackage{epsfig}
\usepackage{amsmath, amsthm, amssymb}
\usepackage[ plainpages=false, colorlinks=true, anchorcolor=blue, linkcolor=blue, citecolor=blue]{hyperref}
\usepackage{color,times,microtype,float,bm,ulem}
\usepackage{multirow}

\newcommand{\beqn}{\begin{equation}}
\newcommand{\eeqn}{\end{equation}}
\newcommand{\beqy}{\begin{eqnarray}}
\newcommand{\eeqy}{\end{eqnarray}}

\newcommand{\fig}[1]{Figure \ref{#1}}
\newcommand{\figs}[1]{Figures \ref{#1}}

\newcommand{\tab}[1]{Table \ref{#1}}
\newcommand{\csec}[1]{\S \ref{#1}}

\newcommand{\Msun}{M_\sun}
\newcommand{\e}[1]{\times 10^{#1}}
\newcommand{\nifs}{$^{56}$Ni }
\newcommand{\gcm}{g cm$^{-3}$}
\newcommand{\Nign}{$N_{\mathrm{ign}}$}
\newcommand{\Rsph}{$R_\mathrm{sph}$}

\newcommand{\pdone}{PD-s128-n0063}
\newcommand{\pdtwo}{PD-s128-n0150}
\newcommand{\pdthree}{PD-s256-n0128}
\newcommand{\pdfour}{PD-s256-n1100}
\newcommand{\pdfive}{PD-s384-n1700}
\newcommand{\pdsix}{PD-s384-n3500}

\newcommand{\Fsims}{simulations with $N_{\mathrm{ign}}\sim100$}
\newcommand{\Msims}{simulations with $N_{\mathrm{ign}}\sim10^3$}

\newcommand{\MB}{$M_{\rm B}$}
\newcommand{\Mbol}{$M_{\rm bol}$}
\newcommand{\dmfB}{$\Delta m_{15}^{\rm B}$}
\newcommand{\dmfbol}{$\Delta m_{15}^{\rm bol}$}
\newcommand{\mdmplaneBB}{\MB -- \dmfB\ plane}
\newcommand{\mdmplaneBbol}{\MB -- \dmfbol\ plane}
\newcommand{\mdmplanebolbol}{\Mbol -- \dmfbol\ plane}
\newcommand{\mdmplanebolB}{\Mbol -- \dmfB\ plane}

\slugcomment{Draft version} 

\begin{document}

\title{Three-dimensional Simulations of Pure Deflagration Models for Thermonuclear Supernovae}
\author{Min Long\altaffilmark{1,2}, George C. Jordan IV\altaffilmark{1, 2}, Daniel R. van Rossum \altaffilmark{1,2}, Benedikt Diemer\altaffilmark{1, 2, 3}, Carlo Graziani\altaffilmark{1,2}, Richard Kessler\altaffilmark{1,2,3}, Bradley Meyer\altaffilmark{5}, Paul Rich\altaffilmark{1,2,6}, Don Q. Lamb\altaffilmark{1, 2, 4}}

\altaffiltext{1}{Flash Center for Computational Science, University of Chicago, Chicago, IL, 60637; long@flash.uchicago.edu}
\altaffiltext{2}{Department of Astronomy \& Astrophysics, University of Chicago, Chicago, IL 60637 USA}
\altaffiltext{3}{Kavli Institute for Cosmological Physics, University of Chicago, Chicago, IL 60637 USA}
\altaffiltext{4}{Enrico Fermi Institute, University of Chicago, Chicago, IL 60637, USA}
\altaffiltext{5}{Department of Physics \& Astronomy, Clemson University, Clemson, South Carolina 29634}
\altaffiltext{6}{Argonne Leadership Computing Facility, Argonne National Laboratory, Argonne, IL 60439}

\shorttitle{Deflagration Simulations of Thermonuclear Supernovae}
\shortauthors{Long et al.}

\begin{abstract}

We present a systematic study of the pure deflagration model of Type Ia supernovae using three-dimensional, high-resolution, full-star hydrodynamical simulations, nucleosynthetic yields calculated using Lagrangian tracer particles, and light curves calculated using radiation transport.  
We evaluate the simulations by comparing their predicted light curves with many observed SNe Ia using the SALT2 data-driven model and find that the simulations may correspond to under-luminous SNe Iax.  
We explore the effects of the initial conditions on our results by varying the number of randomly selected ignition points from 63 to 3500, and the radius of the centered sphere they are confined in from 128 to 384 km.
We find that the rate of nuclear burning depends on the number of ignition points at early times, the density of ignition points at intermediate times, and the radius of the confining sphere at late times.  
The results depend primarily on the number of ignition points, but we do not expect this to be the case in general.  
The simulations with few ignition points release more nuclear energy $E_{\mathrm{nuc}}$, have larger kinetic energies $E_{\mathrm{K}}$, and produce more $^{56}$Ni than those with many ignition points, and differ in the distribution of $^{56}$Ni, Si, and C/O in the ejecta.  
For these reasons, the simulations with few ignition points exhibit higher peak B-band absolute magnitudes $M_\mathrm{B}$ and light curves that rise and decline more quickly;  their $M_\mathrm{B}$ and light curves resemble those of under-luminous SNe Iax, while those for simulations with many ignition points are not.


\end{abstract}

\section{Introduction}
\label{sec:intro}

\begin{figure*}
\centering
\includegraphics[width=6.25in]{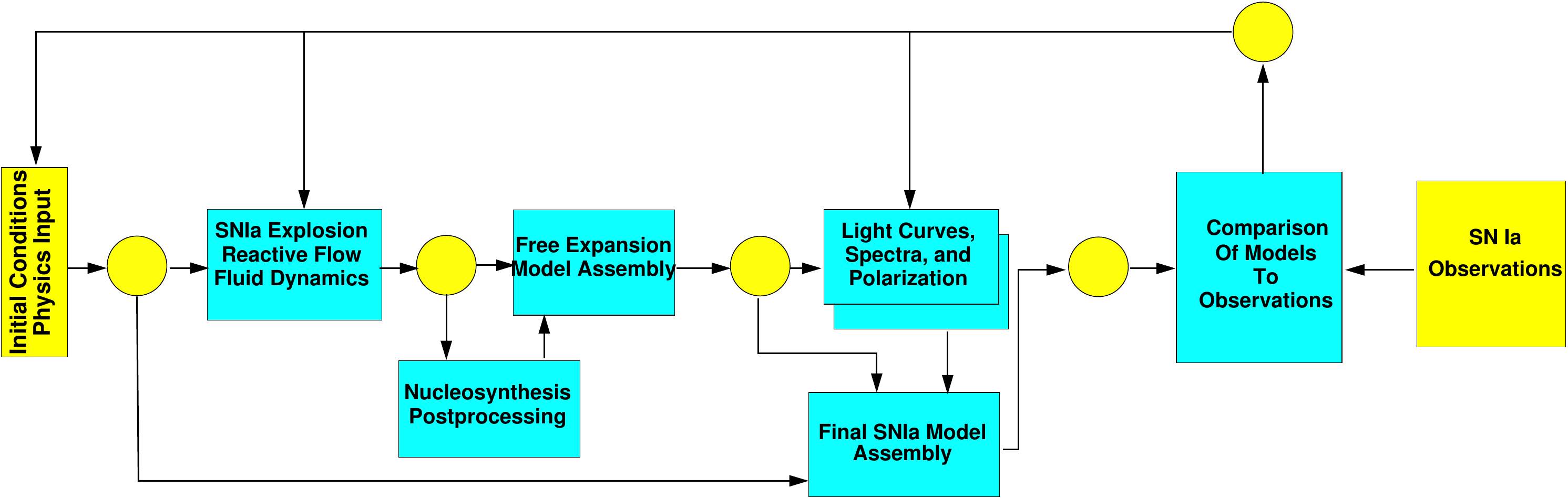}
\caption{
Schematic diagram of the Flash Center SNe Ia simulation and validation pipeline. 
Yellow circles represent data archives.
Blue boxes are computational codes and algorithms. 
Yellow boxes are initial conditions and observations. 
\texttt{FLASH} is used to perform hydrodynamic simulations of the explosion phase (the blue box labeled "SNe Ia Explosion Reactive Flow Dynamics").  
In this paper, \texttt{PHOENIX} is used to perform calculations of the radiative transfer phase (the blue box labeled ``Light Curves, Spectra, and Polarization"). 
A methodology that incorporates \texttt{SNANA} and data-driven models is used to confront the light curves predicted by the simulations with observations (the blue box labeled ``Comparison Of Models To Observations"). 
See \csec{sec:models:pipeline} for further details.
}
\label{fig:pipeline}
\end{figure*}

Type Ia Supernovae (SNe Ia) are classified from their observed spectra, which lack hydrogen lines but show strong silicon lines.  
Since the 1960s \citep{Hoyle_1960}, SNe Ia have widely been thought to be a result of the thermonuclear explosion of a degenerate star in a binary system.  
However, the explosion mechanism and the origin of the observed range of luminosities are not fully understood.  
Numerous different models of the progenitor system and explosion mechanism have been proposed to explain and predict their observed properties \citep[see e.g.,][for recent reviews]{Wang_2010, Hillebrandt_2013}.

The two most popular scenarios for SNe Ia are the single-degenerate (SD) and the double-degenerate (DD) scenarios.  
In the SD scenario, a carbon-oxygen (CO) white dwarf (WD) star undergoes a thermonuclear runaway when its mass approaches the Chandrasekhar limit as a result of accreting hydrogen-rich or helium-rich material from its companion \citep{Whelan_1973,Nomoto_1982}. 
In the DD scenario, the thermonuclear explosion is caused by the merger of two CO WDs \citep{Iben_1984,Webbink_1984}.  
Both scenarios can account for many of the observed properties of SNe Ia \citep{Kasen_2009, Roepke_2012, pakmor_2012}.  
On the other hand, models for both scenarios have trouble explaining why the observed properties of SNe Ia span a relatively small region of parameter space.

Among SD scenarios, the pure deflagration (PD) model is special as it involves only a subsonic deflagration flame that releases nuclear energy, causing the star to expand.  
Therefore, most of the nuclear burning happens at relatively low densities, producing intermediate-mass elements (IME); that is, nuclei with atomic masses higher than O and lower than Fe.  
Because {\nifs} is produced by nuclear burning at high densities, PD simulations produce much less {\nifs} than is needed to explain the peak brightness of normal SNe Ia events.  
Consequently, the PD model has great difficulty in explaining normal SNe Ia.

To produce a more energetic explosion and more {\nifs}, SD models have been proposed in which a detonation follows the deflagration phase, such as the deflagration-to-detonation transition (DDT) model \citep{Khokhlov_1991, Kasen_2009, Krueger_2012, Seitenzahl_2013} and the gravitationally confined detonation (GCD) \citep{Plewa_2004,Townsley_2007,Jordan_2008, Meakin_2009, Jordan_2012}.
The detonation wave is supersonic and quickly incinerates the entire star, releasing additional nuclear energy and producing significant amounts of both iron group elements (IGE) and IME.
It is important to note that the ignition and the deflagration phase are identical in the PD model and the DDT or GCD models up to the point where the detonation occurs.
Thus, simulations of the PD model are not only interesting in themselves, but serve as simulations of the deflagration phase of the DDT or GCD models, and provide the initial conditions for the detonation phase.

Another motivation for simulating the PD model is the observation of SNe Ia that have low expansion velocities and/or unusually low luminosities.  
The number of such known under-luminous SNe Ia is increasing, and it has been suggested that they form a sub-class, Type Iax Supernovae \citep[SNe Iax;][]{Foley_2013}.
Although the PD model has difficulty in producing enough {\nifs} to explain normal SNe Ia, it has the potential to explain under-luminous SNe Ia. 
In this paper, we examine this possibility by simulating the PD model for several different initial conditions and comparing the predicted peak B-band absolute magnitude $M_B$ and decline rate {\dmfB} of their light curves with those of SNe Iax.

Normal luminosity SNe Ia are known to follow the Phillips relation \citep{Phillips_1993, Phillips_1999} in the \mdmplaneBB.  
SNe Iax have lower luminosities, redder spectra, and light curves that decline faster than normal SNe Ia.
Whether the SNe Iax lie on an extension of the Phillips relation, lie on a relation like the Phillips relation but different from it, or are otherwise distributed, has not been established to our knowledge.  
We therefore do not know the distribution in the \mdmplaneBB\ that a model of these events (including the PD model) should predict.  

In the PD model, the final nuclear yields and the distribution of elements in the ejecta -- and thus the light curves and spectra -- depend on density, temperature and electron fraction in the unburned material (here after ``fuel"), and on how the flame evolves.
All of these depend on the initial conditions (ICs).
Study of the PD model therefore requires the exploration of a variety of ICs.
One way to do this is to vary the number and locations of the ignition points in a small region around the center of the  WD \citep[e.g.,][]{Reinecke_2002,Gamezo_2003,Roepke_2007b}.
These ignition points represent flames that arise from fluctuations in the density and temperature in the convective core of the WD.  
High-fidelity simulations of the PD model require three-dimensional (3D) calculations because off-center ignition points are not points but tori in two-dimensional (2D) simulations; and turbulence, which sets the effective flame speed -- and therefore the rate of nuclear burning -- during the deflagration phase, behaves very differently in 2D than in 3D.

A number of of studies of the deflagration phase of SNe Ia have used 3D simulations \citep[e.g.,][]{Reinecke_2002,Gamezo_2003,Roepke_2007b}.
Recently, \citet{Kromer_2013, Roepke_2012,Seitenzahl_2013} extended these studies to simulations with different ignition configurations and a flame model based on the turbulent sub-grid model developed by \citet{Schmidt_2006a, Schmidt_2006b}.
These simulations use an expanding grid \citep{Roepke_2005a} with a resolution of $512^3$ cells, which improves the spatial resolution at the center of the WD relative to that in the outer regions of the computational domain.
The simulations use a monopole gravity solver to treat self-gravity \citep{Kromer_2013, Seitenzahl_2013} and one million Lagrangian tracer particles to calculate the final nucleosynthetic yields.

In this paper, we present a systematic study of the PD model, using 3D, high-resolution, full-star hydrodynamical simulations performed using the \texttt{FLASH} code with adaptive mesh refinement (AMR) and a multipole Poisson solver for self-gravity, nucleosynthetic yields calculated using 10 million Lagrangian tracer particles weighted by both mass and volume, and light curves calculated using the radiation transport code \texttt{PHOENIX}, enhanced to treat the temporal evolution of the ejecta. We perform six PD simulations to explore the dependence of the evolution of the nuclear burning,  the nucleosynthesis yields and their distributions in the ejecta, on ICs (i.e., the number and location of the ignition points).

In \csec{sec:models}, we describe in detail the hydrodynamic simulations, the radiation transport calculations, and our method for validating them using observations of numerous SNe Ia.
The results of the simulation are presented and discussed in \csec{sec:results}.
In \csec{sec:lightcurves}, we present and discuss the light curves predicted by the simulations. 
In \csec{sec:comparisons}, we compare them with the light curves of numerous observed SNe Ia using the data-driven SALT2 model.
In \csec{sec:conclusions}, we compare our results with earlier work, summarize our conclusions, and suggest directions for future work.

\section{Simulation and validation of SN Ia models}
\label{sec:models}

\subsection{Flash Center SN Ia simulation and validation pipeline}
\label{sec:models:pipeline}

We have built an integrated computational pipeline to simulate, validate and evaluate SNe Ia models, shown schematically in \fig{fig:pipeline}.
This paper is the first to present results from the complete pipeline.
The pipeline constitutes data defining the initial conditions and data from observations of SNe Ia (yellow rectangles), computer codes (blue rectangles), and permanent storage of simulation data (yellow circles).  
The major elements of the pipeline are the following:

1. Initial Conditions and Physics Input.  
We take as initial conditions 
(a) the mass of the white dwarf progenitor star, 
(b) whether it is rotating or not, 
(c) its C/O composition, 
(d) the convective flow pattern in its core, 
(e) the number and location of ignition points, and 
(f) whether a detonation is posited (in the DDT model) or followed via simulations (in the GCD model). 
The initial conditions also include simulation parameters such as the finest spatial resolution, the adaptive mesh refinement criteria, and the choice of parameters for the sub-grid flame model.

2. SN Ia Explosion Reactive Flow Fluid Dynamics.  
We use the \texttt{FLASH} hydrodynamics/reactive flow code \citep{Fryxell_2000, Dubey_2009} to predict the amount of nuclear energy that is released in the explosion, and the density and velocity distributions of the resulting freely expanding ejecta.  
The release of nuclear energy is over after 3-10 sec; after 50-100 sec, the expansion becomes nearly homologous (and can thus be computed analytically).  
Therefore, it suffices to calculate the density and velocity distributions only up to this time.  
The simulations contain Lagrangian tracer particles, and the temperatures and densities of these particles are saved as the simulation progresses.

3. Nucleosynthesis Post-Processing.  
We use the nuclear reaction network code Webnucleo v1.6 (Meyer et al., http://www.webnucleo.org/), and the motion and thermodynamic evolution of the Lagrangian tracer particles computed by \texttt{FLASH}, to predict the chemical composition throughout the ejecta \citep{Brown_2005}.

4. Free Expansion Model Assembly.  
We use the density and velocity distribution at 50-100 sec in the Eulerian form provided by \texttt{FLASH} and the final chemical composition in Lagrangian form provided by the Lagrangian tracer particles to assemble the free-expansion model of the ejecta.  

5. Light Curves, Spectra, and Polarization.  
In this paper, we use the radiation transport code \texttt{PHOENIX} \citep{Hauschildt_1999, Hauschildt_2004}, enhanced to treat the temporal evolution of the ejecta \citep{Rossum_2012} to calculate multi-waveband light curves and spectra as a function of time and viewing angle.

6. SNe Ia Model Assembly.  
We combine the initial conditions, free expansion models, and observational properties of a SNe Ia model (velocity, density, chemical composition, light curves, spectra, and polarization) into a single dataset and permanently store it using the data archiving software package {\tt SMAASH} \citep{Hudson_2011a,Hudson_2011b}.

7. Comparison of Models to Observations.  
We evaluate the light curves predicted by the simulations employing a methodology based on the data-driven model SALT2 \citep{Guy_2007, Guy_2010} to represent the heterogeneous family of SNe Ia light curves \citep[][]{Diemer_2013}. 
See \ref{sec:comparisons:lcevaluation} for a detailed discussion.

In the following sections, we discuss each element of the pipeline in more detail.

\subsection{Initial conditions:  white dwarf and configuration of ignition points}
\label{sec:models:input}

\begin{deluxetable}{crccc}
\tablewidth{0cm}
\tablecaption{List of Simulations and Their Initial Conditions }
\tablehead{
\colhead{Simulation}     & 
\colhead{$N_{\mathrm{ign}}$}   &  
\colhead{$R_{\mathrm{sph}}$ (km)}     &
\colhead{$f$}  &
\colhead{$M{_\mathrm{b}}(\Msun)$}
}
\startdata
\pdone & 63 & 128 & 0.12 &0.14$\e{-2}$\\
\pdtwo & 150 & 128 & 0.29 &0.30$\e{-2}$\\
\pdthree & 128 & 256 & 0.03 &0.27$\e{-2}$\\
\pdfour & 1100 & 256 & 0.27 &2.0$\e{-2}$\\
\pdfive & 1700 & 384 & 0.12 &2.9$\e{-2}$\\
\pdsix & 3500 & 384 & 0.25 &5.4 $\e{-2}$\\
\enddata
\tablecomments{
Various numbers \Nign\ of hot, burning bubbles with a radius of 16 km were placed randomly within a sphere of radius \Rsph\ at the start of each simulation to represent ignition points.  
$f$ is the fraction of the confining sphere occupied by the initial bubbles.
$M_{\mathrm{b}}$ is the total burned mass in the initial bubbles.  
}
\label{tab:init}
\end{deluxetable}

The progenitor for the 3D simulations presented here is a cold ($T = 3 \e{7}$ K), non-rotating white dwarf star of mass $M_{\rm WD} = 1.365 \Msun$, composed of 50\% $^{12}C$ and $50\%$ $^{16}$O by mass, and with no convection in its core.  
The Helmholtz equation of state of \citet{Timmes_2000} is used to describe the thermodynamic properties of the stellar plasma, including contributions from blackbody radiation, ion, electrons of arbitrary degeneracy, and the Coulomb contributions due to the interactions between ions and electrons.  
Self-gravity is calculated through a multipole Poisson solver by expanding the mass density field into multipole moments, which are used to construct the scalar gravitational potential in the simulation grids.  
The gravitational acceleration is then calculated from the derivative of the scalar potential at each location in the domain.

In the SD scenario, accretion of matter from the companion star increases the central density and temperature of the white dwarf, leading to the onset of carbon burning as the white dwarf approaches the Chandrasekhar mass limit.  
This smoldering phase ends when Carbon burning runs away, igniting a nuclear flame at one or more points.
Recent work by \citet{Zingale_2011} and \citet{Nonaka_2012} suggests that ignition likely occurs at a single point offset by roughly 40-75 km from the center of the star. 
However, a definitive answer is not yet available, and numerous studies have been done assuming both a single ignition point \citep[e.g.][]{Reinecke_2002a, Gamezo_2003} and multiple ignition points \citep[e.g.][]{Reinecke_2002, Roepke_2005b, Roepke_2006}.

In this paper, we treat the number and the locations of the ignition points as free parameters, as did \citet{Reinecke_2002}, \citet{Roepke_2005b}, and \citet{Roepke_2006}.  
We model the ignition points as small spherical bubbles with radii of 16 km inside of which the matter has burned to nuclear statistical equilibrium (NSE).  
The spherical bubbles are very hot with a temperature of $8\e{9}$ K. The nuclear flame that corresponds to the surface of the bubbles quickly becomes unstable due to the interchange instability that results from the cold fuel overlying the hot, burned matter (hereafter ``ash") in the bubble, and buoyancy-driven turbulent nuclear combustion ensues \citep{Khokhlov_1995,Townsley_2008}.

We performed a relatively small number of simulations because of the large computational cost of 3D hydrodynamic simulations using AMR.  
To systematically explore the effects of the number and locations of the ignition points, we chose spheres of three different radii (\Rsph=128, 256, and 384 km), centered on the center of the star, within which the ignition points are confined.   
We then place the ignition points at randomly selected Cartesian coordinates within the spheres, obeying the requirement that none of the ignition bubbles overlap.  
Thus, if a choice of Cartesian coordinates caused a new ignition bubble to overlap with any of the previous bubbles, this choice of coordinates was discarded and a new choice was made until the desired number of ignition points was placed.  
In order to explore a large dynamic range in the number of ignition points, we performed simulations with a relatively modest number of ignition points, and with nearly the largest number of ignition points possible, for each of the three spheres.  

\tab{tab:init} lists the ICs for the six simulations of the PD model we performed.
The number of ignition points \Nign\ ranges from 63 to 3500.  
The fraction of the volume of the confining sphere occupied by the bubbles ranges from 0.03 to 0.29, and the fraction of the mass of the white dwarf star contained in the bubbles ranges from 0.001 to 0.04, i.e. a small fraction of the overall mass.  

Our choice of ICs for the models \pdone\ and \pdtwo, models \pdthree\ and \pdfour, and models \pdfive\ 
and \pdsix\ allows us to explore the effect of different numbers (and densities) of ignition points on 
the outcome of the simulations, since in each of these comparisons, \Rsph\ is the same, while \Nign\ and the density of ignition points differ.
Also, our choice of ICs for models \pdtwo\ and \pdthree\ allows us to explore the effect of different confining radii (and different densities of ignition points), since in this comparison, \Rsph\ and the density of ignition points differ, while \Nign\ is nearly the same.
However, these choices of ICs make \pdthree\ an outlier in the fraction of the volume of the confining sphere occupied by the bubbles that represent ignition points, and result in ICs that fall into two groups with regard to \Nign:  \Fsims\ and \Msims.

\subsection{Reactive flow fluid dynamics}
\label{sec:models:reactive}

We performed full-star hydrodynamical simulations using \texttt{FLASH} \citep{Fryxell_2000, Dubey_2009}.  
The AMR capability in \texttt{FLASH} enabled us to achieve a spatial resolution of 4 km in the highest refined regions of the computational domain.  
The number of grid cells reached $5 \times 10^9$ in some of the simulations.  
We used the advection-diffusion-reaction (ADR) thermonuclear flame model described in \citet{Calder_2007} and \citet{Townsley_2007}.  
This flame model uses three reaction progress variables to describe the principal burning stages that occur when Carbon and Oxygen burn to iron-group nuclei: carbon burning, relaxation to nuclear quasi-static equilibrium (NQSE), and relaxation to nuclear statistical equilibrium (NSE).  
We follow the explosion and the subsequent expansion of the ejecta until the velocity field becomes nearly homologous, 50-100 sec after the start of the simulations.

\subsection{Lagrangian tracer particles}
\label{sec:models:particles}

To compute the final nuclear products, we use passive Lagrangian particles that trace the motion of the 
fluid elements and record their thermodynamic histories.
The tracer particles are placed in the part of the computational domain that lies inside the star at the 
beginning of the simulation, and are post-processed using a large nuclear network after the simulation 
has completed.
Tracer particles are usually chosen proportional to the density of the local fluid element; in this case, each 
particle represents the same mass. 
We call particles of this type mass-weighted (M-weighted) particles; they provide 
good spatial resolution in high-density regions.

However, as pointed out by \citet{Seitenzahl_2010}, M-weighted tracer particles poorly capture the nucleosynthetic yields in the outer layers of the ejecta, where the density is low and the number of particles is small.
These outer layers are important when calculating light curves and spectra, because they contain most of the IME, and the wavelength-dependent photosphere resides in the outer layers through maximum light.
Thus, we introduce a second type of tracer particle that is chosen proportional to the volume of the local fluid element. 
We call particle of this type volume-weighted (V-weighted) particles. 
An appropriate combination of M-weighted and V-weighted particles can provide good spatial resolution in both high-density and low-density regions of the ejecta.

\begin{figure}
\centering
\includegraphics[width=3in]{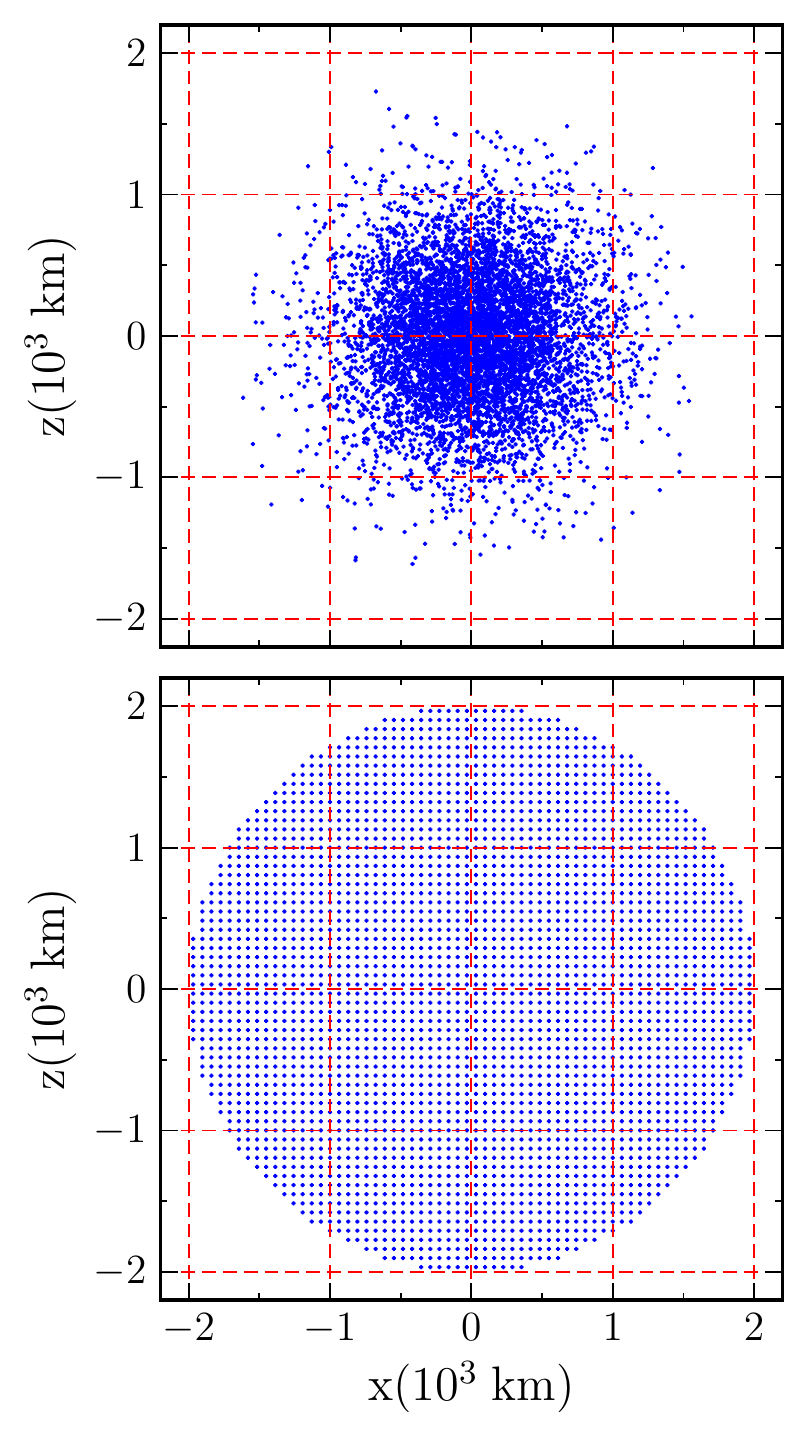}
\caption{
Example initial distribution of Lagrangian tracer particles used in the simulations. 
The upper panel shows the distribution of mass-weighted (M-weighted) particles, while the lower panel shows the distribution of volume-weighted (V-weighted) particles within a slab of thickness 64 km centered on the center of the white dwarf star and projected onto the $x-z$ plane.  
The V-weighted particles provide a much higher initial spatial resolution at radii greater than $1.0 \e{3}$ km than the M-weighted particles.
}
\label{fig:particles}
\end{figure}

\fig{fig:particles} illustrates an initial distribution for both types of passive tracer particles in a 3D simulation. 
The figure shows an example distribution with 125,000 M-weighted and 125,000 V-weighted particles, since it is not practical to plot the distribution of the full 10 million particles.
\fig{fig:particles} shows the particles in a slab through the center of the star with a thickness of 64 km, which is the volume represented by the V-weighted particles shown. 
There are 7097 M- and 3024 V-weighted particles in the slab. 
The M-weighted particles sample the fluid well inside a radius of $1.0\e{3}$ km.  
Even though the number of M-weighted particles is approximately 2.5 times the number of V-weighted particles, the M-weighted particles represent the fluid poorly outside of this radius, where the density is low and the deflagration flame produces primarily IME. 
In contrast, the V-weighted particles represent the low-density fluid well, and thus complement the M-weighted particles.
  
All of the 3D simulations presented in this paper use a superposition of both types of particles, distributed like the example in \fig{fig:particles}.
Each simulation uses $2\e{6}$ M-weighted particles and $8\e{6}$ V-weighted particles.
The M-weighted particles correspond to fluid elements with mass $1.365\Msun/2\e{6} = 6.8\e{-7}\Msun$ each, while the V-weighted particles correspond to fluid elements with masses {ranging from $1.5\e{-10}\Msun$ to $6.3\e{-6}\Msun$}.

\subsection{Nucleosynthesis post-processing}
\label{sec:models:nuclear}

We post-process the thermal histories of the 10 million tracer particles in each simulation with a state-of-the-art isotopic nuclear network, using a customized version of the Webnucleo code v1.6 (Meyer et al, http://www.webnucleo.org/). This procedure yields the abundances of the elements for each particle.

\subsection{Reconstruction of the abundances of the elements on the simulation grid}
\label{sec:models:reconstruction}

We use the computed chemical abundances of the elements for all 10 million tracer particles to determine the abundances of the chemical elements on the simulation grid. 
We calculate the abundances of chemical elements $X_\mathrm{i}$ for each grid cell using a mass-weighted average over all of the particles in the cell.
For cells that do not contain any tracer particles at the end of the simulation, we find the tracer particles in each 3D octant that are nearest to the cell, and use these particles to evaluate the mass-weighted average of $X_\mathrm{i}$ in the cell.
This process is performed with a {\tt FLASH} unit called Nuc2Grid.

\subsection{Calculation of light curves and spectra}
\label{sec:models:light curves}

In order to compare the results of our simulations with observations, we calculate light curves and spectra using the stellar atmosphere code \texttt{PHOENIX} \citep{Hauschildt_1999, Hauschildt_2004, Rossum_2012}, which is described in more detail in \csec{sec:lightcurves}.
Although 3D radiation transport (RT) is desirable and {\tt PHOENIX} has the ability to do 3D calculations \citep[e.g.][]{Hauschildt_2006, Hauschildt_2010}, self-consistent calculations of SNe Ia light curves and spectra in 3D with sufficient spatial, angular, and wavelength resolution are computationally very challenging and a work in progress.
Therefore, we present light curves obtained from one-dimensional (1D) spherically symmetric radiation transport calculations in this paper, and defer 3D calculations to the future.

We perform the 1D RT calculations on a regular velocity grid with 128 cells.  
We determine the distribution of elemental abundances using 20,000 tracer particles drawn from the 10 million tracer particles in the hydrodynamic simulation.  
{Because of the sparse sampling of original V-weighted particles near the center of the ejecta (due to the small volume there) and original M-weighted particles in the outer part of the ejecta (due to the low density there), we use all of the tracer particles that are available in these regions.  The intermediate region between these two regions contains an abundance of both original V-weighted and M-weighted particles, and we select a fraction of both to complete the sub-sample.}
We calculate the abundances of chemical elements $X_\mathrm{i}$ for each RT cell using a mass-weighted average over all of the particles in the cell.

\subsection{Comparison of predicted light curves to observations}
\label{sec:models:comparison}

We evaluate the light curves predicted by the simulations with the methodology of \citet{Diemer_2013, Kessler_2009} which uses the data-driven model SALT2 \citep{Guy_2007,Guy_2010} to represent the heterogeneous family of SNe Ia light curves. 
This method allows us to compare our simulations to the overall population of observed SNe Ia rather than individual objects. We describe the procedure in detail in \csec{sec:comparisons:lcevaluation}.

\section{Results for the hydrodynamic explosion phase}
\label{sec:results}

In this section, we explore the dependence of important properties of the nuclear burning process, and final nuclear products, on the ICs (i.e., the number and location of the ignition points).
These properties include the propagation of the deflagration flame and the nuclear burning rate; the evolution of internal, gravitational, and kinetic energies; the overall nucleosynthetic yields; and the distribution of IGE, IME, and unburnt C/O in the resulting ejecta.

\subsection{Evolution of nuclear burning and various energies}
\label{sec:results:evolutionall}

The six initial conditions for which we have performed simulations allow us to identify several factors that determine the rate and total amount of nuclear burning, and therefore the total amount of {\nifs} (which determines the peak luminosity of the SNe Ia); and the kinetic energy $E_\mathrm{K}$, which is a factor in determining the rate of rise and the rate of decline of the bolometric light curve.

\subsubsection{Evolution of nuclear burning}
\label{sec:results:evolutionall:burning}

\begin{figure}
\centering
\includegraphics[trim = 0mm 21mm 0mm 0mm, clip, scale=0.45]{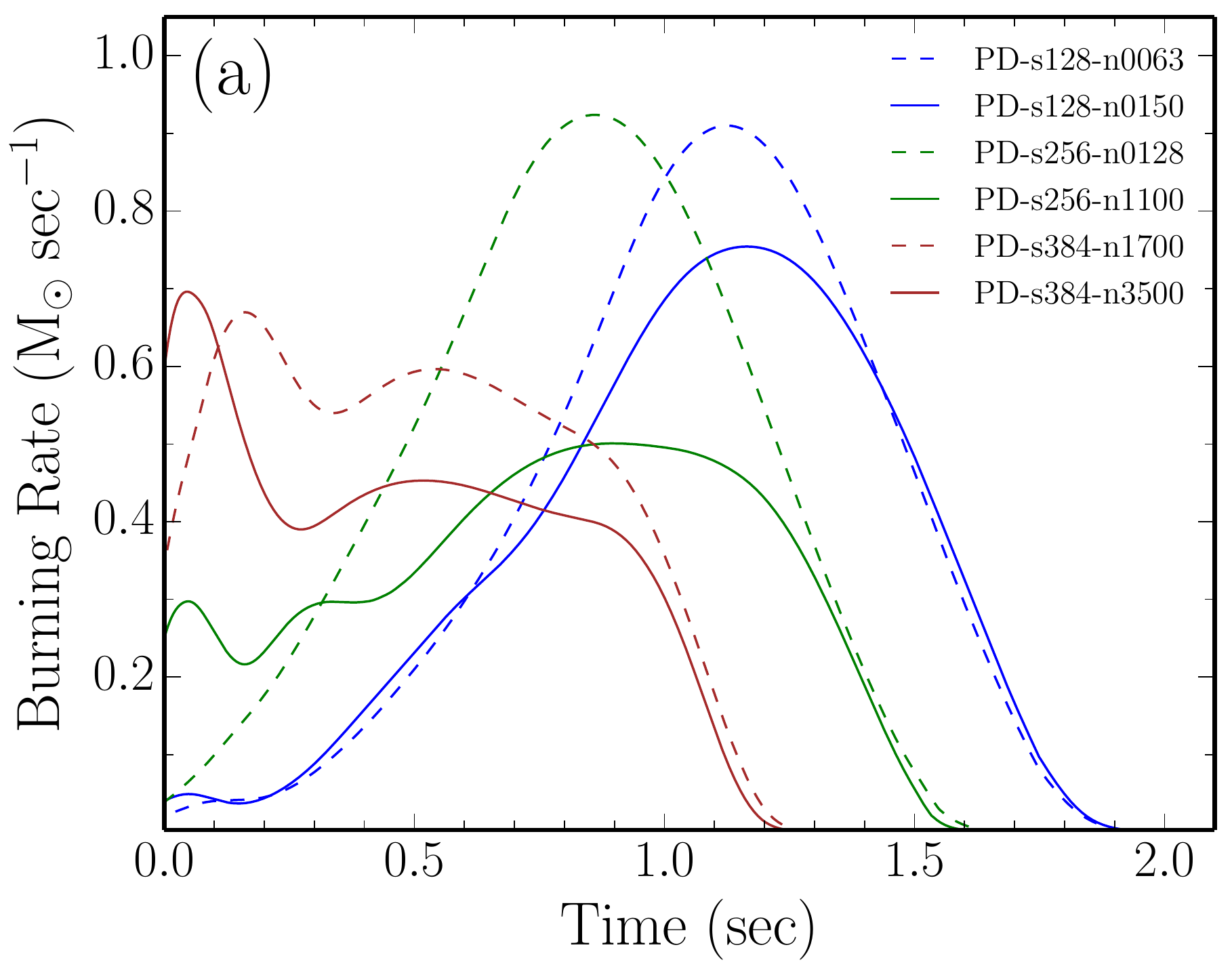}
\includegraphics[trim = 0mm 0mm 0mm 0mm, clip, scale=0.45]{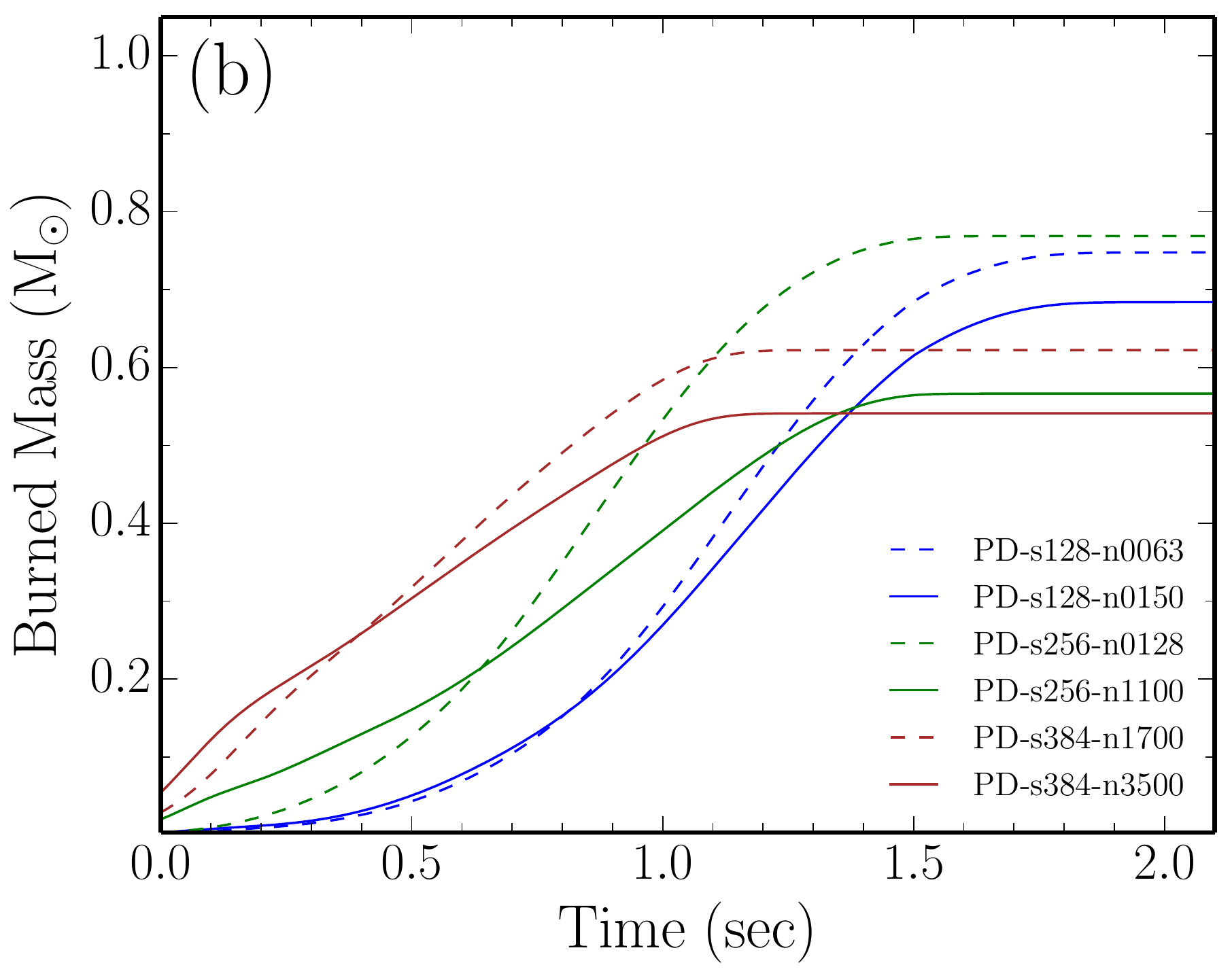}
\caption{
a): The nuclear burning rate as a function of time. (b): The burned mass $M_\mathrm{b}$ as a function of time. Blue, green, and brown lines correspond to the pairs of simulations with confining spheres of radii \Rsph= 128, 256 and 384 km. The simulations with a small fraction $f$ of the volume of the confining sphere filled by ignition point bubbles are shown as dashed lines, while those with nearly the largest possible fraction are shown as 
solid lines.
}
\label{fig:burnedmass}
\end{figure}

\figs{fig:burnedmass}(a) and (b) show the nuclear burning rate and the burned mass $M_\mathrm{b}$ as a function of time for all six simulations.  
The final $M_\mathrm{b}$ is also listed in column three in \tab{tab:final}.  
\fig{fig:burnedmass}, especially panel (a), enables us to disentangle the effects of the number of ignition points \Nign, the density of ignition points, and the radius of the confining sphere \Rsph\ on the rate of nuclear burning.

Specifically, \fig{fig:burnedmass}(a) shows that, {\it at early times}, the \Fsims\ have much lower nuclear burning rates and those \Msims\ have much larger burning rates.  
In fact, the rank order of the simulations based on the \Nign\ and the initial burning rate is the same.  
The explanation of this behavior is simple:  there are many more ignition bubbles with identical sizes in these simulations, and at early times the nuclear burning rate $R_\mathrm{nuc}\propto N_\mathrm{ign}$.  
One can see this directly by comparing the nuclear burning rate at early times between simulations \pdone\ and \pdtwo, \pdthree\ and \pdfour, and \pdfive\ and \pdsix.

\fig{fig:burnedmass}(a) also shows that, {\it at intermediate times}, the nuclear burning rates increase rapidly and reach the highest values in the \Fsims\, while the burning rates saturate or even decline in the \Msims.  
This behavior is due primarily to the difference in the density of ignition points:  the ignition bubbles in the \Fsims\ have a lot of room to grow, whereas the ignition bubbles in the \Msims\ do not, and soon merge, reducing the flame area and therefore the burning rate.  

However, another factor is also important at intermediate times.  
The hot, low-density bubbles that represent ignition points quickly become interchange unstable, after which the effective flame speed $S_\mathrm{eff} = S_\mathrm{t}\sim0.5\sqrt{AgL}$.  
In this expression, $A\equiv(\rho_0-\rho_1)/(\rho_0+\rho_1)$ is the Atwood number, which measures the density contrast between the fuel ($\rho_0$) and the ash ($\rho_1$), $g$ is the local gravitational acceleration and $L$ is the length scale of interest in the problem, which is here the radius of the 
bubble.  
The mean effective gravitational acceleration felt by the flame front is $Ag$, which grows rapidly with radius from the center of the star \citep[see, e.g., Figure 9 in][]{Roepke_2006}.  
Therefore the mean buoyant force $Ag$ felt by a hot burning bubble increases rapidly with distance $r$ from the center of the star.  
For these bubbles, $S_\mathrm{eff}$, and therefore the nuclear burning rate generated, is much larger.  
Since we place the ignition points randomly within the confining spheres of radius \Rsph, simulations with the largest \Rsph\ have the most ignition bubbles at large $r$.  
Thus in simulations with the same \Nign\ but different \Rsph\, the burning rate increases more quickly for the simulation that has the larger \Rsph.  
This accounts for the higher burning rate in simulation \pdthree\ compared to simulation \pdtwo\ in the time interval $\sim0.2-0.5$ sec.

Finally, \fig{fig:burnedmass}(a) shows that, {\it at late times}, the nuclear burning rate declines soonest in the simulations for which the sphere is largest and latest in the simulations for which the sphere is smallest.  
The decline is due to the fact that, at late times, the majority of the ignition bubbles have merged in all of the simulations, forming a large, amorphous, burning bubble.  
When this large hot bubble reaches the outer regions of the star where the density is low, the burning is quenched and the burning rate rapidly declines.  
This happens soonest in the simulations with the largest \Rsph\ because these simulations have more ignition points far from the center of the star (and thus closer to the stellar surface), and the buoyant force felt by these ignition points is much greater, as discussed above.


Our discussion of \fig{fig:burnedmass} shows that the number of ignition points \Nign, the density of ignition points, and the radius of the confining sphere \Rsph\ all play a role in determining the amount of burned mass (and therefore the mass of \nifs\ and the peak luminosity of the supernova).  
In the six simulations of the PD model we performed, \Nign\ is the most important factor governing the outcome of the simulations.  
This is evident from the values of the final amounts of burned mass $M_\mathrm{b}$, mass of \nifs, nuclear energy released $E_{\mathrm{nuc}}$, and kinetic energy $E_{\mathrm{K}}$ in \tab{tab:final}.  
It will also be evident when we discuss the time evolution of these energies, and the distribution of chemical elements in the ejecta and the properties of the predicted light curves, below.

In fact, the results of the six simulations of the PD model we performed fall into two groups with respect to each of these properties:  the three simulations with a modest number of ignition points and those with many ignition points.  
However, it seems certain that simulations having intermediate numbers of ignition points (i.e., $N_\mathrm{ign}\sim150-1100$) would have properties that fill in the gap between these two groups.  
It is also clear that, had we been able to study a much larger range of possible ICs, the relative importance of \Nign, the density of ignition points, and \Rsph\ to the outcome of the simulations would vary greatly; and the outcome of the simulations would cover a much larger range in all of the properties, including the observed properties, of the resulting SNe Ia.

\subsubsection{Evolution of internal, gravitational, kinetic, and total energies}
\label{sec:results:evolutionall:energy}

\begin{figure*}
\mbox{
\includegraphics[trim = 0mm 22mm 0mm 0mm, clip, scale=0.35]{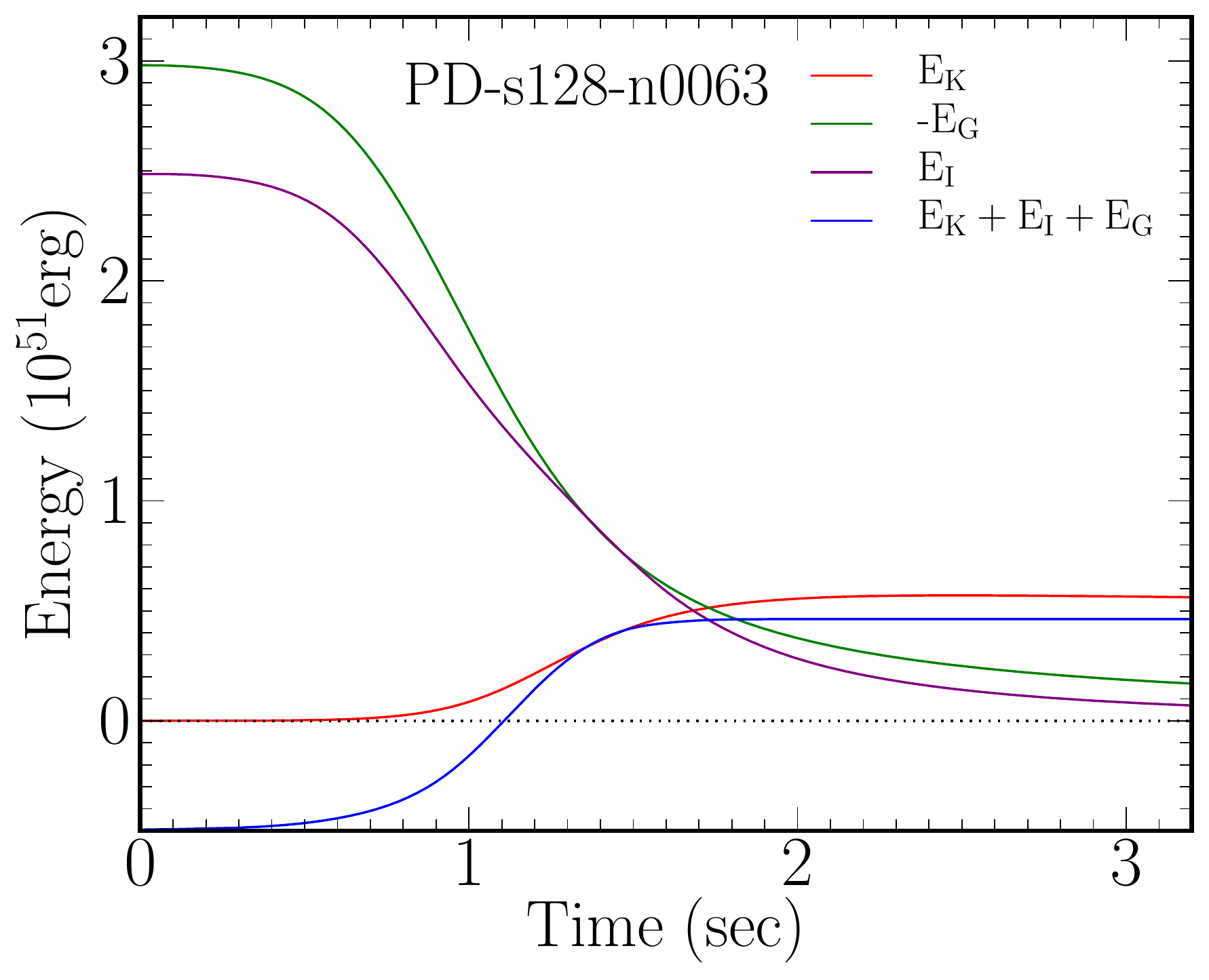}
\includegraphics[trim = 15mm 22mm 0mm 0mm, clip, scale=0.35]{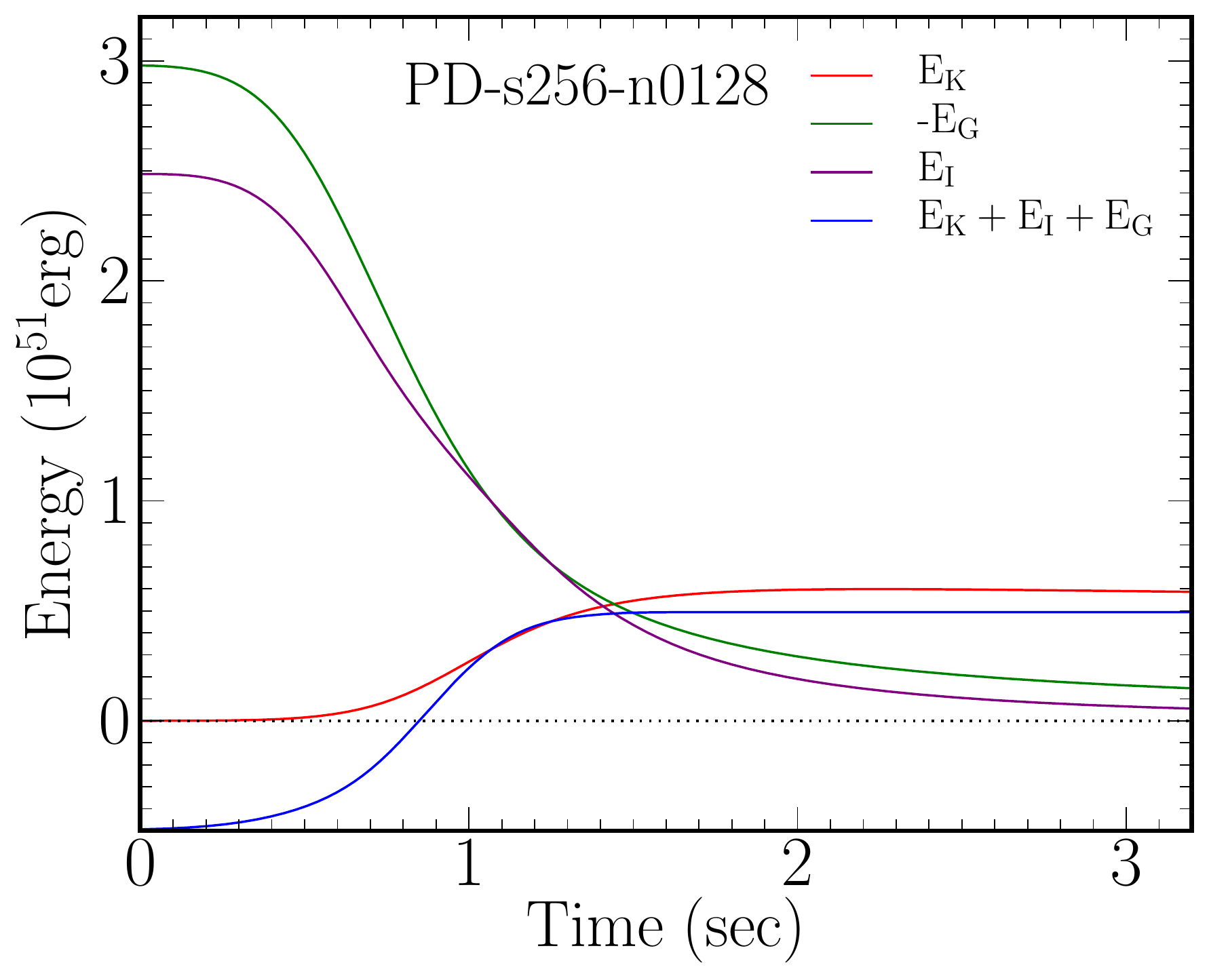}
\includegraphics[trim = 15mm 22mm 0mm 0mm, clip, scale=0.35]{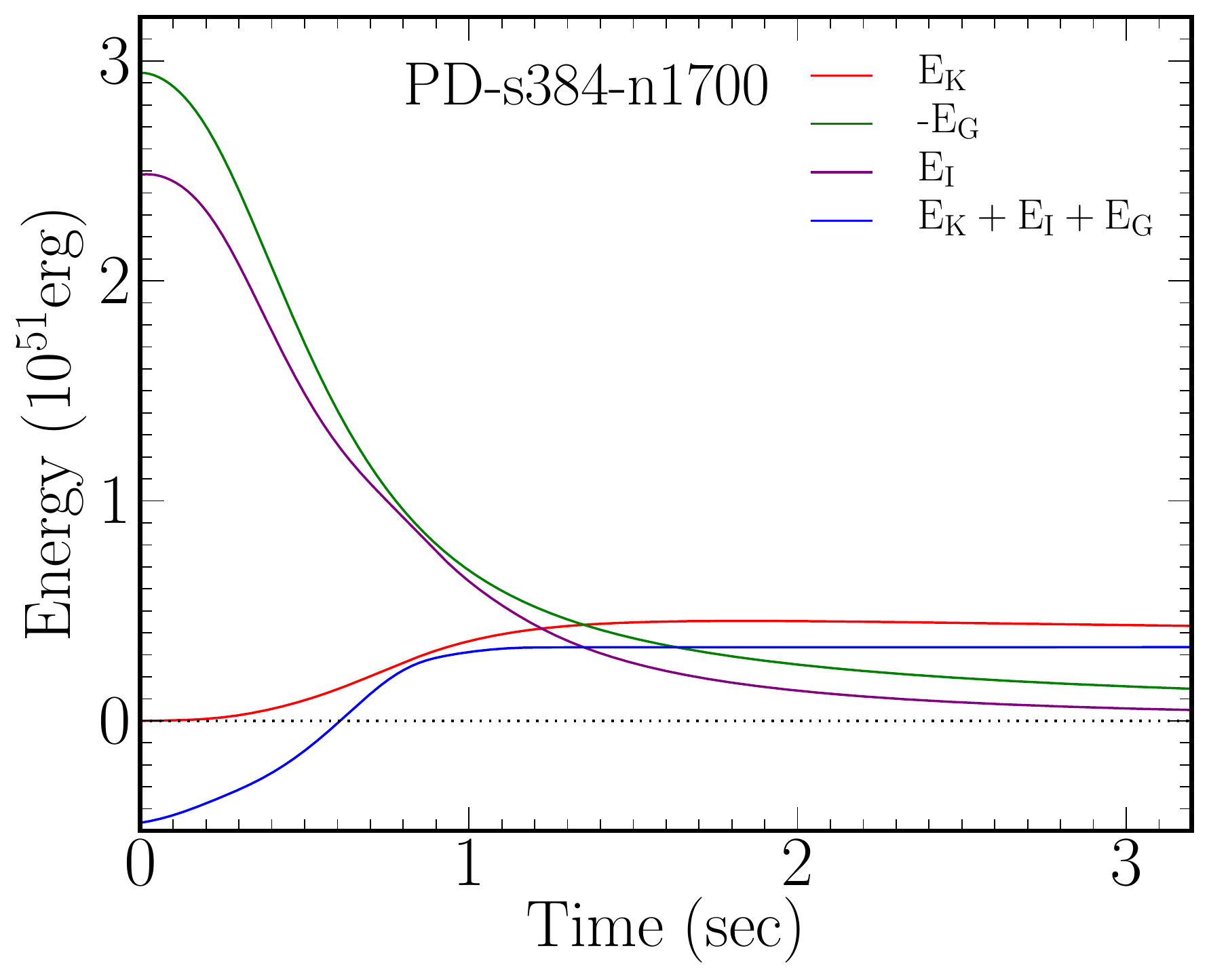}
} \\
\mbox{
\includegraphics[trim = 0mm 0mm 0mm 0mm, clip, scale=0.35]{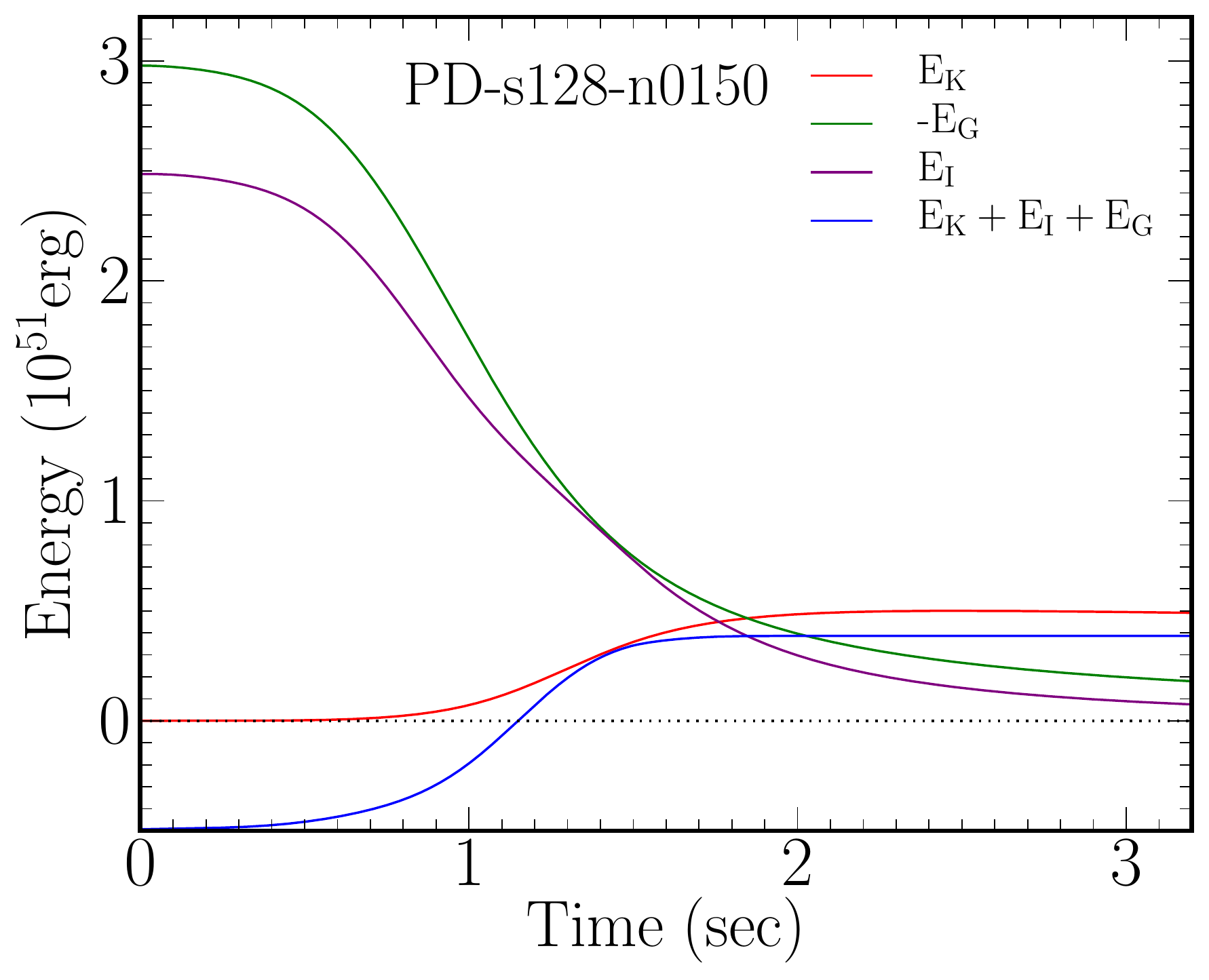}
\includegraphics[trim = 15mm 0mm 0mm 0mm, clip, scale=0.35]{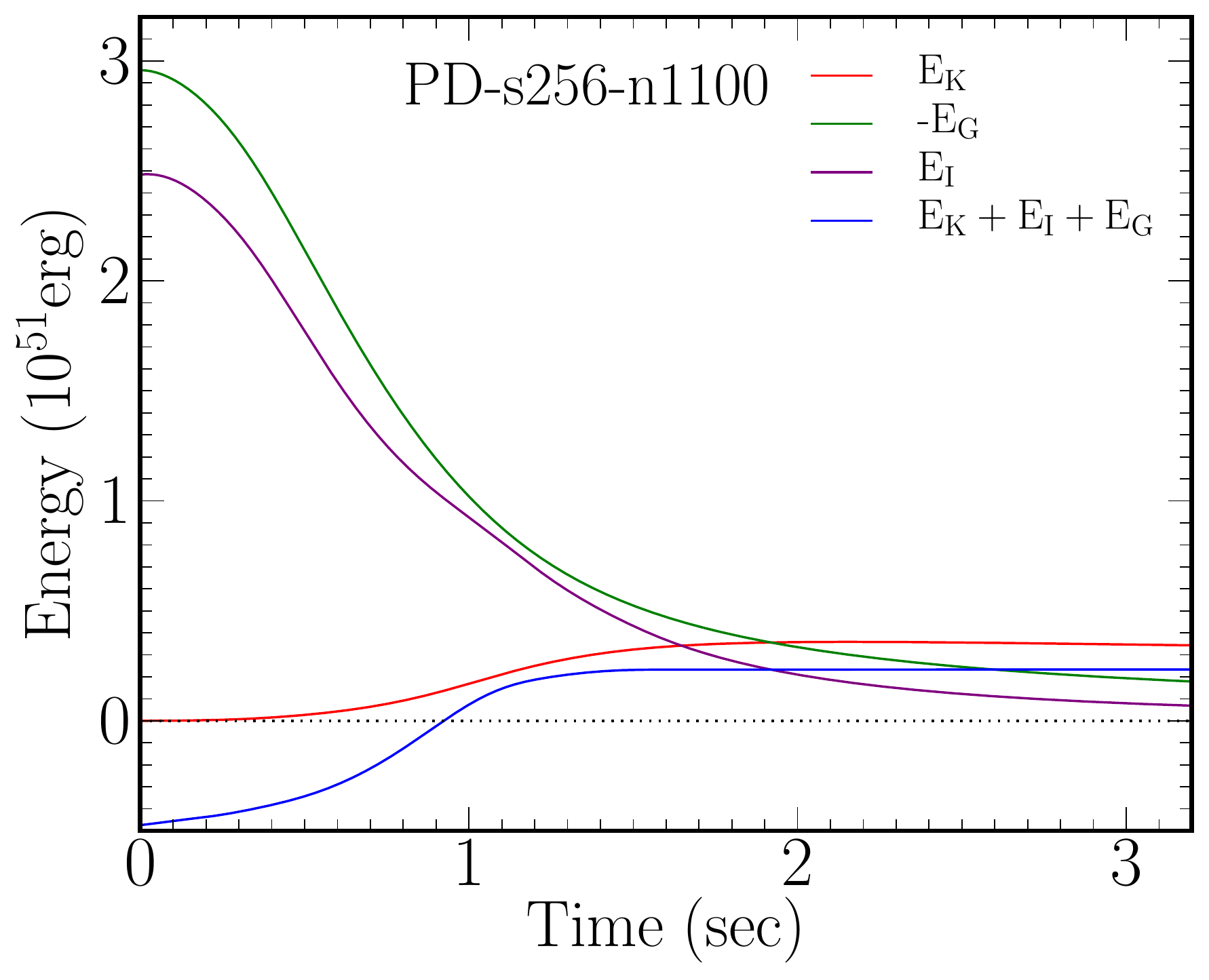}
\includegraphics[trim = 15mm 0mm 0mm 0mm, clip, scale=0.35]{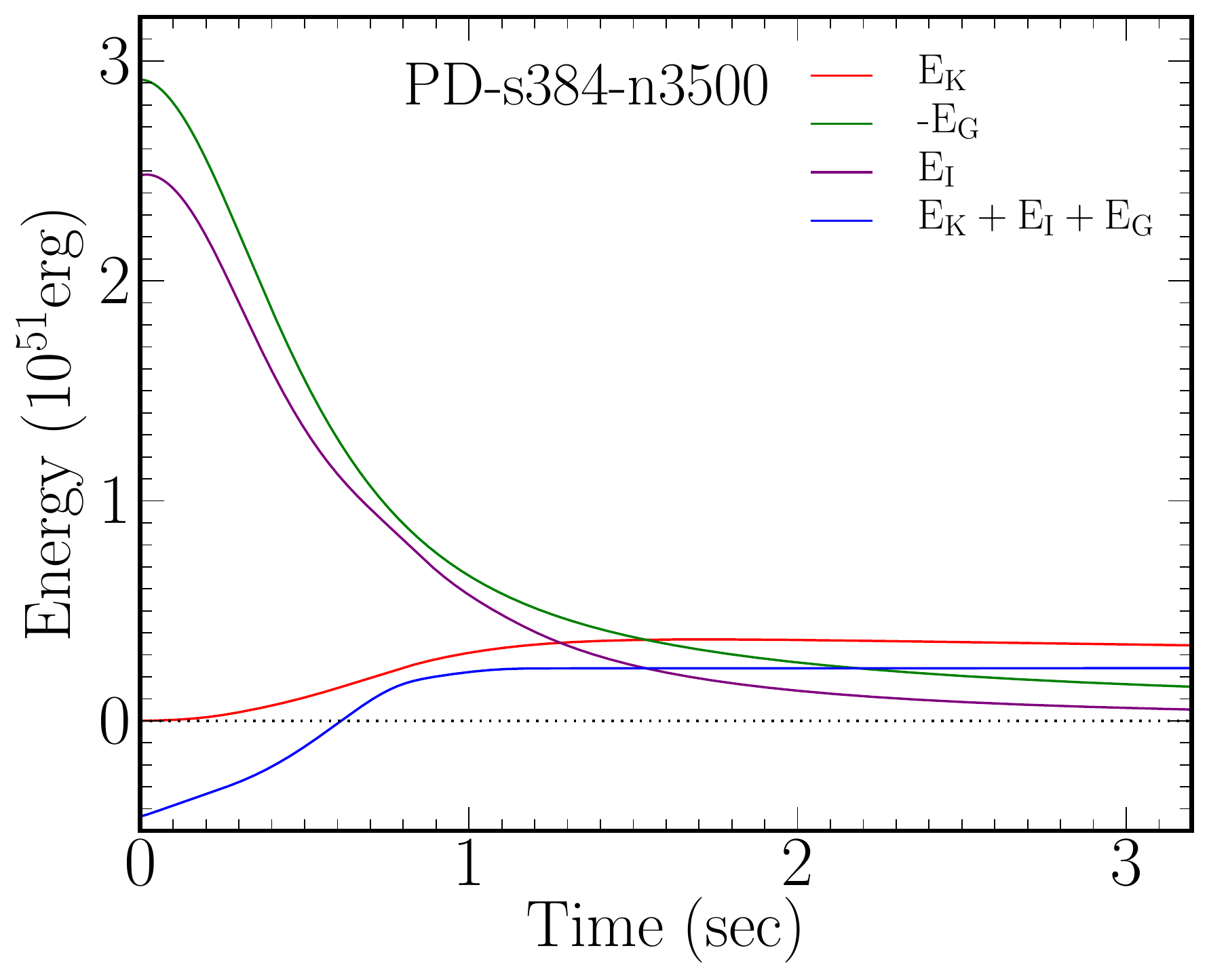}
} 
\caption{
Evolution of the internal energy, $E_\mathrm{I}$, the negative of the gravitational potential energy, $-E_\mathrm{G}$, and the kinetic energy, $E_\mathrm{K}$, for all six simulations. In the \Fsims, $E_\mathrm{I}$ and -$E_\mathrm{G}$ decrease slowly at first, reflecting the initially low burning and the low rate of expansion of the star in these simulations. The opposite is the case in the \Msims.
}
\label{fig:energy}
\end{figure*}

\fig{fig:energy} shows the time evolution of the kinetic, internal and gravitational potential energy for all six simulations. These results provide further support for the above discussion.  
In the \Fsims, the internal energy $E_\mathrm{I}$ and the negative of the gravitational potential energy $-E_\mathrm{G}$ decrease very little at first, demonstrating that these stars expand slowly in this phase.
However, at late times ($t >1.0$ sec), $E_\mathrm{I}$ and $-E_\mathrm{G}$ decrease rapidly.  
In contrast, in the \Msims, $E_\mathrm{I}$ and $-E_\mathrm{G}$ decrease quickly, showing that these stars expand rapidly at early times.

Similarly, in the \Fsims, the kinetic energy $E_\mathrm{K}$ and the total energy $E_\mathrm{total} = E_\mathrm{K} + E_\mathrm{I} + E_\mathrm{G}$ increase slowly, with $E_\mathrm{total}$ crossing zero at $t = 0.9 - 1.2$ sec, but their final values are large (i.e.,$E_\mathrm{K}=0.5-0.6\e{51}$ erg, $E_\mathrm{total}=0.4-0.5\e{51}$ erg).
In contrast,  in the \Msims, $E_\mathrm{K}$ and $E_\mathrm{total}$ increase rapidly, with $E_\mathrm{total}$ crossing zero at $t = 0.6-0.9$ sec, but their final values are small (i.e., $E_\mathrm{K}=0.3-0.4\e{51}$ erg, $E_\mathrm{total}=0.2-0.35\e{51}$ erg).

It is interesting to compare our \fig{fig:energy} with Figure 8 in \citet{Meakin_2009}, which shows the time evolution of $E_\mathrm{I}$, $E_\mathrm{G}$,  $E_\mathrm{K}$, and $E_{\rm total}$ in the GCD model, a delayed detonation model of SNe Ia. 
In the PD model, $E_\mathrm{I}$ and $-E_\mathrm{G}$ decrease more rapidly, and $E_\mathrm{K}$ and $E_{\rm total}$ are smaller at late times compared to the GCD model with its delayed detonation.

\subsection{Evolution of the nuclear flame}
\label{sec:results:evolutionall:flame}

\begin{figure*}
\centering
\includegraphics[width=6.2in]{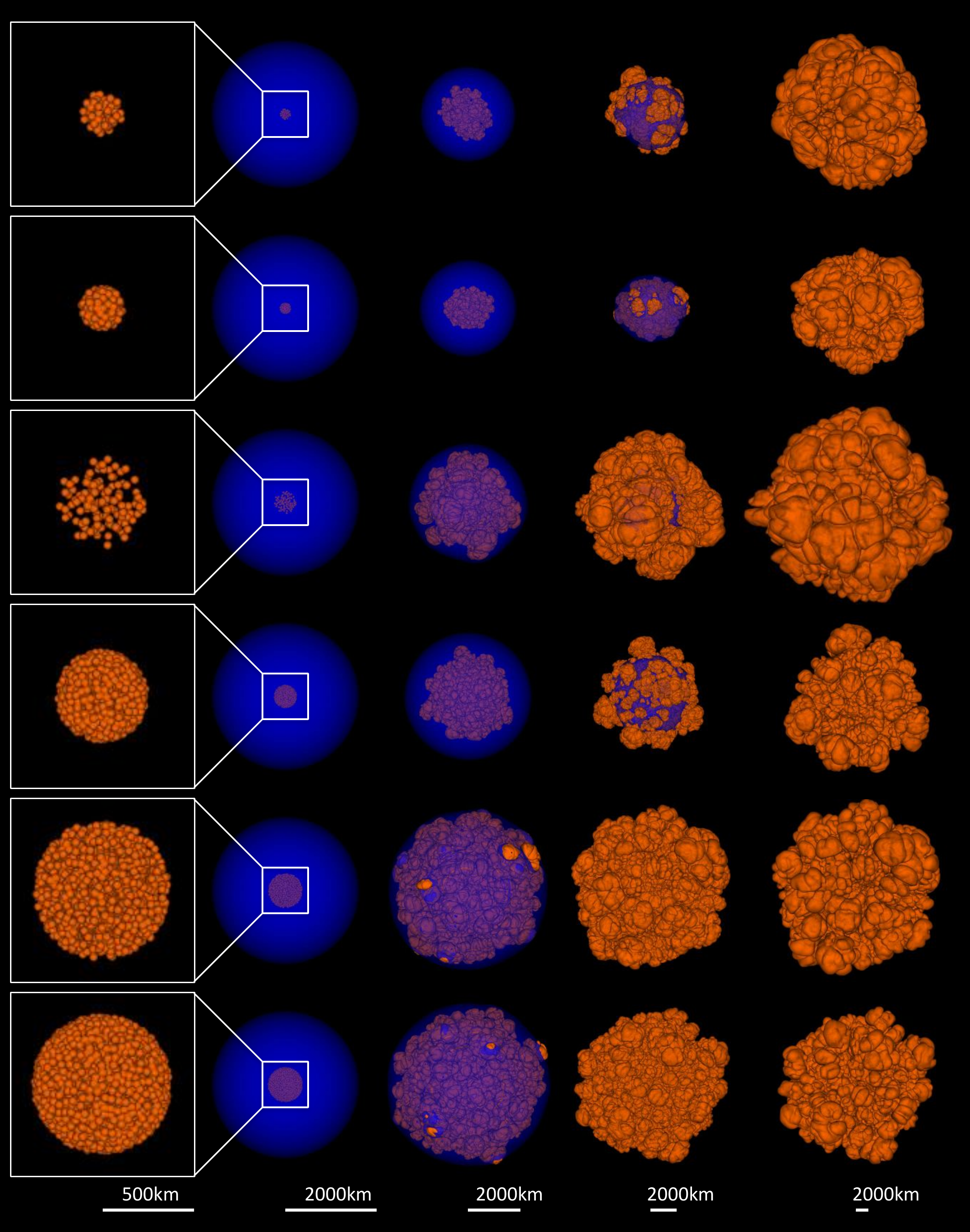}
\caption{
3D snapshots of the ignition points and nuclear flame fronts in all six PD simulations at times $t =$ 0.0, 0.8, 1.2 and 2.0 sec (from left to right). 
The order of the simulations from top to bottom corresponds to that in \tab{tab:init}. 
The left end of the legend bars indicates the position of the center of the white dwarf star.  
The blue contour shows the isosurface of density $1.0\e{7}$\gcm, the approximate density at which the nuclear flame quenches; the orange contour corresponds to the location in the nuclear flame front at which $^{12}$C burns.  
The figures shows that the white dwarf and the nuclear flame front expand at different rates, and that the nuclear flame front breaks through the stellar surface at different times, depending on the number of ignition points.}
\label{fig:flame}
\end{figure*}

\begin{figure*}
\centering
\includegraphics[width=6.2in]{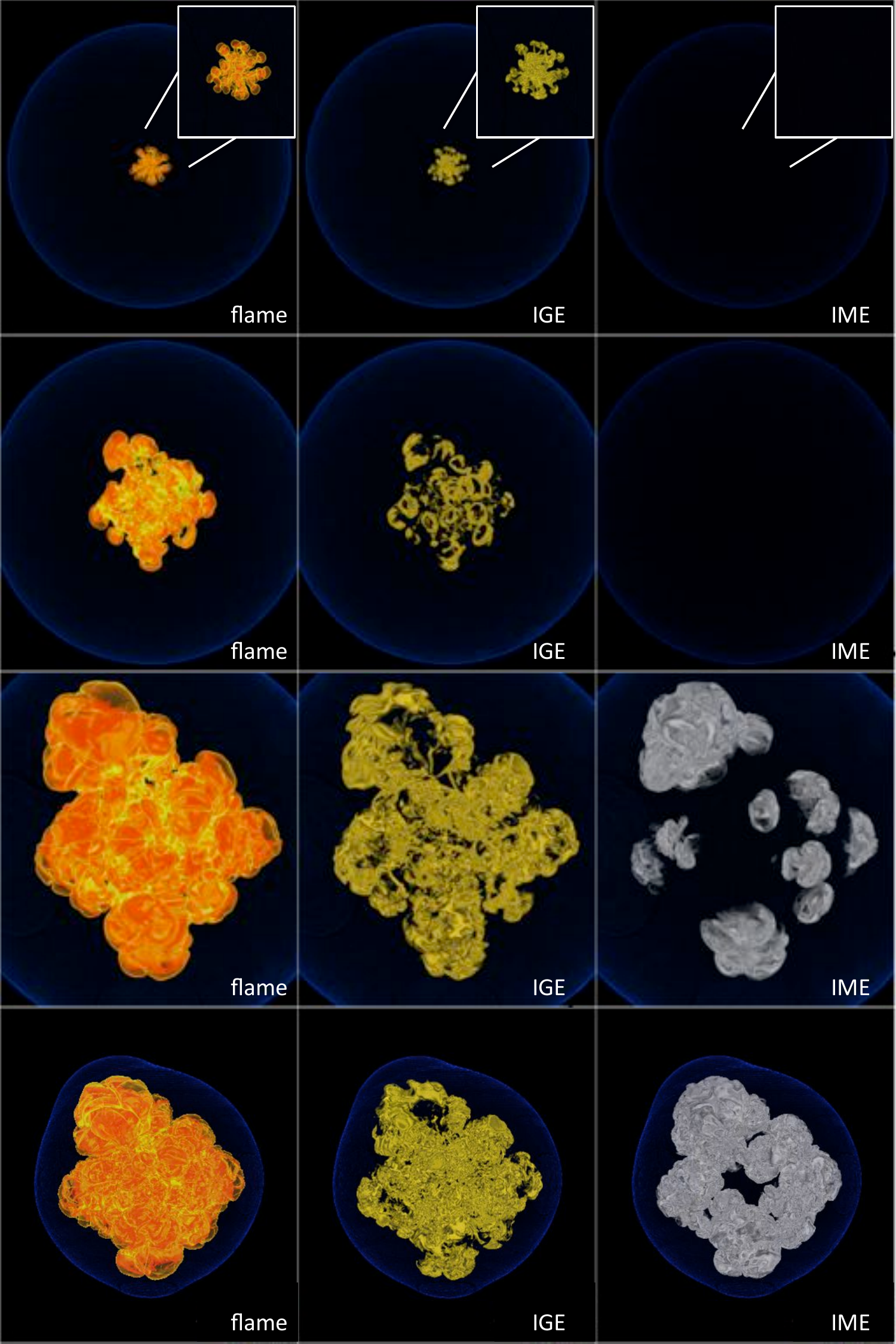}
\caption{
Volume-rendered images of the evolution of the nuclear flame front, and {the regions} where IGE and IME have been produced, at $t=$ 0.37, 0.80, 1.05, and 1.20 sec (from top to bottom) in the {\pdone} simulation.  
{The ratio of length scales for rows from top to bottom is, 1:1:1:2}.
The blue contour shows the isosurface of density $1.0\e{7}$\gcm (as in \fig{fig:flame}).
The blue surface corresponds to a density $1.0 \times 10^7$ g cm$^{-3}$ (as in Figure \ref{fig:flame}). 
A 3D animation is provided at http://flash.uchicago.edu/$\sim$long/PureDef/8km-PureDef-DDT-Compositions-1365x768.mp4 {or see the online version of this paper}.
}
\label{fig:volume}
\end{figure*}

\begin{deluxetable*} {ccrccccccc}
\tablecaption{Nucleosynthetic yields and energies for the six simulations listed in \tab{tab:init}. }
\tablehead{
\colhead{No.}  &
\colhead{Ignition Model} &
\colhead{$N_{\mathrm{ign}}$} &
\colhead{$R_{\mathrm{sph}}$ (km)} &
\colhead{$M_{\mathrm{b}}^\star(\Msun)$} &
\colhead{$M_{\mathrm{IGE}}^\dagger(\Msun)$}&
\colhead{$M^\dagger(^{56}\mathrm{Ni})/\Msun$} &
\colhead{$M^\dagger(\mathrm{Si})/\Msun$} &
\colhead{$E_{\mathrm{nuc}} (10^{51}$erg)}&
\colhead{$E_{\mathrm{K}} (10^{51}$erg)}
}
\startdata
1 & \pdone & 63 & 128 &0.748  & 0.377 & 0.276 &0.205 & 0.960 & 0.470\\
2 & \pdtwo & 150 & 128 &0.684 &0.353 &0.255 &0.172 & 0.882 & 0.394 \\
3 & \pdthree & 128 & 256 &0.769 &0.409 &0.288 &0.196 & 0.995 & 0.501\\
4 & \pdfour & 1100 & 256 &0.566 &0.263 &0.172 &0.127 & 0.717 & 0.244\\
5 & \pdfive & 1700 & 384 &0.622  &0.281 &0.187 &0.105 & 0.802 & 0.342\\
6 & \pdsix & 3500 & 384 &0.541 &0.212 &0.135 &0.085 & 0.678 & 0.248\\
\enddata
\tablecomments{
$^\star$ Final burned masses are derived using the reaction dynamics and energetics in the \texttt{FLASH} simulations.\\
$^\dagger$ Total IGE, \nifs, and Si masses are computed from the full nuclear network.
}
\label{tab:final}
\end{deluxetable*}

\fig{fig:flame} shows snapshots of the nuclear flame front in all six simulations listed in \tab{tab:init} from top to bottom, at times $t =$ 0.0, 0.8, 1.2, and 2.0 sec. The blue contour shows the isosurface of density $1.0\e{7}$\gcm, the approximate density at which the nuclear flame quenches; the orange contour is an isosurface of one of three reaction progress variables, corresponding to the location in the nuclear flame front at which $^{12}$C burns (see \csec{sec:models:reactive}).

The far-left images in \fig{fig:flame} show the different radii \Rsph\ of the confining sphere, and the different numbers \Nign\ of ignition points at the beginning of the six simulations.    
The second column shows the distribution of ignition points in the WD (blue contour).
Soon after the simulations begin, the nuclear flame corresponding to the surface of the hot, low-density bubbles becomes turbulent due to the interchange instability (see \csec{sec:results:evolutionall:burning}).  
As a result, the flame surface becomes strongly wrinkled, as is evident in the images of the flame front in the third, fourth and fifth columns.  
The rapid increase in the area of the flame surface, as well as the effective burning speed, causes the nuclear burning rate to increase rapidly.  
In the simulations with high densities of ignition points (i.e., with larger values of $f$ in \tab{tab:init}), the rapidly expanding, burning bubbles merge sooner than those in the simulations with low densities (see \csec{sec:results:evolutionall:burning}).  
{An additional factor is the larger number of ignition points at large radii in models with larger confining radii \Rsph, which decreases the rate of burning and leads to a relatively slower expansion and a smaller burning front at late times (e.g., at $t>0.8$ sec) (see \csec{sec:results:evolutionall:burning}).}

\fig{fig:flame} illustrates the dependence of the behavior of the nuclear burning on the number \Nign\ of ignition points that we discussed in \csec{sec:results:evolutionall}.  
The radii of the white dwarfs at 0.8 sec in \fig{fig:flame}, as reflected by the blue surfaces, are in one-to-one correspondence with the amount of burned mass $M_\mathrm{B}$ at 0.8 sec (\fig{fig:burnedmass}), and with the amount by which $E_\mathrm{I}$ and $-E_\mathrm{G}$ have decreased at this time (\fig{fig:energy}).  

Furthermore, the visibility of the blue surface at 1.2 sec (the isosurface of the density where the nuclear flame quenches), is also in one-to-one correspondence with $M_\mathrm{b}$ and the amount by which $E_\mathrm{I}$ and $-E_\mathrm{G}$ have decreased at this time (\fig{fig:energy}). 
This correspondence emphasizes the important role of the slower expansion of the star in burning more matter, and thus producing more {\nifs} which results in higher peak luminosities.

Finally, the relative growth of the radius of the flame surface between 1.2 and 2.0 sec in the images in the fifth (far-right) column in \fig{fig:flame} are in one-to-one correspondence with the relative increase at late times of the burned mass $M_\mathrm{B}$ in \fig{fig:burnedmass}, and the kinetic energy $E_\mathrm{K}$ and total energy in \fig{fig:energy}.

\subsection{Evolution of nucleosynthetic yields and their spatial distribution}
\label{sec:results:evolutionall:yields}

In this section, we discuss the overall nucleosynthetic yields and the spatial distribution of IGE, IME, and unburnt C/O in the ejecta.

\subsubsection{Evolution of production of iron-group and intermediate-mass elements}
\label{sec:results:evolutionall:yields:volume}

\fig{fig:volume} shows volume-rendered snapshots of the evolution of the nuclear flame front, and of {the regions} where IGE and IME have been produced, in the {\pdone} simulation. 
{The left images show the surface on which the reaction progress variable corresponding to Carbon burning is 0.5, and the center and right images show the regions where the mass fraction of IGE produced by the flame exceeds 0.75 and the mass fraction of IME produced by the flame lies between 0.05 and 0.20.
The left column therefore shows the evolution of the nuclear flame, and the center and right columns show the evolution of the regions where IGE and IME have been produced. }
From top to bottom, the rows refer to snapshots at 0.37, 0.80, 1.05, and 1.20 sec; the second and fourth of these times correspond to the second and third times shown in \fig{fig:flame}.

The left image at 0.37 sec shows that the initial hot, low-density bubbles have become unstable due to the interchange instability, and formed mushroom-shaped plumes. 
The center image shows that IGE are being created in these plumes; the mushroom shape of the flame regions creating the IGE is evident. 
The plumes are rising rapidly, leaving cold, dense, unburnt C/O behind.  
At the same time, cold C/O is flowing into the core of the star (this is more evident in the animation in the caption in \fig{fig:volume}). 
At 0.80 sec, the nuclear flame front has become more turbulent (left), and IGE are being created at the caps of the mushroom-shaped burning regions (center). 
By this time, virtually no IME have been created yet (right).

At 1.05 sec, the nuclear flame front has become increasingly turbulent, as have the regions where IGE are being created. 
It is important to note that no IGE are being created in the core of the star which remains mostly unburnt C/O. 
This behavior is more evident in an animation provided on our website (see the caption in \fig{fig:volume}).  
IME are now created in the large regions of the burning plumes at lower densities.

The snapshots at 1.20 sec correspond to the time when nucleosynthesis is ending. 
The length scale in this row of images is twice that in the other three rows in order to show the entire star, which has expanded significantly. 
The nuclear flame and the IGE producing regions have become fully turbulent, with the core of the star still remaining mostly unburnt C/O.

\subsubsection{Nucleosynthetic yields}
\label{sec:results:evolutionall:yields:total}

\begin{figure*}
\centering
\mbox{
\includegraphics[trim = 0mm 15mm 0mm 0mm, clip, scale=0.45]{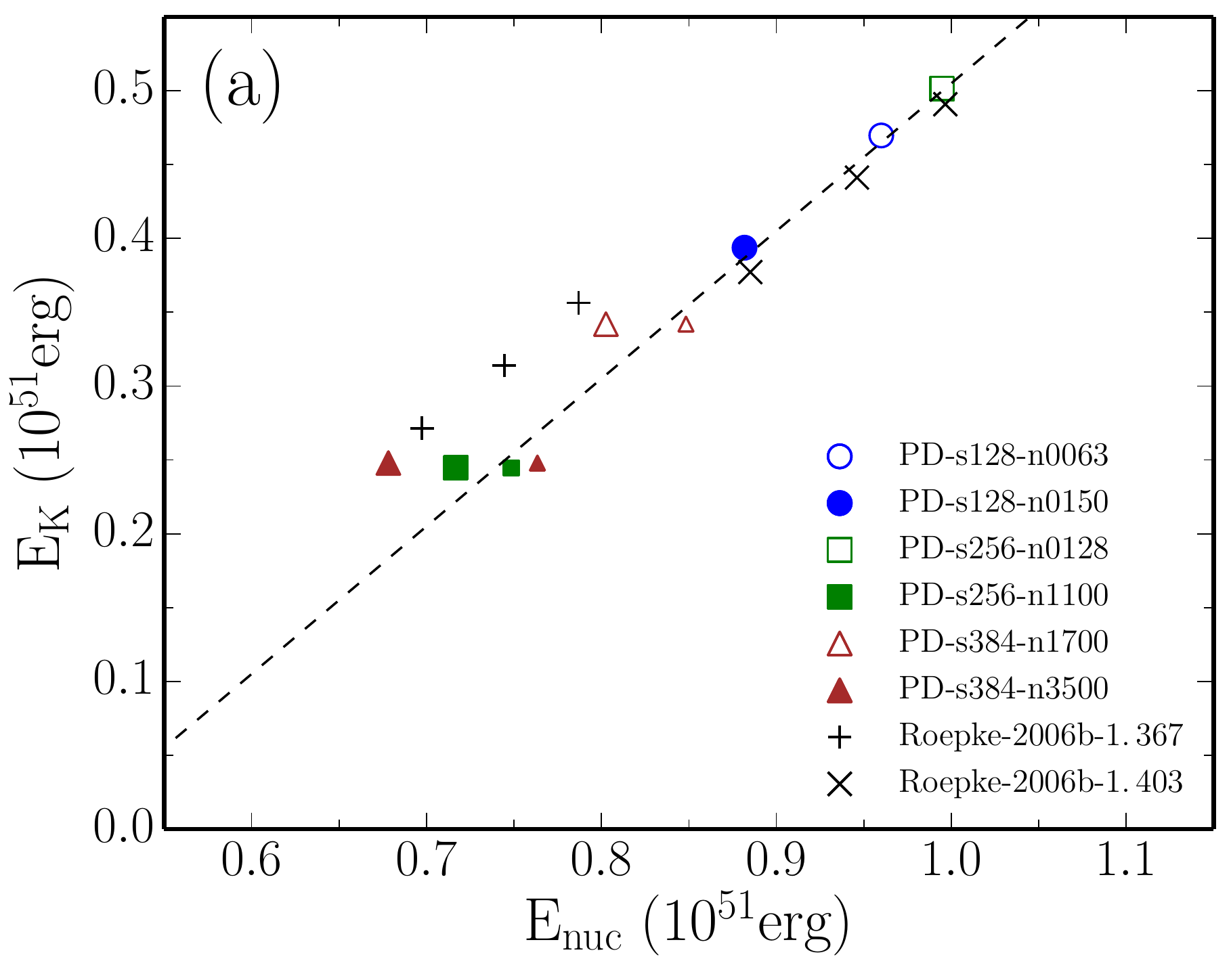}
\includegraphics[trim = 0mm 15mm 0mm 0mm, clip, scale=0.45]{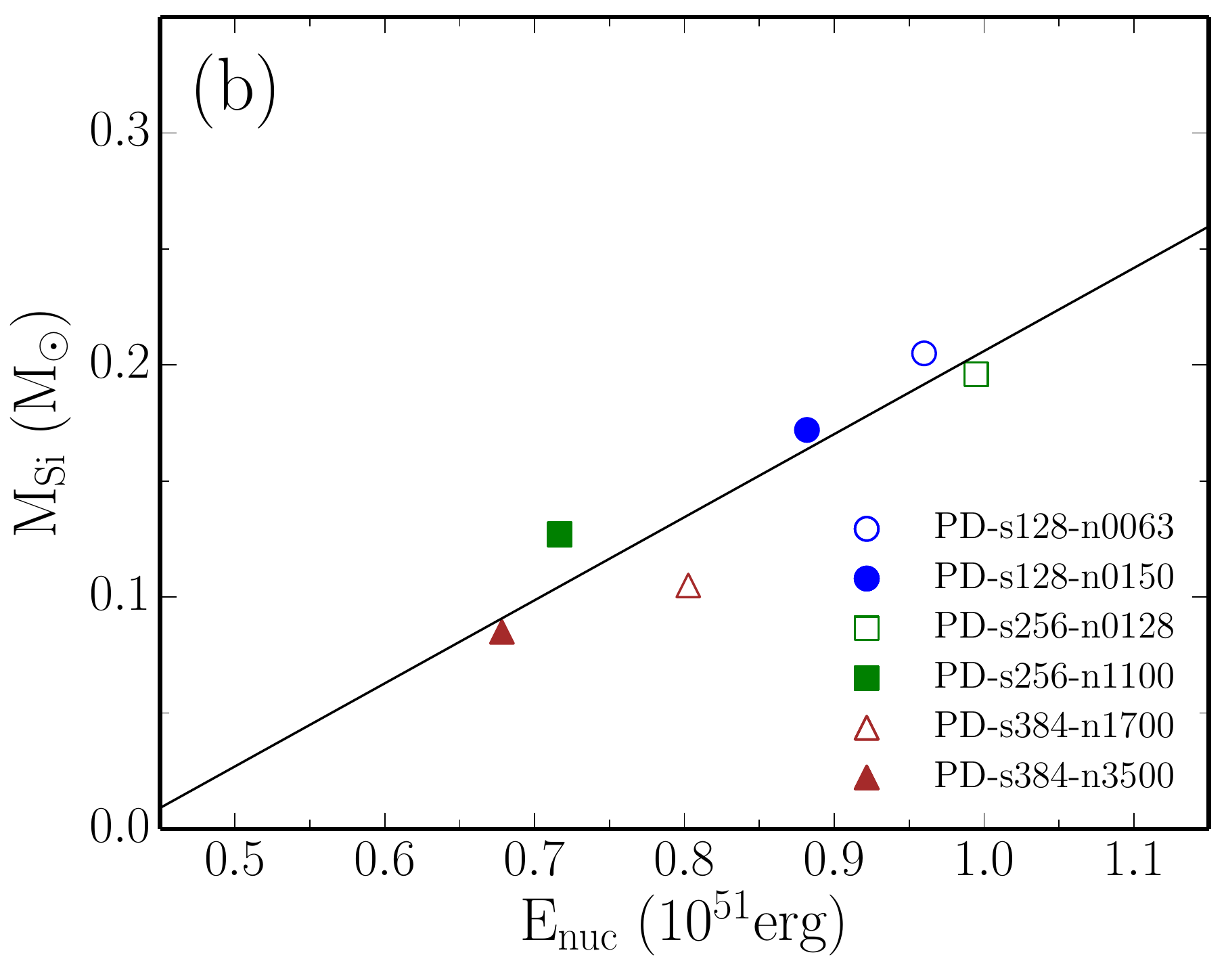}
}
\mbox{
\includegraphics[trim = 0mm 0mm 0mm 0mm, clip, scale=0.45]{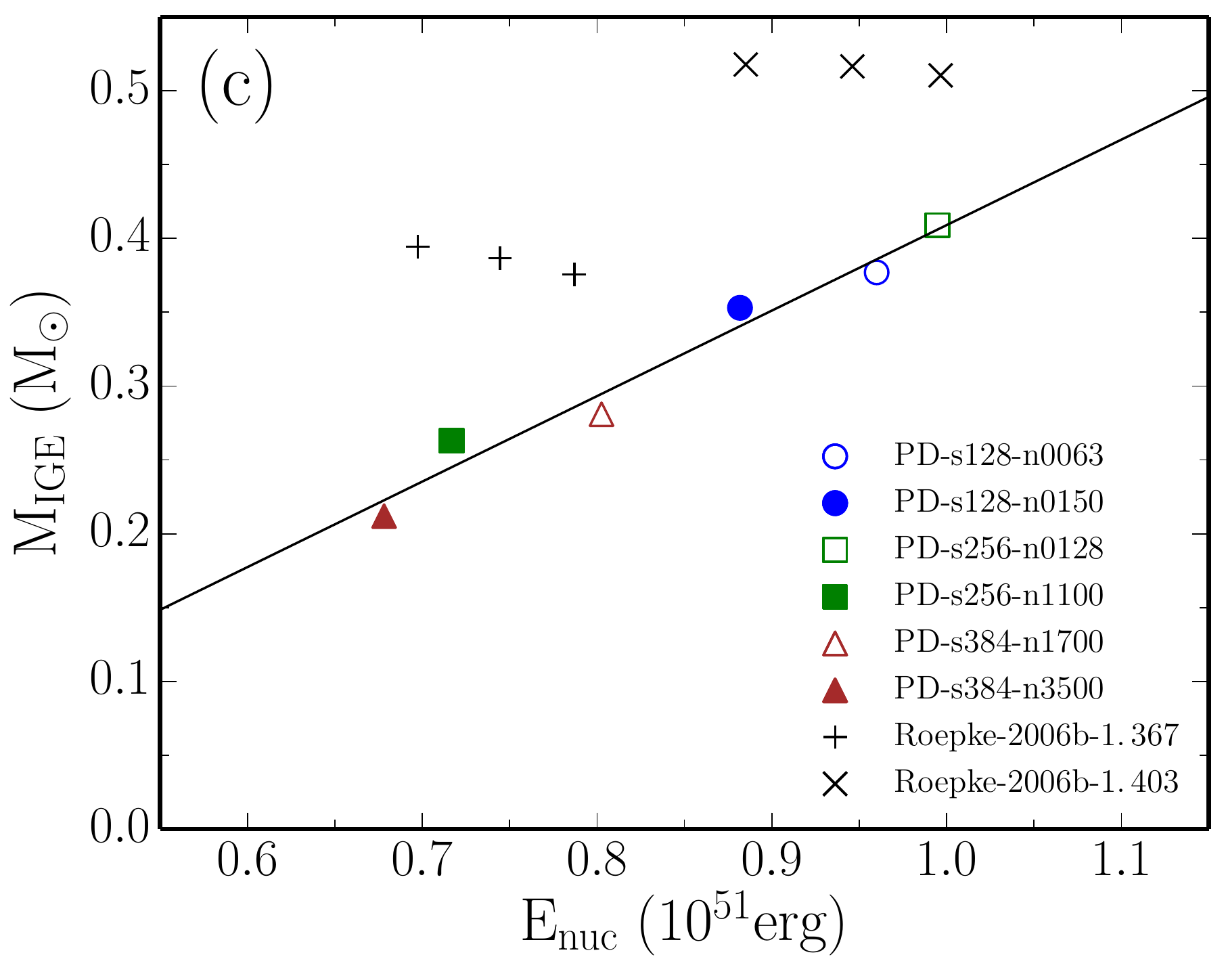}
\includegraphics[trim = 0mm 0mm 0mm 0mm, clip, scale=0.45]{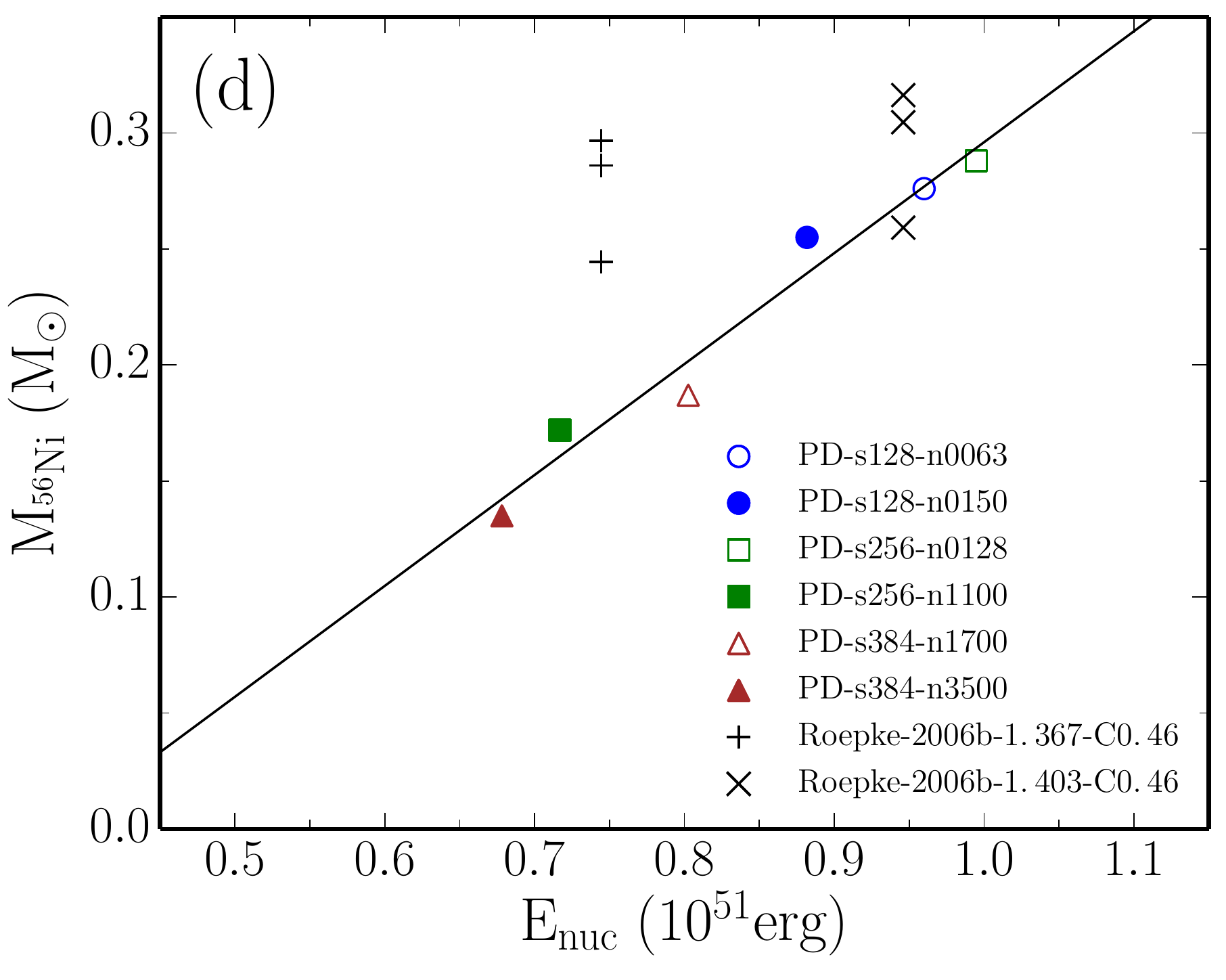}
}
\caption{
(a) Kinetic energy $E_\mathrm{K}$,  (b) Si,  (c) IGE,  and (d) \nifs masses as a function of released nuclear energy ($E_\mathrm{nuc}$). 
Circles, squares and triangles denote the three sets of simulations with \Rsph$=$128, 256 and 384 km respectively.
Filled symbols represent simulations with nearly the largest possible \Nign\ (see \csec{sec:models:input}) while empty symbols are not. 
Black crosses represent the two sets of simulations from Table 3 in \citet{Roepke_2006a} with a 1.367 or 1.403 $\Msun$ white dwarf.
In panel (a), the three smaller symbols represent estimates of the nuclear energies for \Msims\ that include the nuclear energy released by the initial bubbles assuming they had burned to \nifs; the estimates for the other simulations are not shown as the differences are negligible. 
In panel (d), the black crosses represent simulations from \citet{Roepke_2006a} that used an initial Carbon mass fraction of 0.46 with three different initial metalicities and two different white dwarf masses.
The solid lines show linear fits to the results from our six simulations (see \tab{tab:fit}).
The dashed line represents $E_\mathrm{K}=E_\mathrm{nuc}-0.496$.
}
\label{fig:em}
\end{figure*}

\begin{figure*}
\centering
\mbox{
\includegraphics[trim = 0mm 15mm 0mm 0mm, clip, scale=0.45]{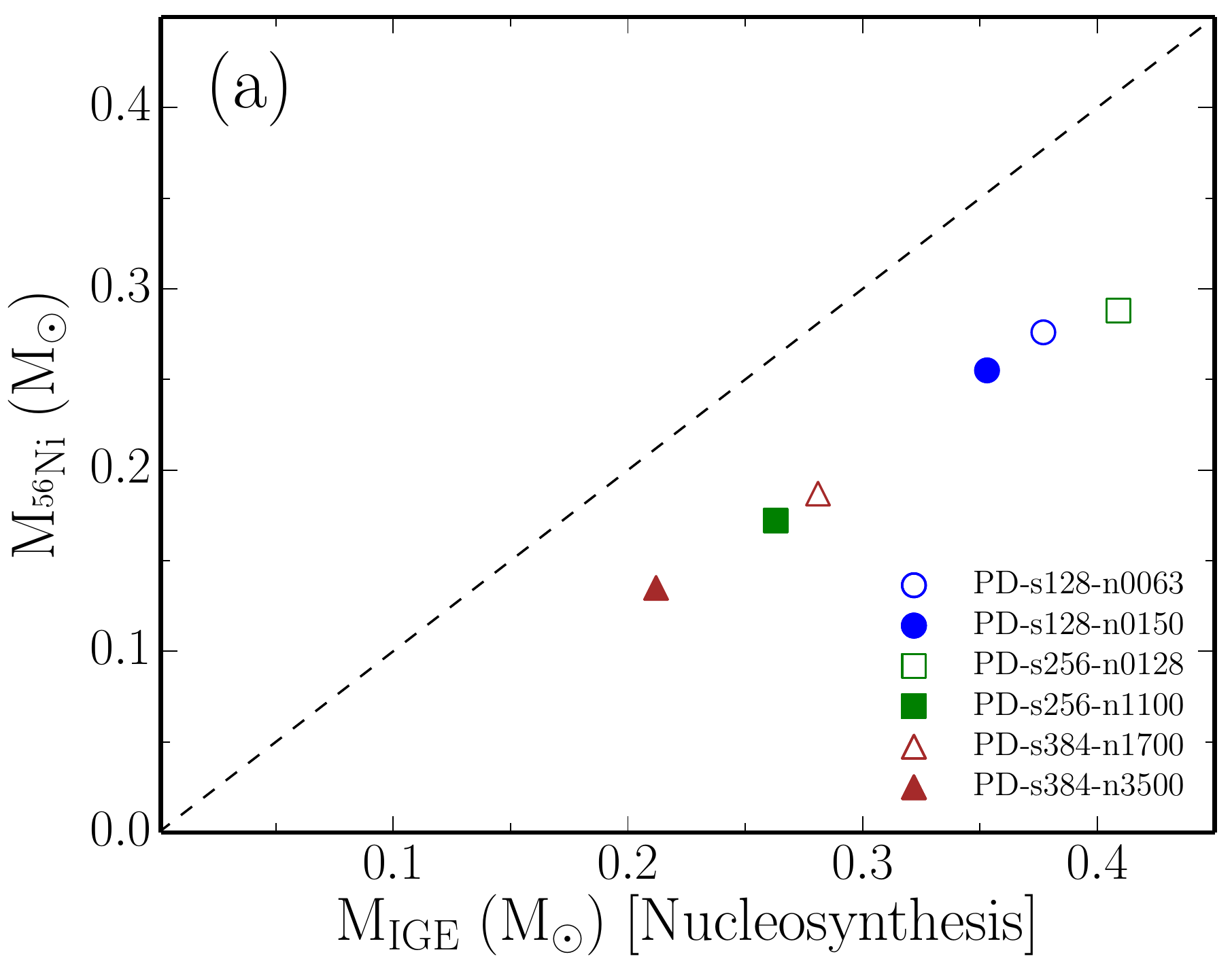}
\hspace{0.3cm}
\includegraphics[trim = 0mm 15mm 0mm 0mm, clip, scale=0.45]{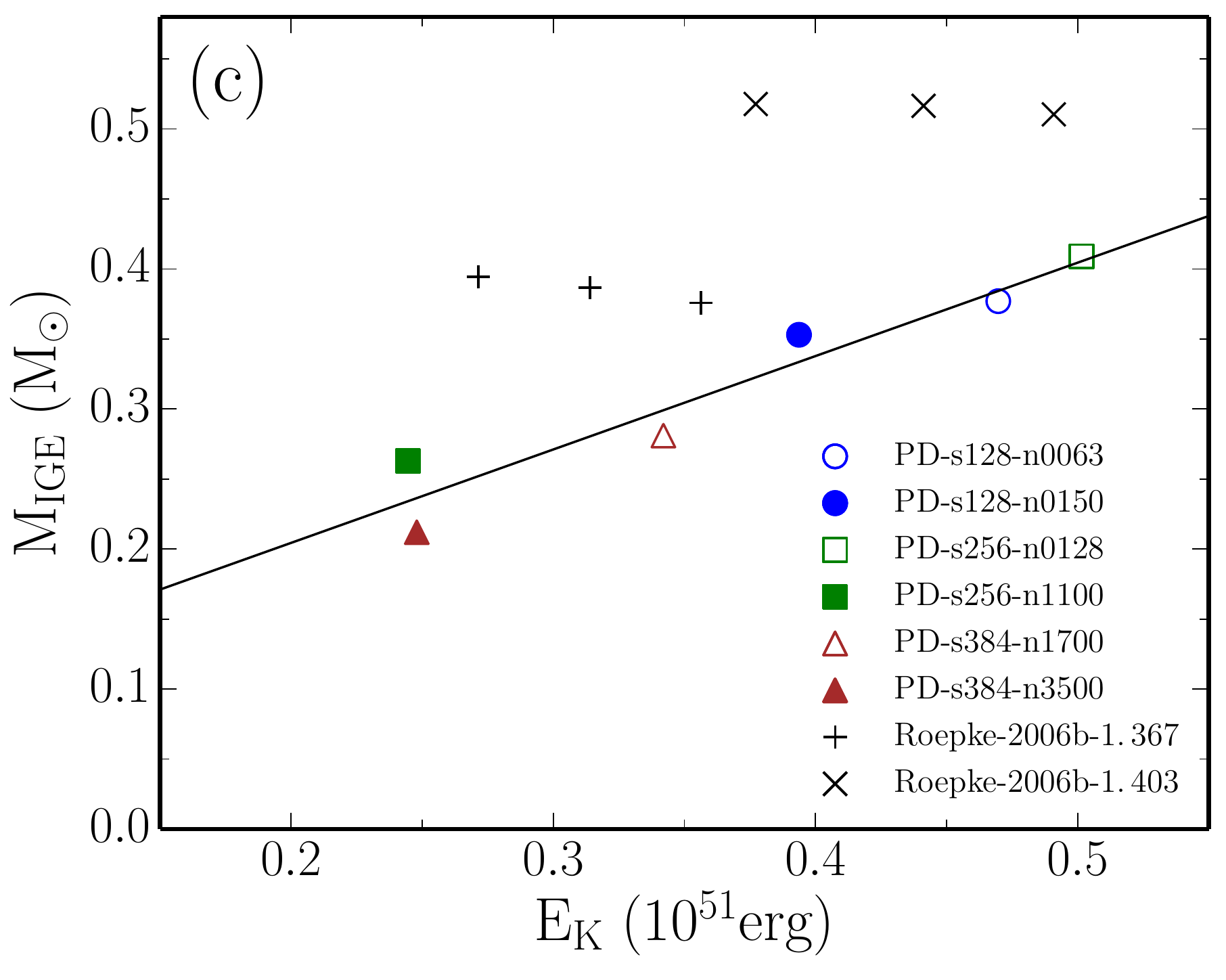}
}
\mbox{
\includegraphics[trim = 0mm 0mm 0mm 0mm, clip, scale=0.45]{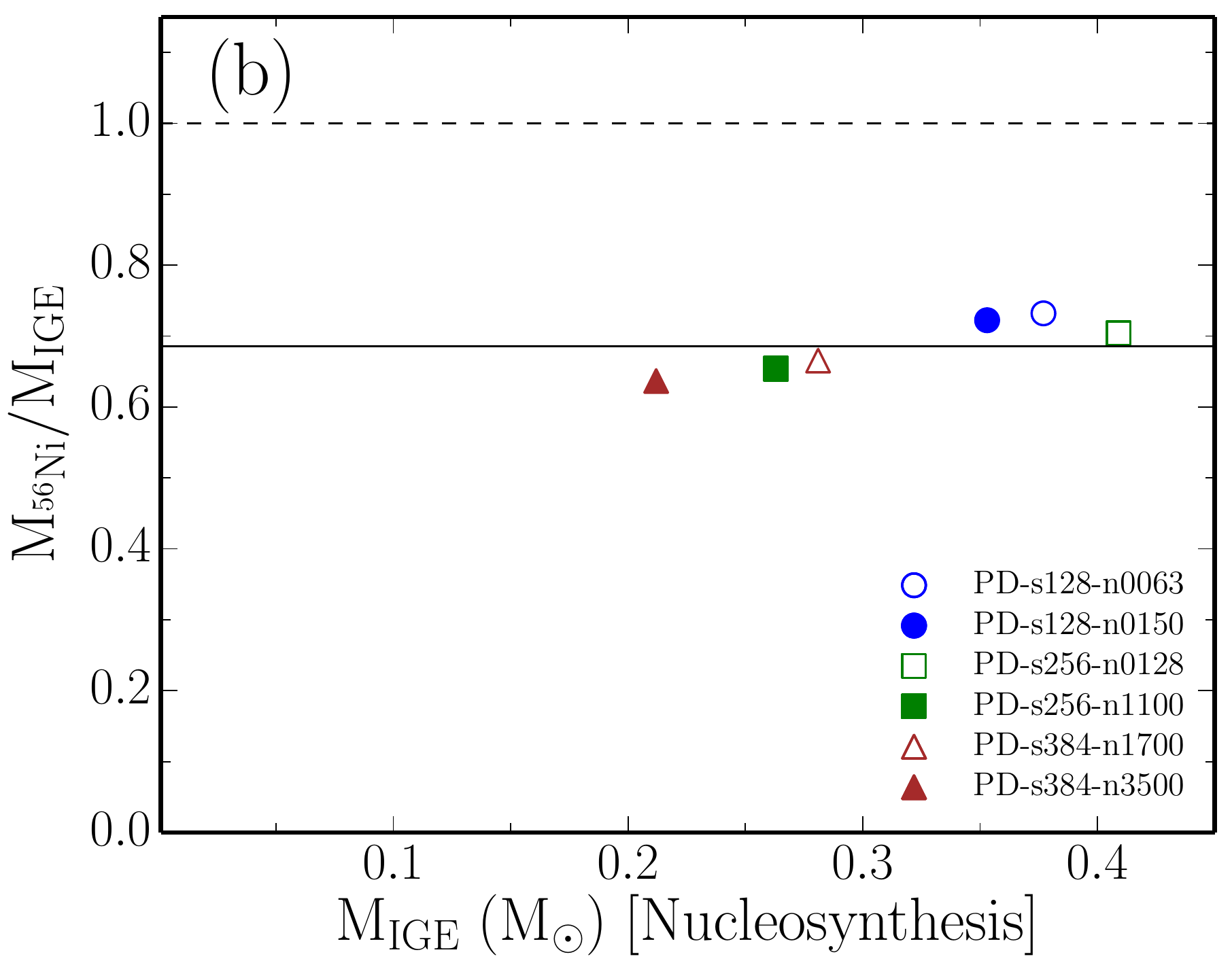}
\hspace{0.3cm}
\includegraphics[trim = 0mm 0mm 0mm 0mm, clip, scale=0.45]{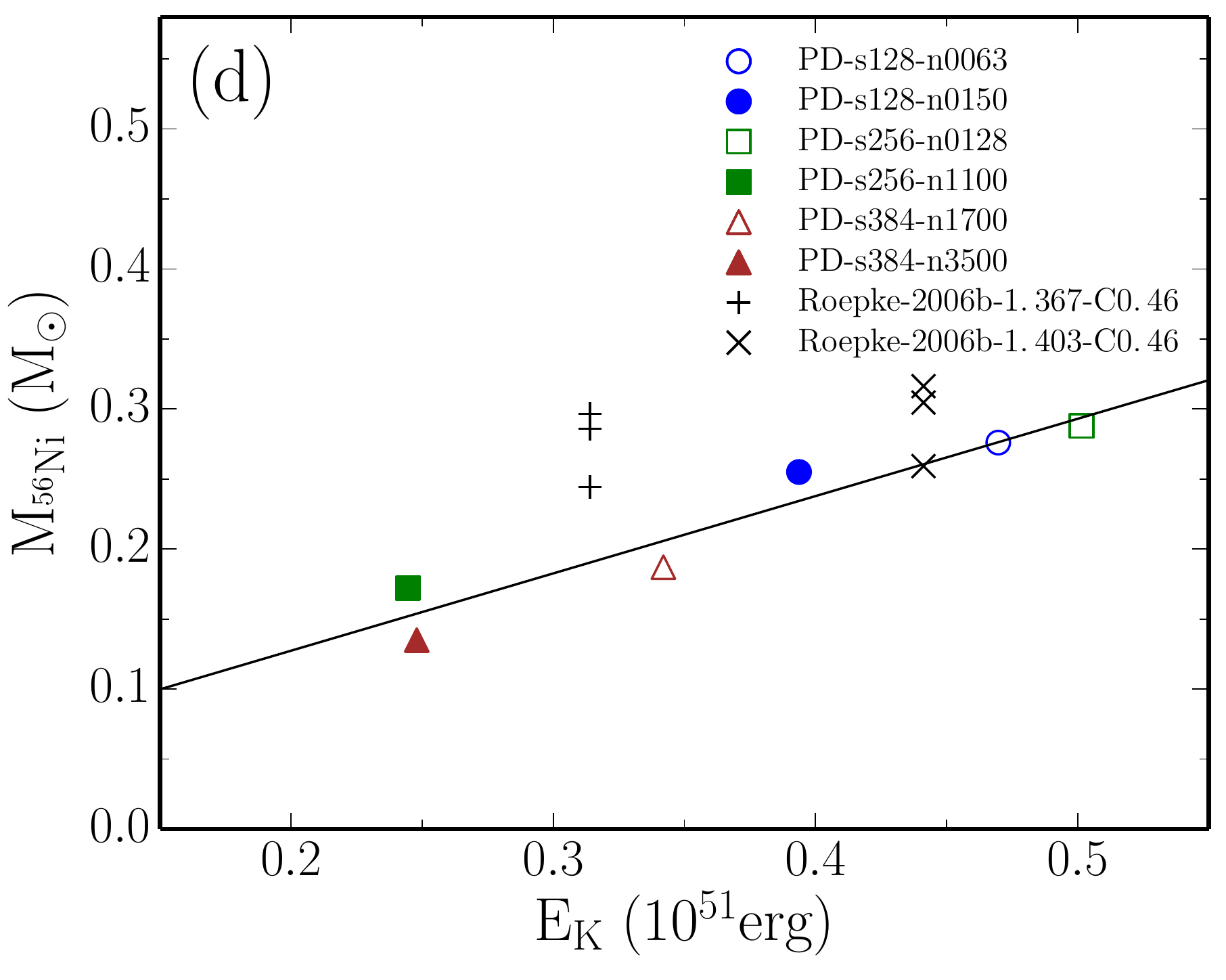}
}
\caption{
(a) \nifs mass as a function of IGE mass;
(b) fraction of \nifs max in IGE mass; 
(c) IGE mass as a function of kinetic energy $E_{\mathrm{K}}$, symbols have the same meaning as in \fig{fig:em}(c);
(d) \nifs mass as a function of $E_{\mathrm{K}}$,  symbols have the same meaning as in \fig{fig:em}(d). 
The solid lines show linear fits to the results from our six simulations (see \tab{tab:fit}). 
The dashed lines represent the relation $M_{56_{\mathrm{Ni}}}=M_{\mathrm{IGE}}$.
}
\label{fig:Mek}
\end{figure*}

\begin{deluxetable} {lcccccc}
\tablecaption{Best-fit parameters to data in \fig{fig:em}(b)-(d) and \fig{fig:Mek} in the equation $M=mE+M_0$ }
\tablehead{
\colhead{ }  &
\colhead{Fig. 7b} &
\colhead{Fig. 7c} &
\colhead{Fig. 7d} &
\colhead{Fig. 8c}&
\colhead{Fig. 8d} 
}
\startdata
$m(\Msun/E_{51})$        & 0.358    & 0.578     & 0.478     &0.667     & 0.552  \\
$M_0(\Msun)$                & -0.152   & -0.170   & -0.182    &0.071      &0.017   \\
$m/M_0(E_{51}^{-1})$    & -2.35     & -3.41      &-2.62       &9.37       &33.2    \\
\enddata 
\label{tab:fit}
\end{deluxetable}

As discussed in \csec{sec:models:nuclear} and \csec{sec:models:reconstruction}, we post-processed 10 million Lagrangian tracer particles to determine the 3D spatial distribution of the nucleosynthetic yields. 
A sub-sample of 20,000 of these particles was used for the 1D RT calculations with \texttt{PHOENIX}.  
This sub-sample was also used for the nucleosynthetic yields given in \tab{tab:final}, which lists the total burned mass $M_\mathrm{B}$; the masses of IGE, {\nifs} and Si; and the final nuclear energy $E_\mathrm{nuc}$ and kinetic energy $E_\mathrm{K}$ in each of the six simulations.  
The simulations produce $(0.135 - 0.288) \Msun$ of {\nifs} and $(0.085-0.205) M_\sun$ of Si; release $(0.678-0.995)\e{51}$ erg of nuclear energy and produce $(0.244-0.501)\e{51}$ erg of kinetic energy.  
All of these values are considerably smaller than those for normal SNe Ia, as expected, and are more typical for under-luminous SNe Ia. We discuss the possible relevance of the simulations to the latter in more detail below.

\figs{fig:em} and \ref{fig:Mek} show that strong correlations exist among the quantities $E_\mathrm{nuc}$, $E_\mathrm{K}$, $M_\mathrm{IGE}$,  $M_{^{56}\mathrm{Ni}}$, and $M_\mathrm{Si}$.  
We briefly discuss each of these correlations in turn. \tab{tab:fit} shows the best-fit parameters for straight-line fits to the data in some of these figures.

\fig{fig:em}(a) shows that the nuclear energy released is proportional to the final kinetic energy in the ejecta.  
This is a direct consequence of the equation $E_\mathrm{K} = (E_\mathrm{I}+ E_\mathrm{G} )_i + E_\mathrm{nuc}$, and of the initial thermal and potential energies $(E_{\mathrm I}+ E_\mathrm{G} )_\mathrm{i} $ which are constant, $-0.496 \times 10^{51}$ ergs (see also \fig{fig:energy}), since the white dwarf progenitor is the same in all simulations.  
All of the simulations release slightly less nuclear energy than required by this equation because the nuclear energy released in the formation of the initial bubbles is not included.  
This effect is more noticeable in the \Msims, since the masses in the initial bubbles are greater.  
Correcting these simulations for the missing nuclear energy moves them to the right, and all three then lie very close to the required relation between $E_\mathrm{K}$ and $E_{\rm G}$.

\figs{fig:em}(b), (c), and (d) show that the masses of Si, IGE, and {\nifs} produced by the simulations are linearly proportional to the nuclear energy released in the simulations.  
The difference between the slopes of the first and the latter two reflects the fact that less nuclear energy is released when C/O burns to Si than to IGE and {\nifs}. 
The difference between the latter two is a result of the electron captures that occur when burning to IGE occurs at very high densities.

The values of $E_\mathrm{nuc}$ in these figures are affected by the fact that the simulations do not include the nuclear energy released in the formation of the initial bubbles, as mentioned above.   
On the basis of the corrections we estimated in \fig{fig:em}(a), the slope in \fig{fig:em}(b) is affected, but not a great deal.
However, the masses of IGE and {\nifs} are also affected, so that the slopes in \figs{fig:em}(c) and (d) are hardly affected at all. 

\fig{fig:Mek}(a) shows that $M_{^{56}\mathrm{Ni}}$ is proportional to $M_\mathrm{IGE}$, but that the slope is not quite unity.  
The reason is that the simulations that burn less mass do so at higher densities, and thus produce less {\nifs} due to electron captures, while those that burn more mass do so at somewhat lower densities, and thus produce more {\nifs}.  
\fig{fig:Mek}(b), which shows the ratio $M_{^{56}\mathrm{Ni}}/M_\mathrm{IGE}$ as a function of $M_\mathrm{IGE}$, demonstrates this more clearly.  
The \Msims, which burn more mass at early times and at higher densities, exhibit the smallest values of $M_{^{56}\mathrm{Ni}}/M_\mathrm{IGE}$, while the \Fsims\ exhibit the largest values of this ratio.

\fig{fig:Mek}(c) and (d) shows that the masses of IGE and {\nifs} are proportional to the kinetic energy $E_\mathrm{K}$. 
This behavior is expected since $E_\mathrm{K} \propto E_{\rm nuc}$, as shown in \fig{fig:em}(a).  
However, it is interesting to show these relations, since the burned mass $M_\mathrm{b}$ is roughly proportional to $M_{^{56}\mathrm{Ni}}$, and $E_\mathrm{K}$ is a major factor affecting the rate of rise and decline of the bolometric light curve.

\subsubsection{Spatial distribution of nucleosynthetic yields}
\label{sec:results:evolutionall:yields:spatial}

\begin{figure*}
\centering
\includegraphics[width=6in]{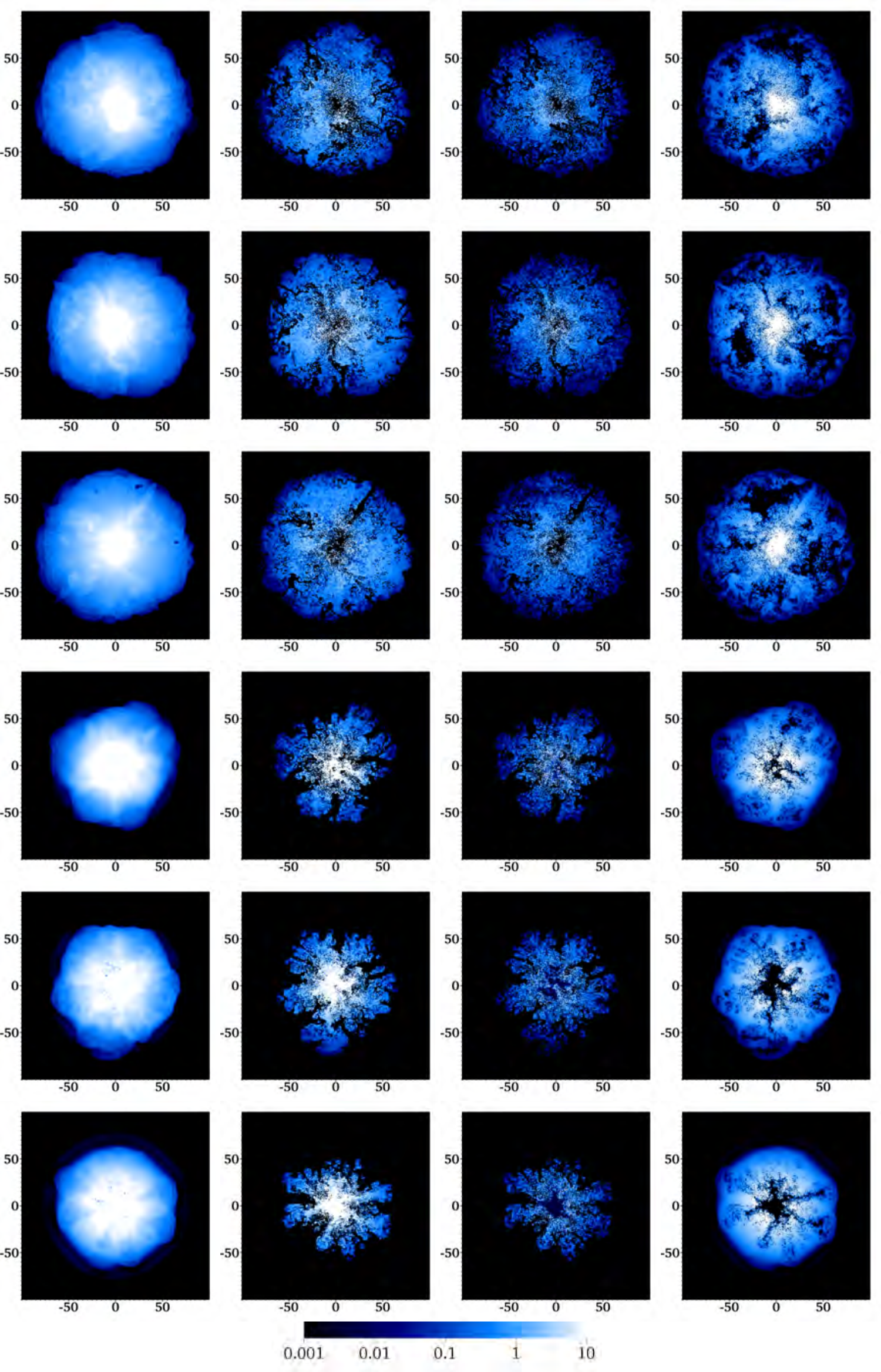}
\caption{
Distribution of the total mass densities (left column), {\nifs} masses (second column), Si masses (third column) and C masses (right column) in the $x-z$ plane, during the homologous expansion stage.
From top to bottom, each row represents the corresponding simulation in \tab{tab:final}.
The axes are in units of $1.0\e{9}$cm.
The density is plotted on a logarithmic scale from $10^{-3}$ to 10 \gcm.
}
\label{fig:slice}
\end{figure*}

\begin{figure}
\centering
\includegraphics[width=3.25in]{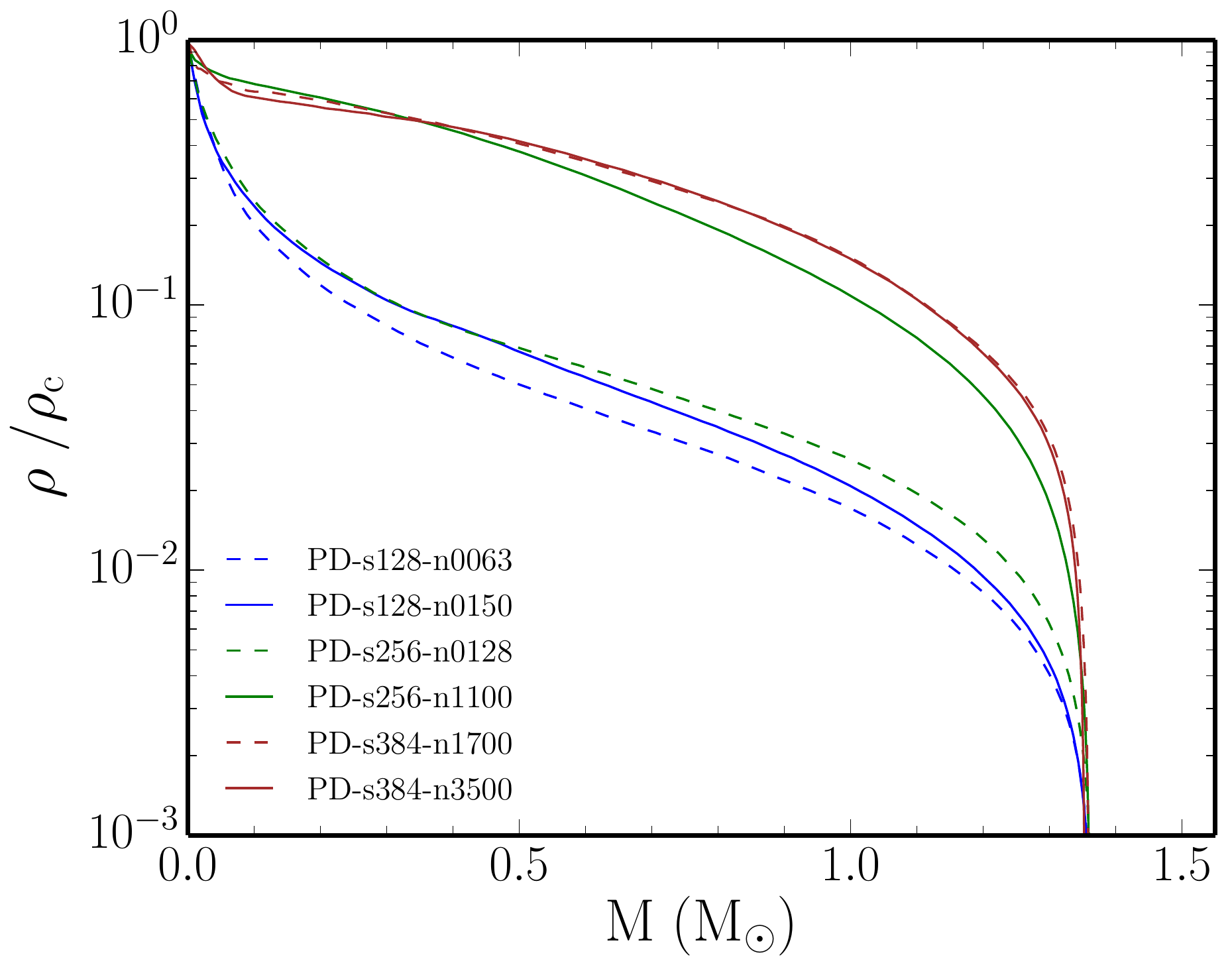}
\includegraphics[width=3.25in]{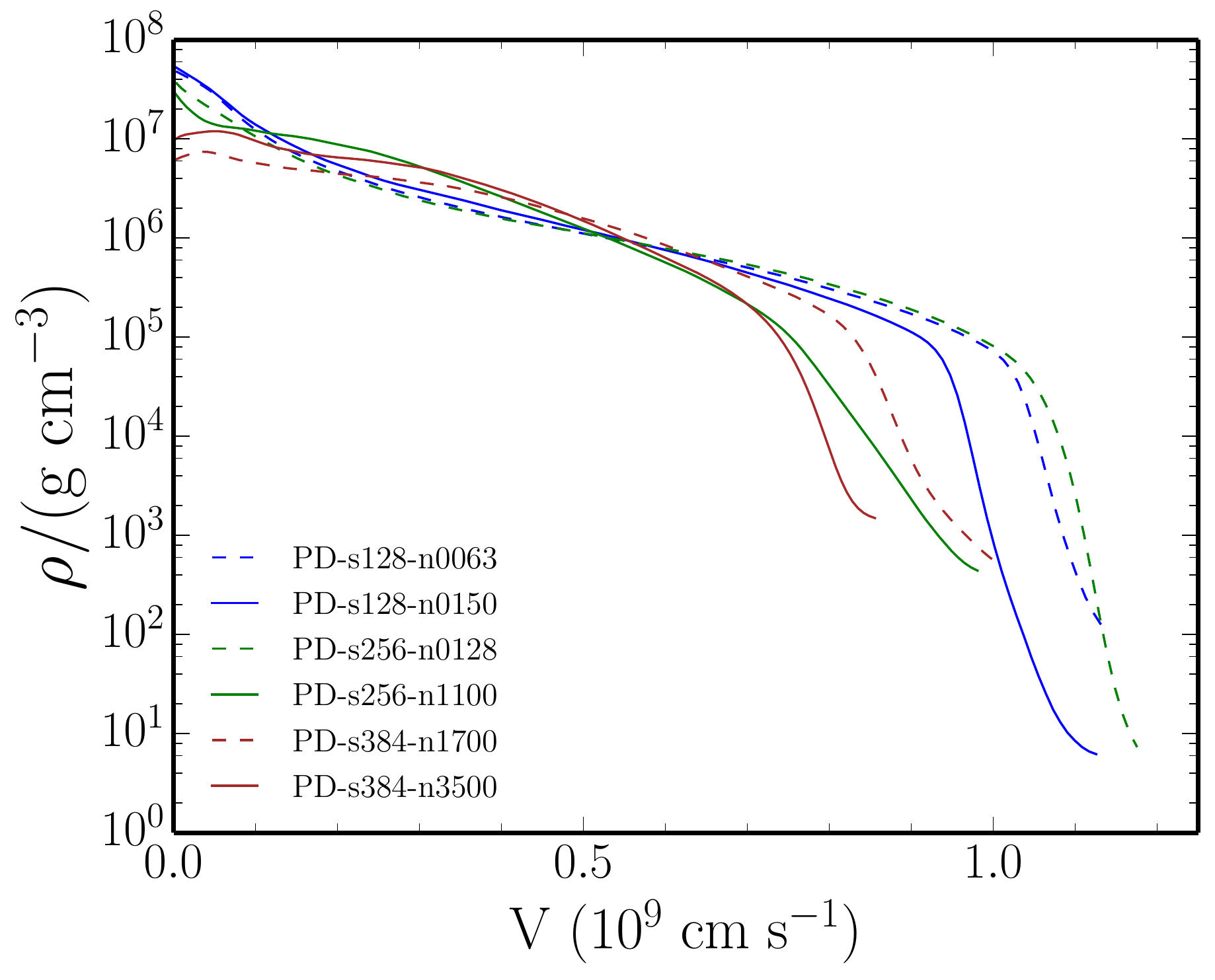}
\includegraphics[width=3.25in]{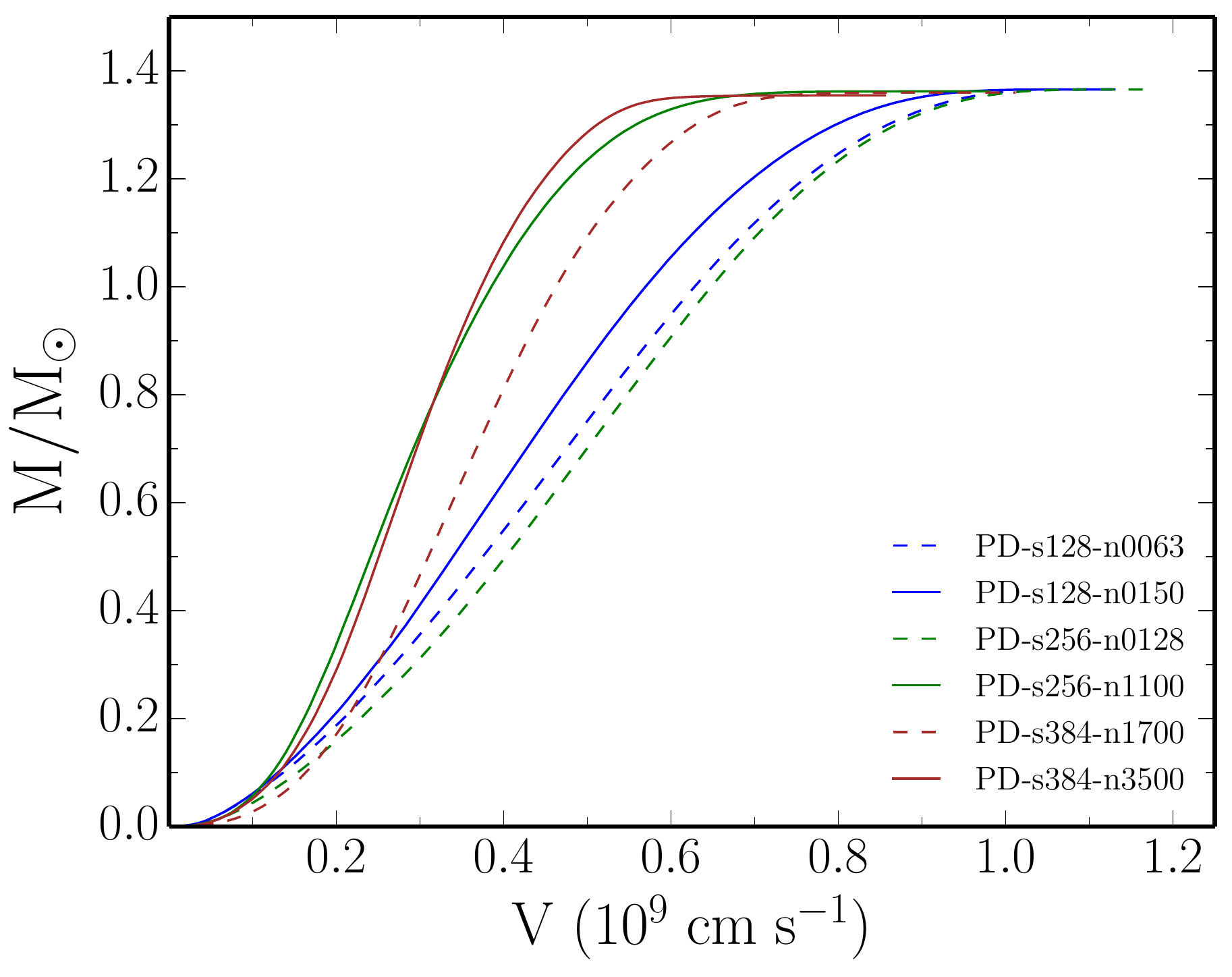}
\caption{
Total mass density profiles at the time shown in \fig{fig:slice} for all six simulations.
Top panel:  density normalized to the central density $\rho_c$ as a function of mass.
The \Fsims\ have a density cusp in the core, much lower densities in an extended region surrounding the core, and material extending to high velocities.  
In contrast, the \Msims\ have an extended, high-density region in their core, a much thinner low-velocity region surrounding it, and no material extending to high velocities.
Middle panel:  Profiles of the total mass density as a function of velocity.  
The \Fsims\ have a density cusp in the core and material extending to higher velocities.
Bottom panel:  Profiles of the cumulative mass as a function of velocity.  
The \Fsims\ have less mass in the core and more mass at higher velocities.
}
\label{fig:density-structure}
\end{figure}

\begin{figure*}
\centering
\mbox{
\includegraphics[trim = 0mm 0mm 0mm 0mm, clip, scale=0.45]{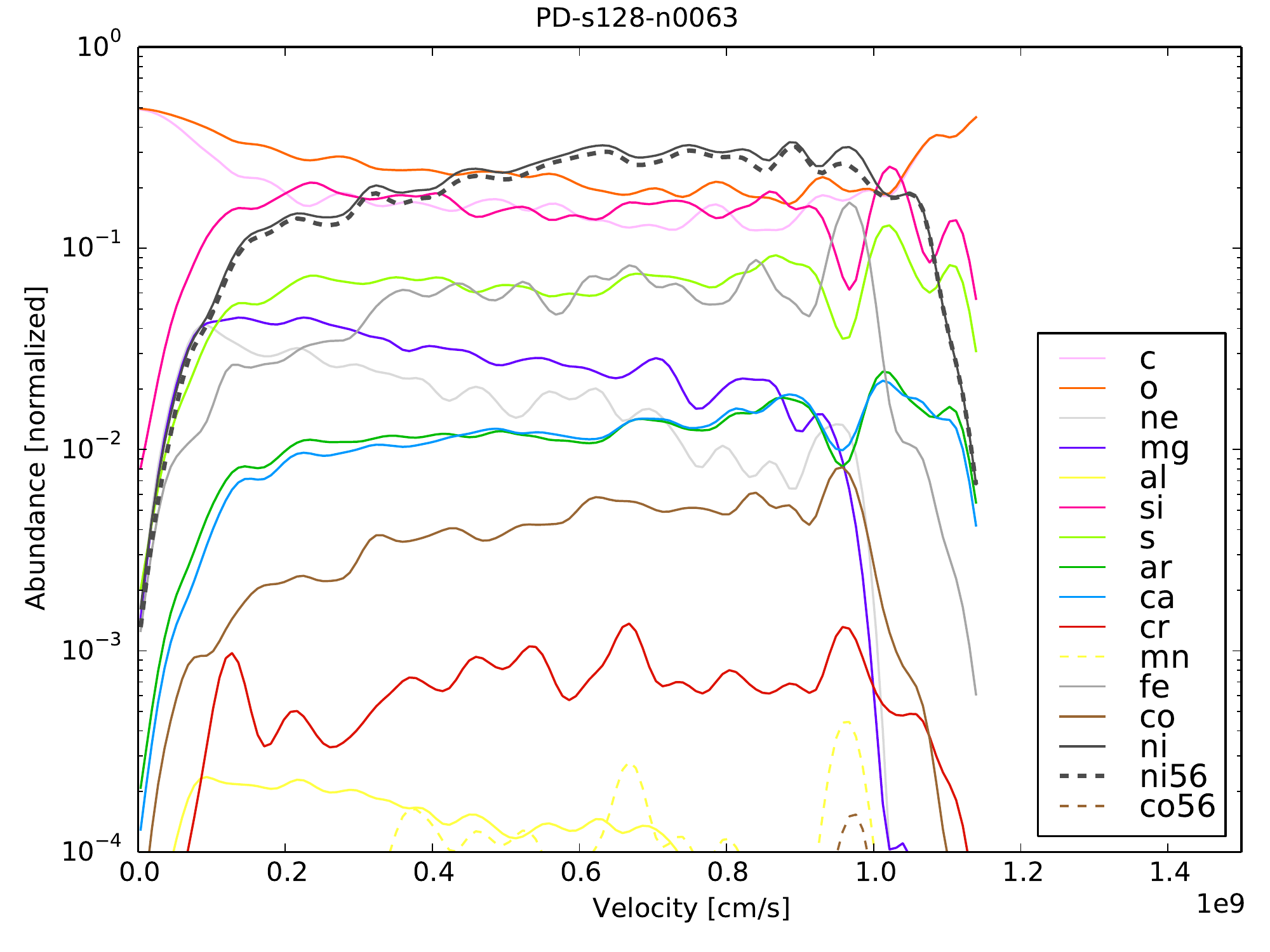}
\hspace{0.2cm}
\includegraphics[trim = 20mm 0mm 0mm 0mm, clip, scale=0.45]{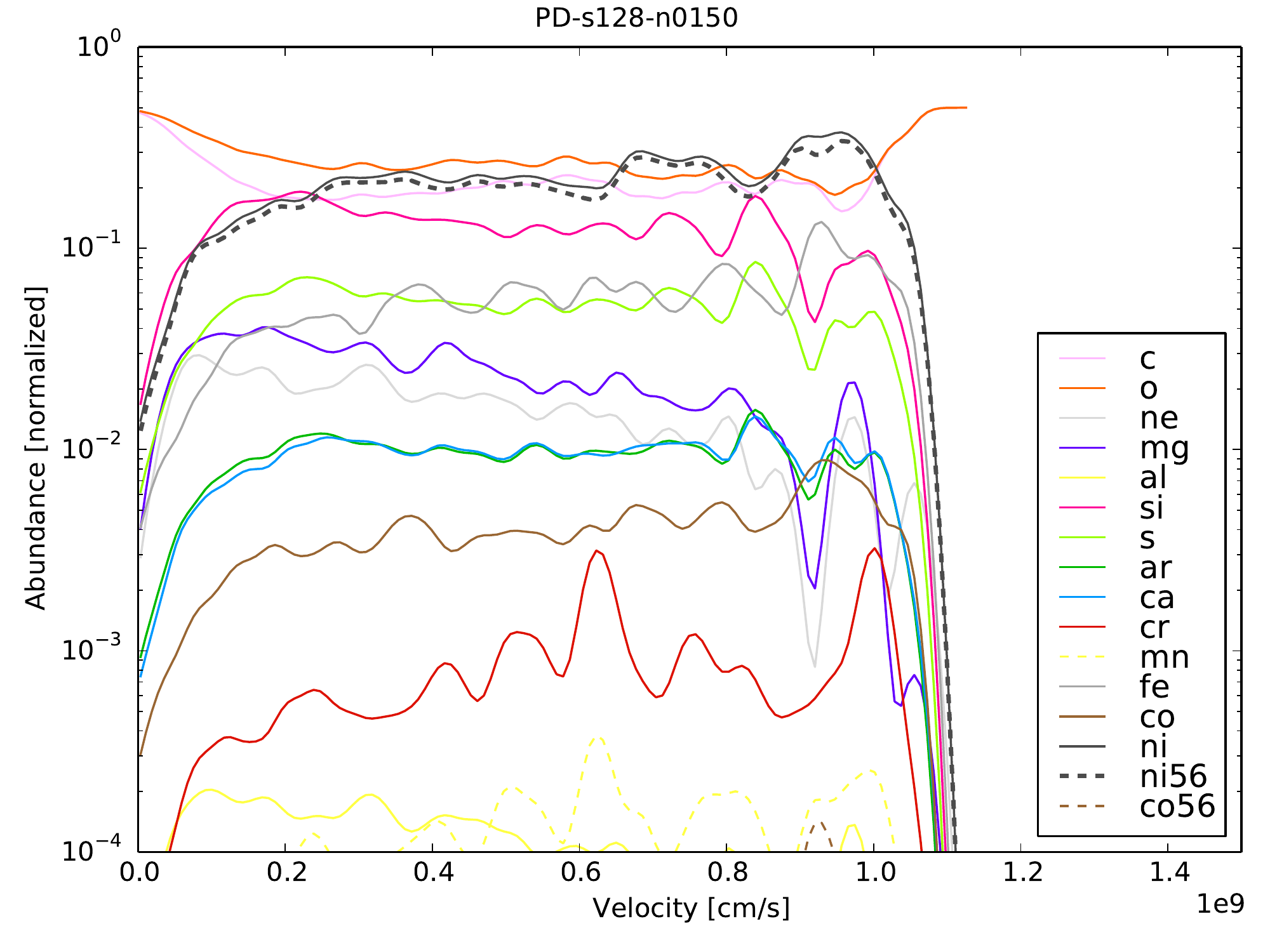}
} \\
\mbox{
\includegraphics[trim = 0mm 0mm 0mm 0mm, clip, scale=0.45]{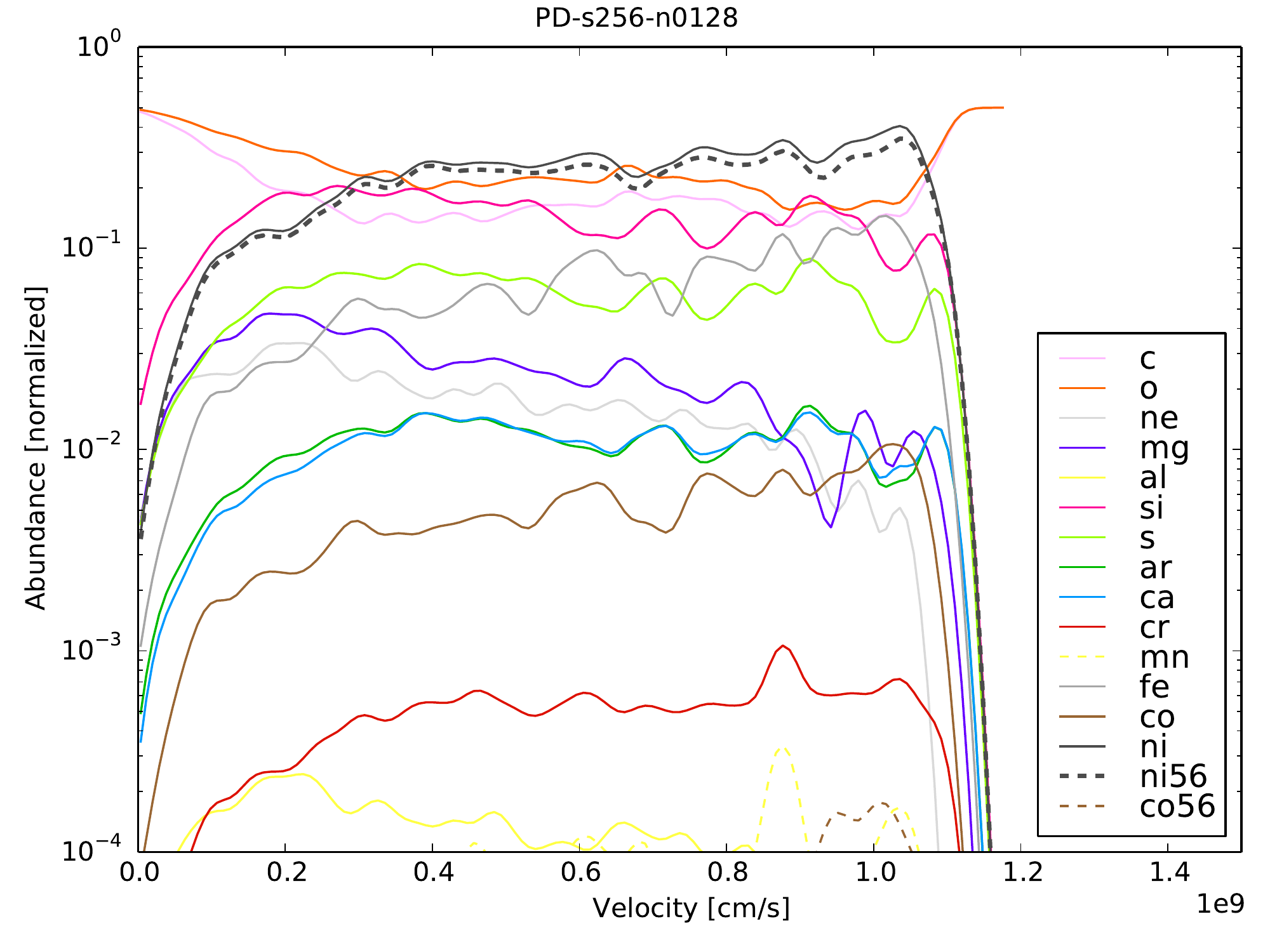}
\hspace{0.2cm}
\includegraphics[trim = 20mm 0mm 0mm 0mm, clip, scale=0.45]{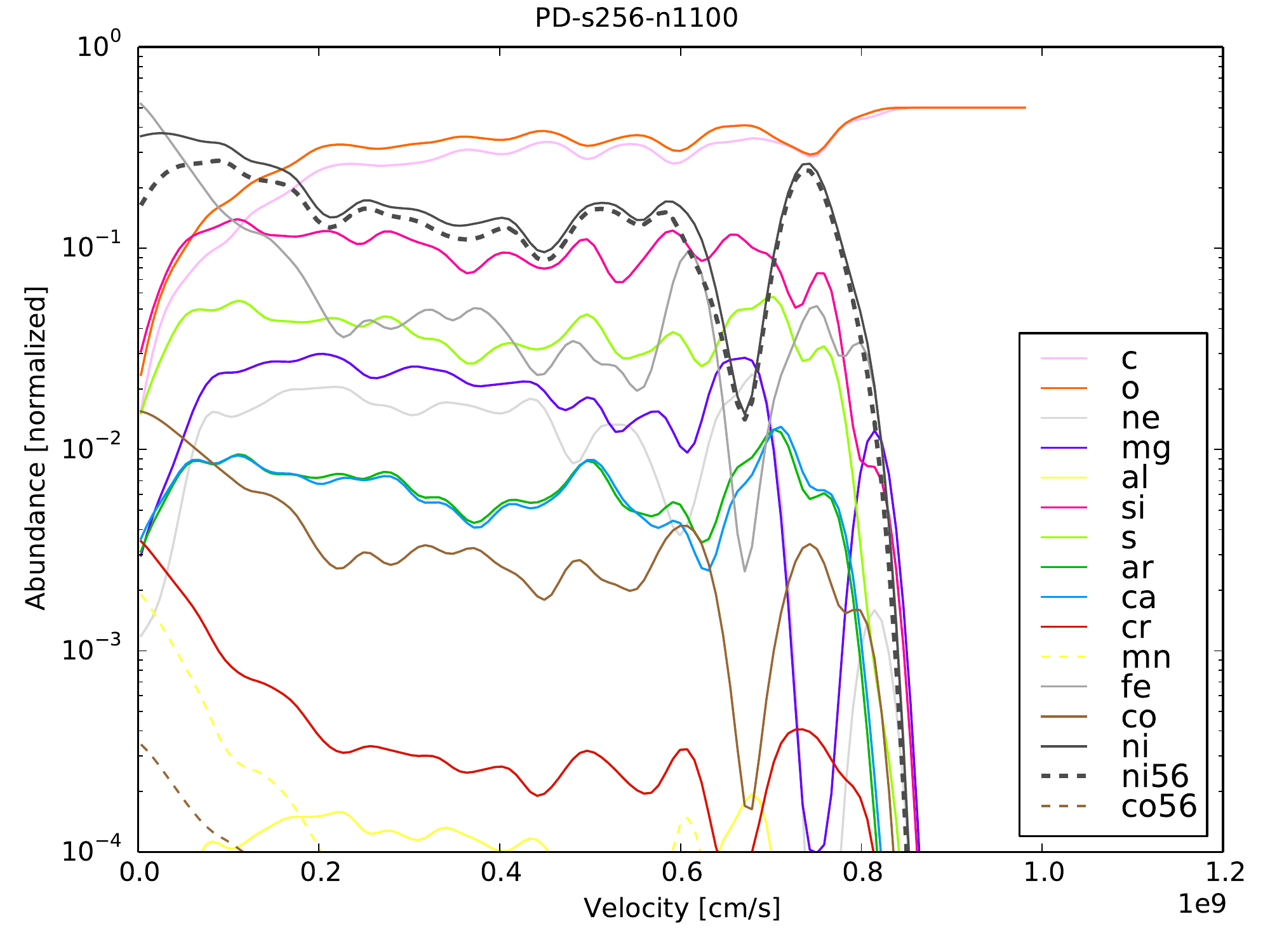}
} \\
\mbox{
\includegraphics[trim = 0mm 0mm 0mm 0mm, clip, scale=0.45]{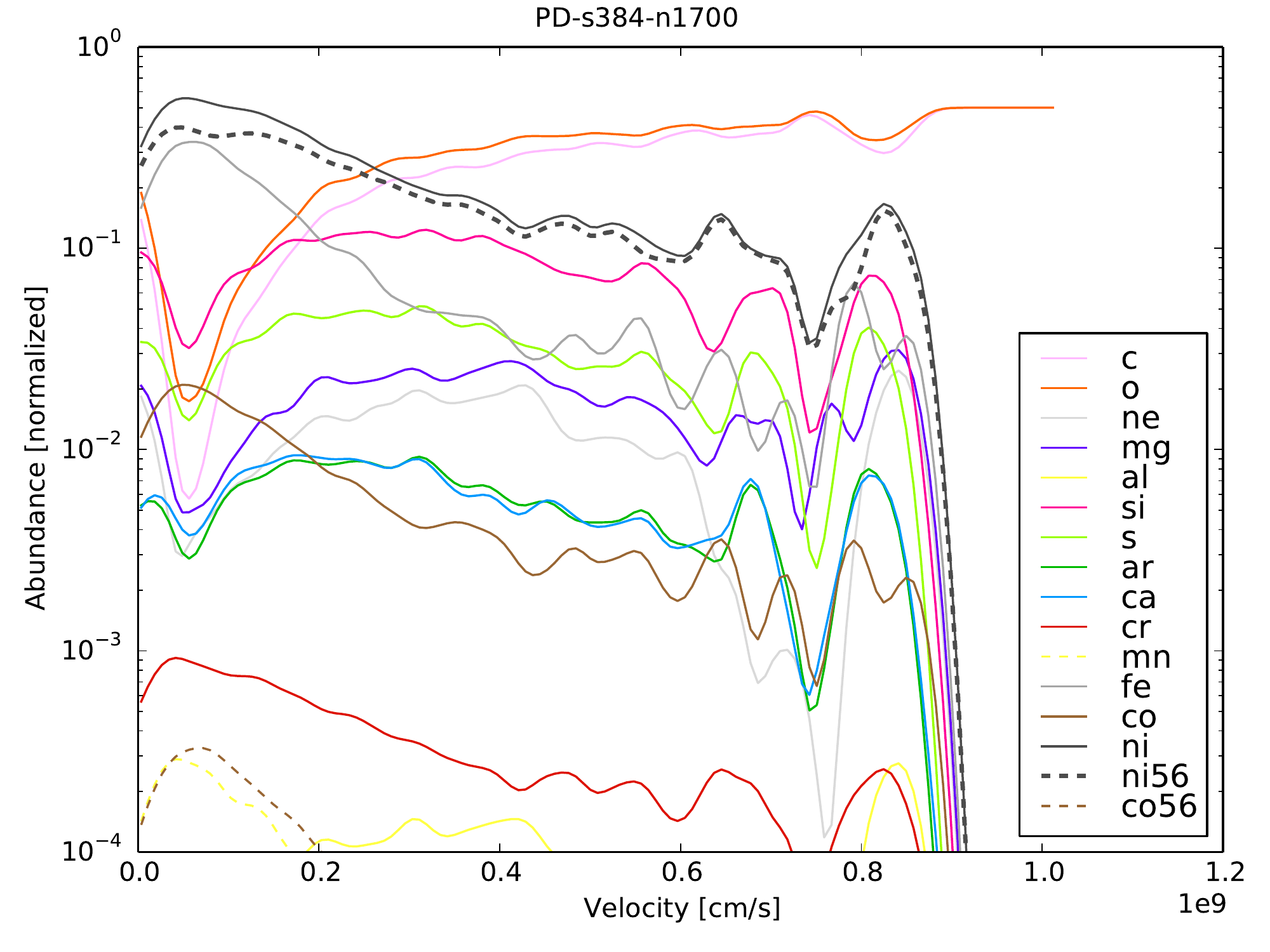}
\hspace{0.2cm}
\includegraphics[trim = 20mm 0mm 0mm 0mm, clip, scale=0.45]{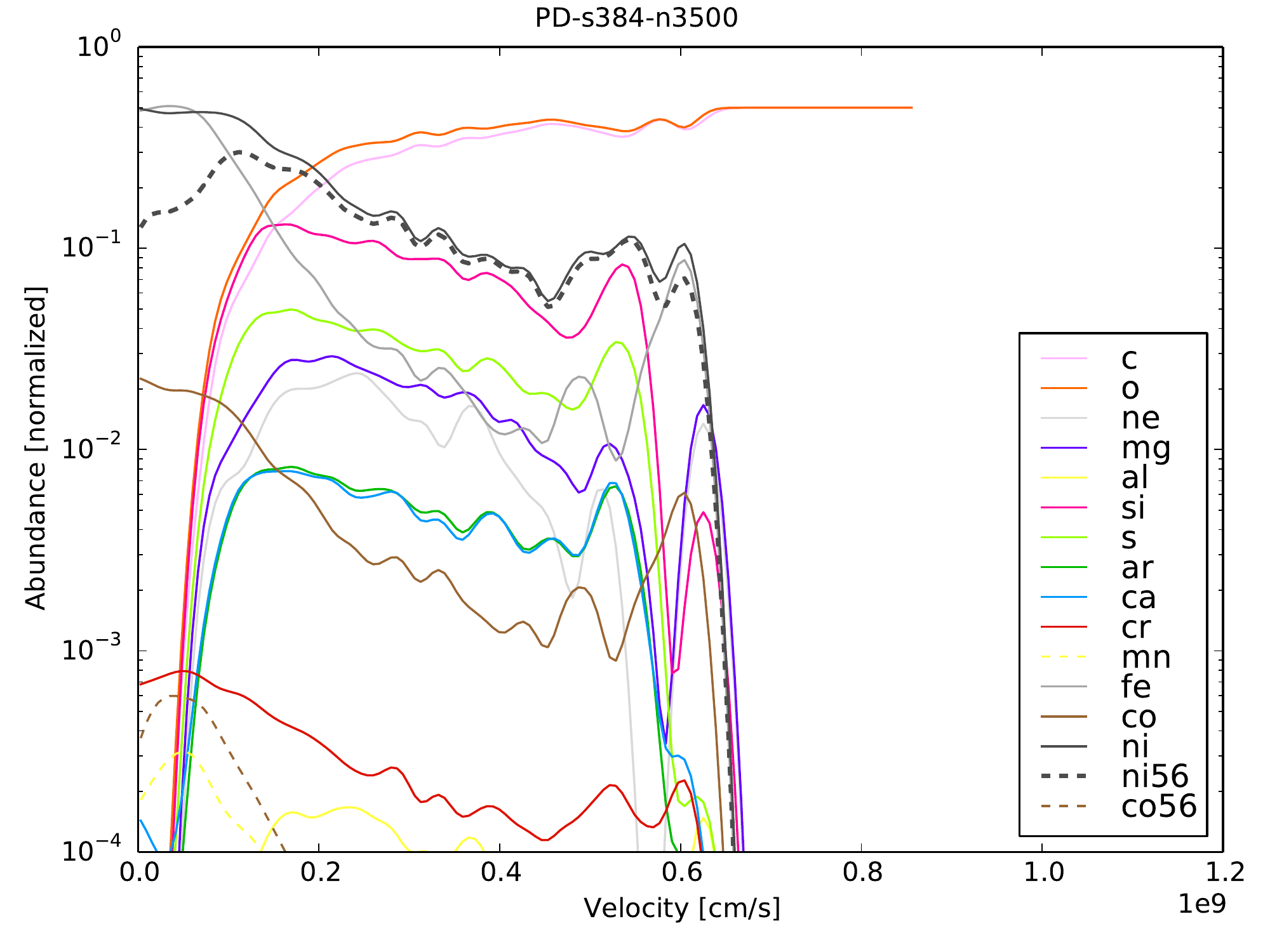}
}
\caption{
Chemical element mass fractions as a function of velocity for all six simulations of the PD model.
The top, middle, and bottom rows show simulations with \Rsph$=$128, 256, and 384 km respectively.
The simulations in the right column have nearly the largest possible \Nign\ in the confining sphere \Rsph.
The \Fsims\ have little \nifs in the core but a considerable amount throughout the rest of the ejecta, except in the outermost region.  
In contrast, the \Msims\ have more \nifs in the core, less throughout the rest of the eject, and very little in the outermost regions of the ejecta. 
The \Msims\ have much less unburned C/O in the core of the star, but more unburned C/O throughout the rest of the ejecta than the \Fsims.
}
\label{fig:abund-vel}
\end{figure*}

The spatial distribution of the elements in the ejecta produced in the explosion phase is important in determining the light curves, as is the total amount of radioactive {\nifs}. 

\fig{fig:slice} shows the spatial distributions of the total mass density, the masses of {\nifs}, Si, and C for the simulations at $t\sim70$ sec. 
At this time, the expansion of the ejecta has become homologous; the distributions therefore remain the same at later times.  
The outermost region of the domain is less well resolved than the inner region due to the smaller number of tracer particles there, even though 8 million of the 10 million tracer particles were placed proportional to the initial local volume of the fluid; i.e., they were V-weighted particles (see \csec{sec:models:particles}).

The far-left images show that the ejecta produced by the \Fsims\ have a high-density core surrounded by a much larger, low-density region.  
Two factors are responsible for this structure.  
First, the hot bubbles corresponding to the ignition points accelerate away from the center of the star due to the buoyant force, leaving behind cold, dense, unburnt C/O. 
Second, as we discussed in \csec{sec:results:evolutionall:burning} and \fig{fig:burnedmass}, the nuclear burning rate rapidly increases at intermediate times in these simulations, and the burning in the outer regions lasts longer than in the \Msims, producing an extended hot, low-density region surrounding the core. 
These features are also evident in the two middle images, which show that these ejecta have an absence of {\nifs} and Si in their core and a great deal of these elements in the region surrounding the core.  
It is also evident in the far right image, which shows that the cold, dense core in the ejecta is comprised of unburned C/O.

In contrast, the far-left images also show that the ejecta produced by the \Msims\ have a core that is less dense and more extended.  
The two factors responsible for this structure are the inverse of the above factors.  
First, the bubbles corresponding to the large numbers of ignition points in these simulations quickly merge, incinerating most of the core.  
Second, as we discussed in \csec{sec:results:evolutionall:burning}, the nuclear burning rate decreases because of the rapidly decreasing density of the star, leaving much of the region surrounding the core unburned.  
Furthermore, the slow rate of nuclear burning allows large plumes of hot ash to form, which rise into the outer regions of the star.  
Around these plumes, cold, unburned C/O sinks toward the core but is unable to penetrate it fully.  
As a result, the distributions of {\nifs} and Si in the ejecta have a star-shaped pattern, while the distribution of unburned C/O has an inverse pattern.

The very different evolution of the nuclear burning in the \Fsims\ and \Msims\ produces strong contrasts in the distributions of the overall mass density, and the mass densities of {\nifs}, Si, and unburned C/O.  
{The difference in the distribution of the mass density is evident in \fig{fig:density-structure} (top panel), which shows the angle-averaged density profile normalized by the central density at the time shown in \fig{fig:slice} for all six simulations.
The profiles for the \Fsims\ have a dramatic cusp at the center and an extended, low-density region outside of it. 
In contrast, the profiles for the \Msims\ are relatively flat.}

\fig{fig:density-structure} also shows the angle-averaged profiles of the mass density (middle panel) and the cumulative mass (bottom panel) as a function of velocity for all six simulations.  
The density profiles for the \Fsims\ again exhibit a cusp.  
In the \Fsims, the velocity of the ejecta extends to $10,000 - 11,000$ km s$^{-1}$, but in the \Msims, it extends only to $8,000 - 9,000$ km s$^{-1}$.  
The profiles of cumulative mass as a function of velocity show the same difference.

\fig{fig:abund-vel} shows the abundance mass fractions of the chemical elements as a function of velocity for all six simulations.  
The \Fsims\ have little \nifs\ in the core of the ejecta, but a considerable amount throughout the rest of it, except in the outermost layers.  
They also have a large amount of unburned C/O throughout the ejecta, especially in the core and the outermost layers.  
In contrast, the \Msims\ have mostly \nifs\ in the core, less throughout the rest of the ejecta, and very little in the outermost layers.
The \Msims\ have much less unburned C/O in the core of the star, but more unburned C/O throughout the rest of the ejecta than the \Fsims.

These differences produce clear observational signatures in their light curves and spectra, as we now discuss.  

\section{Synthetic Light Curve Results}
\label{sec:lightcurves}

\begin{figure*}
\centering
\mbox{
\includegraphics[trim = 0mm 20mm 0mm 0mm, clip, scale=0.48]{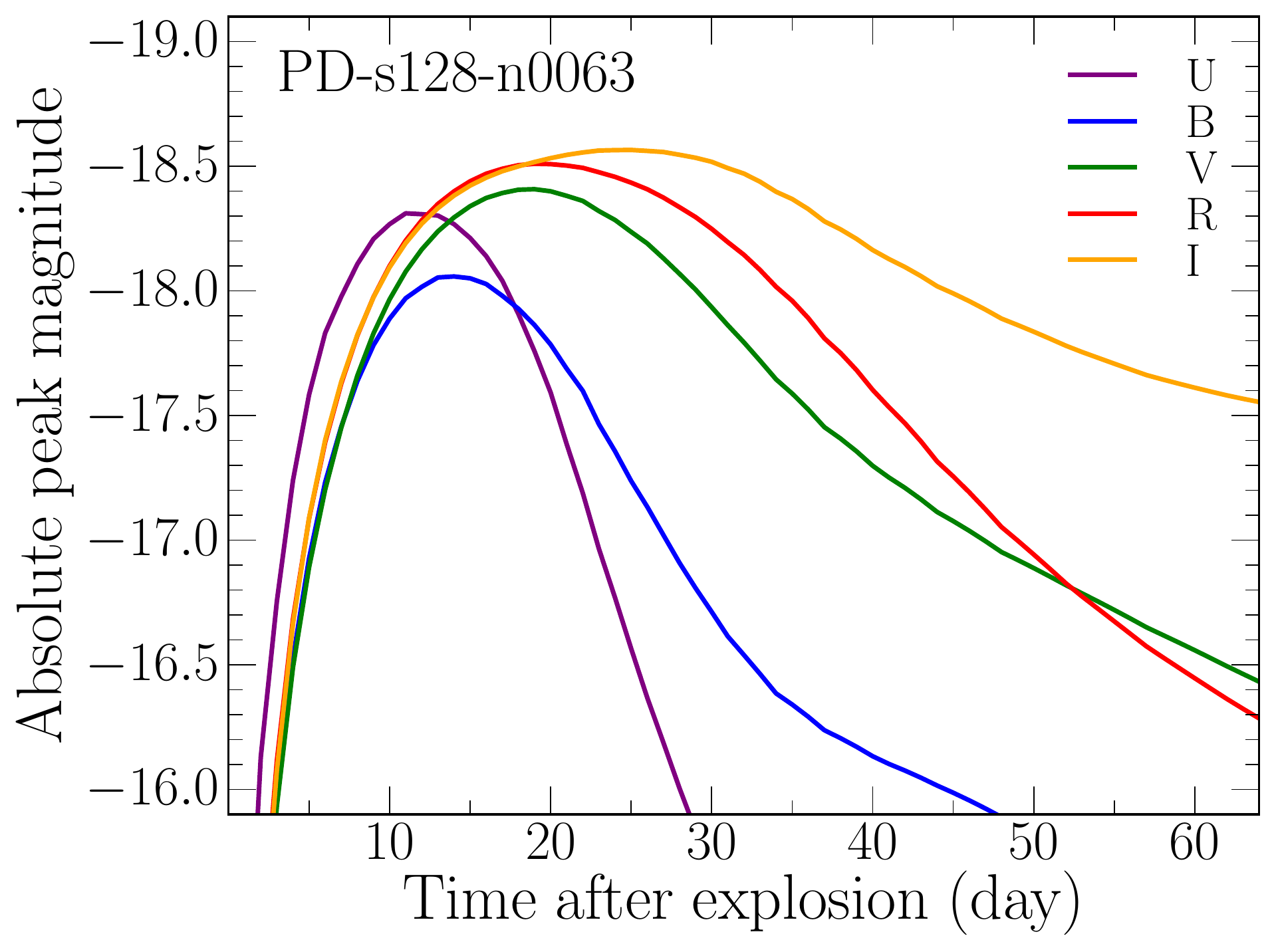}
\includegraphics[trim = 34mm 20mm 0mm 0mm, clip, scale=0.48]{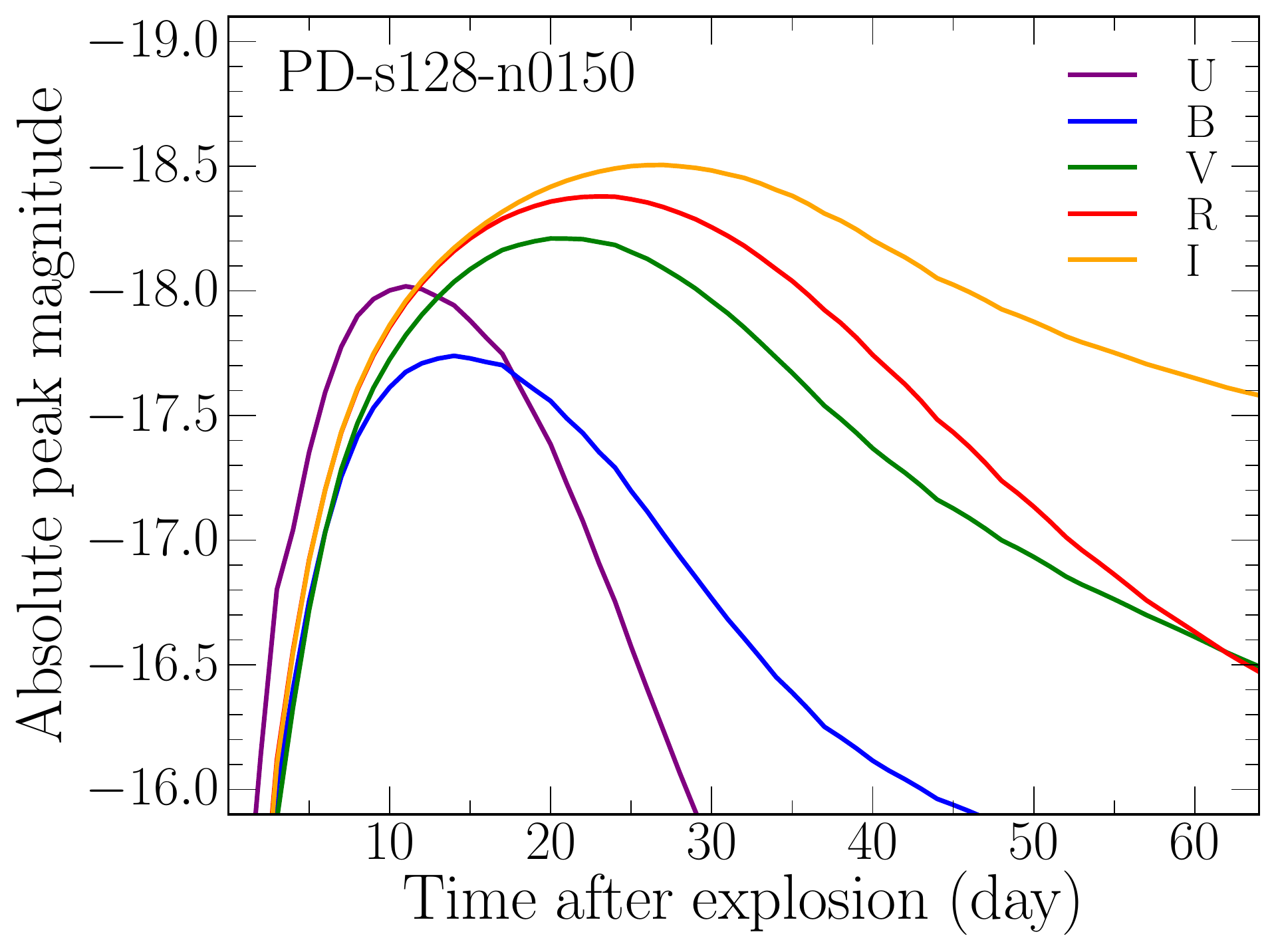}
}  
\mbox{
\includegraphics[trim = 0mm 20mm 0mm 0mm, clip, scale=0.48]{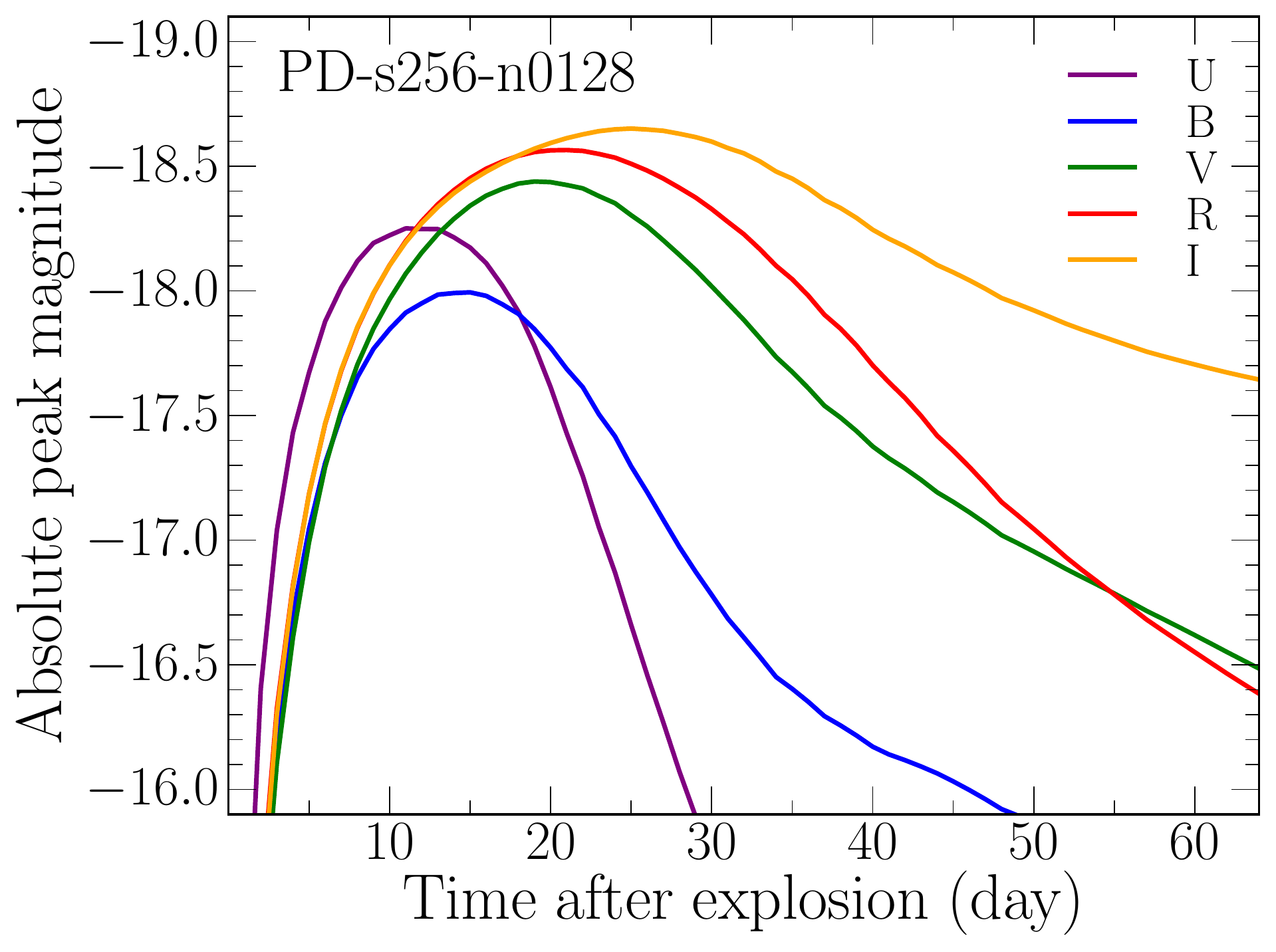}
\includegraphics[trim = 34mm 20mm 0mm 0mm, clip, scale=0.48]{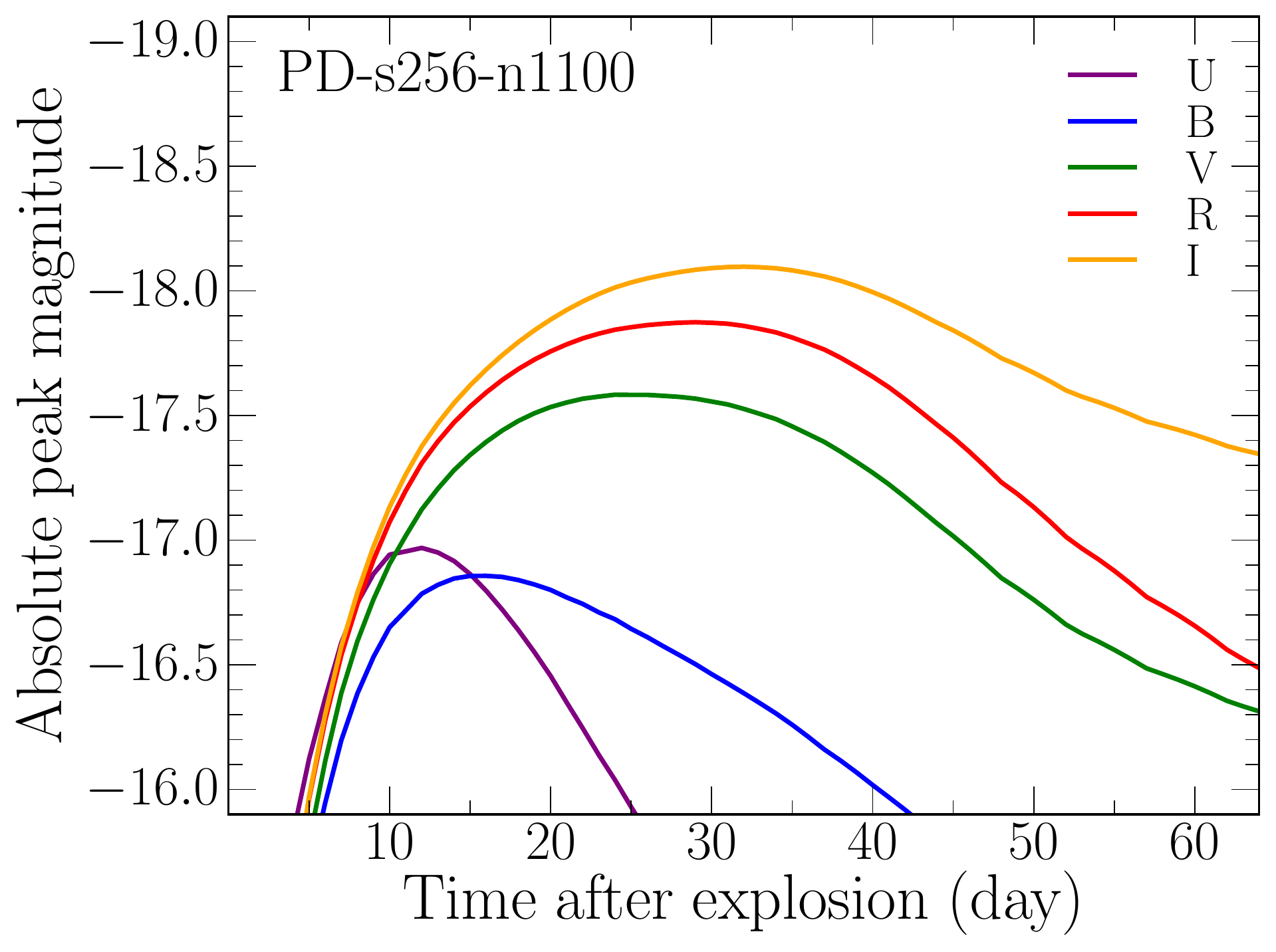}
} 
\mbox{
\includegraphics[trim = 0mm 0mm 0mm 0mm, clip, scale=0.48]{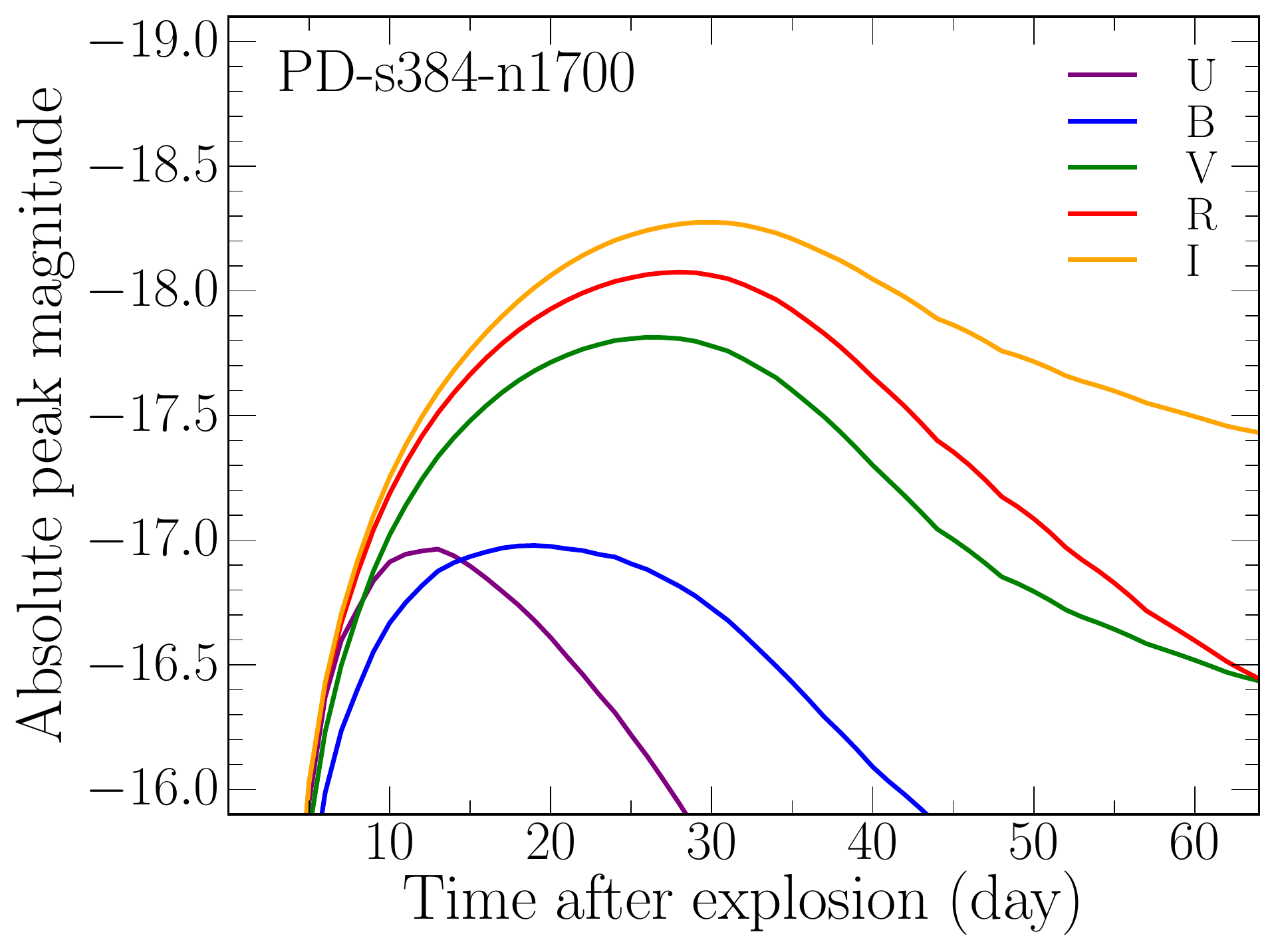}
\includegraphics[trim = 34mm 0mm 0mm 0mm, clip, scale=0.48]{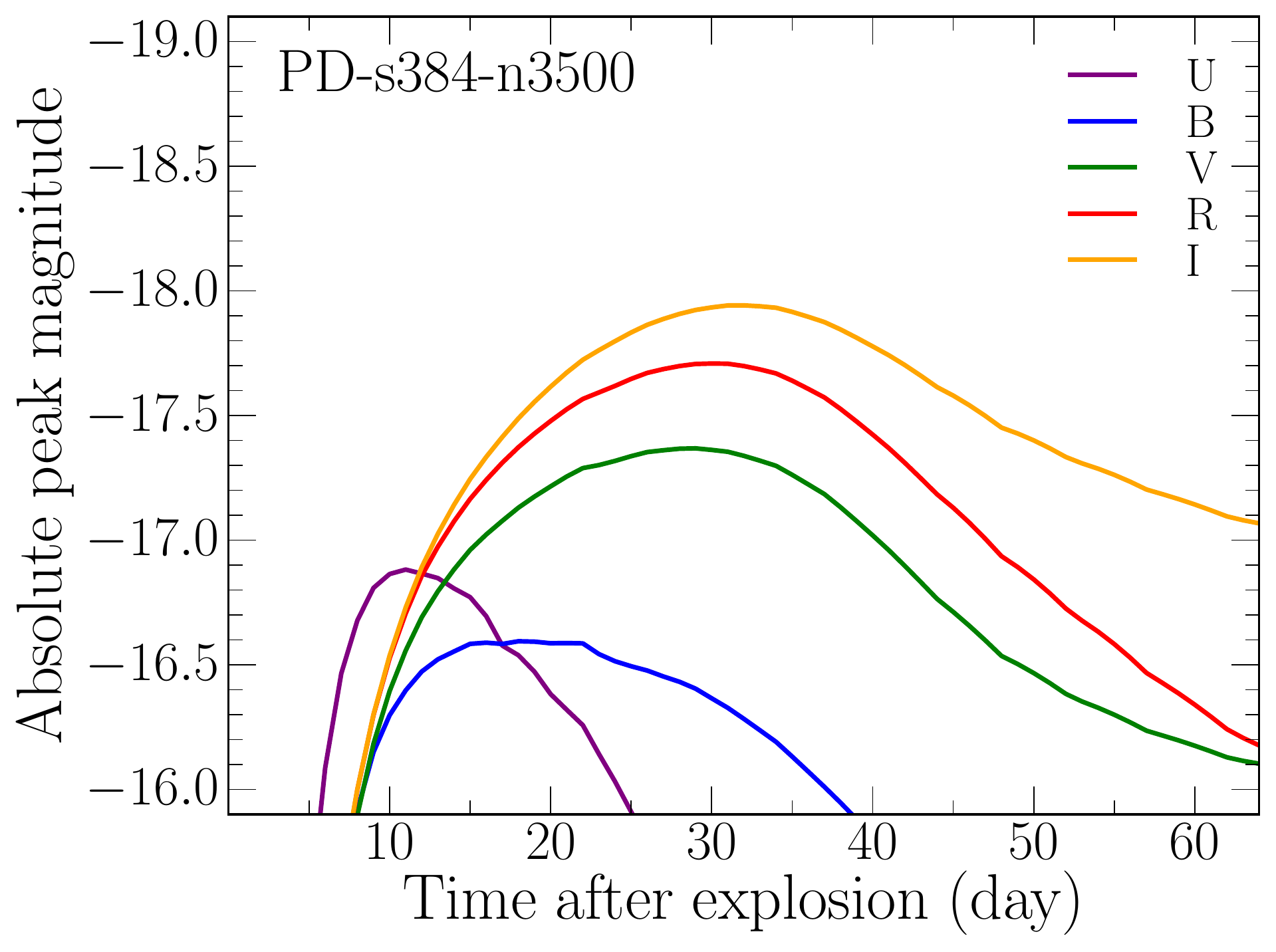}
}
\caption{
Synthetic light curves in UBVRI bands for all six simulations. 
The top, middle, and bottom rows show simulations with \Rsph=128, 256, and 384 km respectively.
The simulations in the right column have nearly the largest possible \Nign\ in the sphere while those in the left column do not.
The \Msims\ rise more slowly, have larger absolute peak magnitudes, decline more slowly, and are redder than the \Fsims.
}
\label{fig:lc1}
\end{figure*}

\begin{figure}
\centering
\includegraphics[width=3.15in]{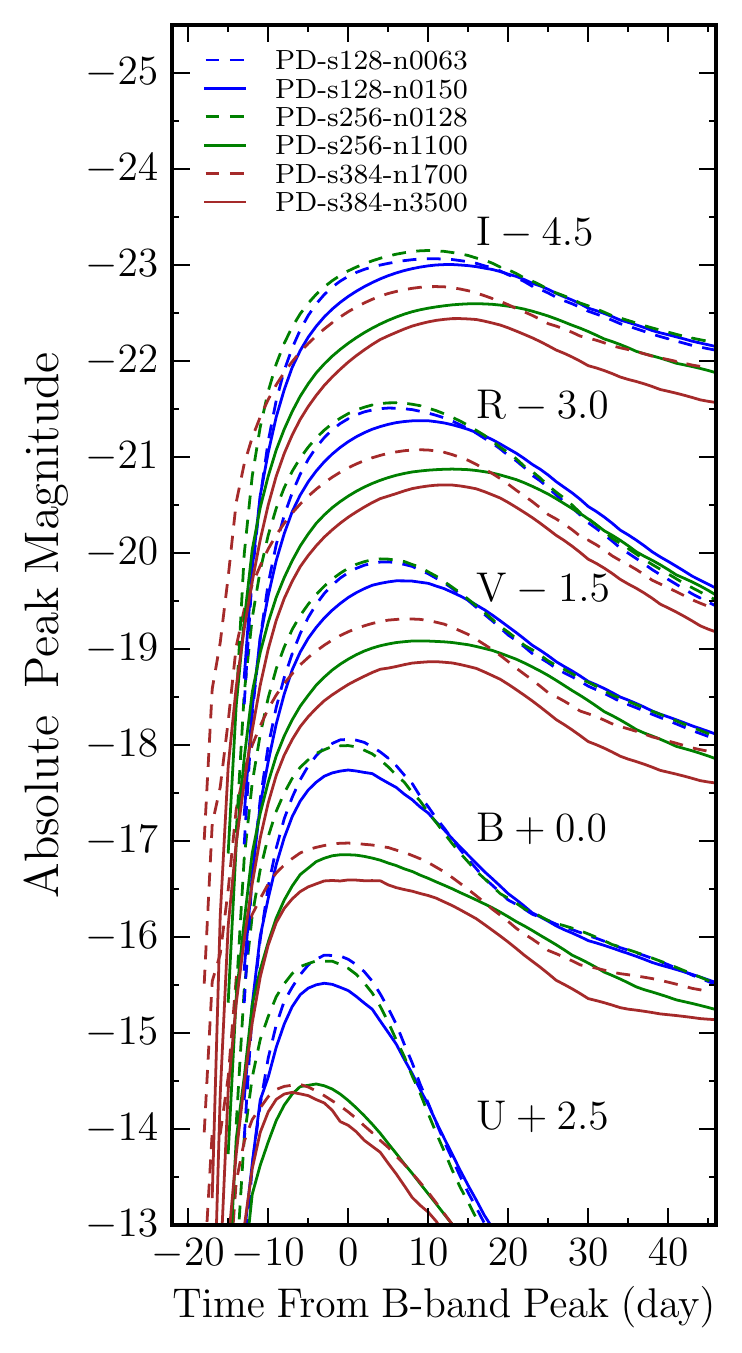}
\caption{
The same light curves as shown in \fig{fig:lc1}, but grouped by filter band, and with the peaks of their B-band light curves aligned at $t=0$. 
The \Msims\ (dashed and solid red lines, solid green line) have significantly higher stretch and smaller absolute peak magnitudes than the \Fsims\ (dashed green line and dashed and solid brown lines), particularly in the $U$ and $B$ bands.
}
\label{fig:lc2}
\end{figure}

Light curves and spectra are computed for the simulations using {\tt PHOENIX}, a general purpose stellar atmosphere and radiation transport code \citep{Hauschildt92, Hauschildt_1999, Hauschildt_2004}, assuming spherical symmetry.  
The time-evolution in the post-hydrodynamic, radiation-dominated, explosion phase is followed using the Radiation Energy Balance method \citep{Rossum_2012}.  
In contrast to Monte Carlo methods that are commonly used for SNe Ia light curve calculations \cite[e.g.][]{Lucy99, Kasen06, Kromer09}, {\tt PHOENIX} numerically solves the special relativistic radiative transfer equation using highly accurate short characteristics and operator splitting methods \citep{Olson87}.  
It does not use the Sobolev approximation, diffusion approximations, or opacity binning.

It is clear from \fig{fig:slice} that there is significant structure in the spatial distributions of the chemical elements throughout the ejecta. 
The \Fsims\ have a large amount of unburned C/O in their cores, as discussed in the previous section, while the \Msims\ develop strong individual plumes that leave traces of ash on their way to the surface, in between which are regions of unburned C/O.  
The high spatial resolution and the large number of Lagrangian tracer particles used in the simulations reveal many small-scale structures in addition to the large-scale asymmetries.

Both the small-scale and the large-scale structures can have significant effects on the shapes of the light curves and spectra.  
To determine how good the {\tt PHOENIX} approximation of spherical symmetry is, one needs to compare the light curves and spectra calculated here using 1D radiation transport with light curves and spectra calculated using 3D radiation transport calculations using the same spatial resolution as the {tt FLASH} hydrodynamic explosion phase simulations.  
This task is far beyond the ability of current RT codes and the performance of current computers, and must be left for future work.  
In spite of this limitation, it is well worth presenting the light curves and spectra calculated using {\tt PHOENIX}.

\fig{fig:lc1} shows $UBVRI$ light curves for all six simulations.  
The light curves for the \Fsims\ rise more quickly, have greater peak magnitudes, and decay more quickly than the light curves for the \Msims.  
The first difference is due to the fact that the former have larger kinetic energies $E_\mathrm{K}$ and the {\nifs} extends closer to the surface of the ejecta.  
The second difference is due to the larger amounts of {\nifs} produced in them, while the third difference is again due to their larger kinetic energies $E_\mathrm{K}$, which cause their ejecta to expand more rapidly and therefore become optically thin sooner.

Simultaneously showing the light curves in all filter bands for each simulation reveals the light curve colors (i.e., where the light curves in each of the bands peak relative to each other).  
The \Fsims\ are bluer than the \Msims. 
\fig{fig:lc2} shows the same light curves as \fig{fig:lc1}, but now grouped by filter band, and with the B-band peak magnitudes aligned, showing the differences between the light curves in terms of magnitude and stretch.  
The \Fsims\ are brighter and have significantly smaller stretch than the \Msims. 
Overall, the six simulations of the PD model produce dimmer and broader light curves than do normal SNe Ia, with absolute B-band magnitudes ranging from $-16.5$ to $-18.0$ and values of the SALT2 stretch parameter ranging from -1 to +4.

\section{Comparisons with Observations}
\label{sec:comparisons}

In this section, we compare the light curves predicted by our simulations with observations. In particular, we evaluate the overall light curve properties using a method based on data-driven models, and compare particular properties such as peak magnitude, stretch, and color evolution with observed values.

\subsection{Light curve evaluation with the SALT2 model}
\label{sec:comparisons:lcevaluation}

\begin{figure*}
\centering
\includegraphics[trim = 0mm 0mm 0mm 0mm, clip, scale=0.47]{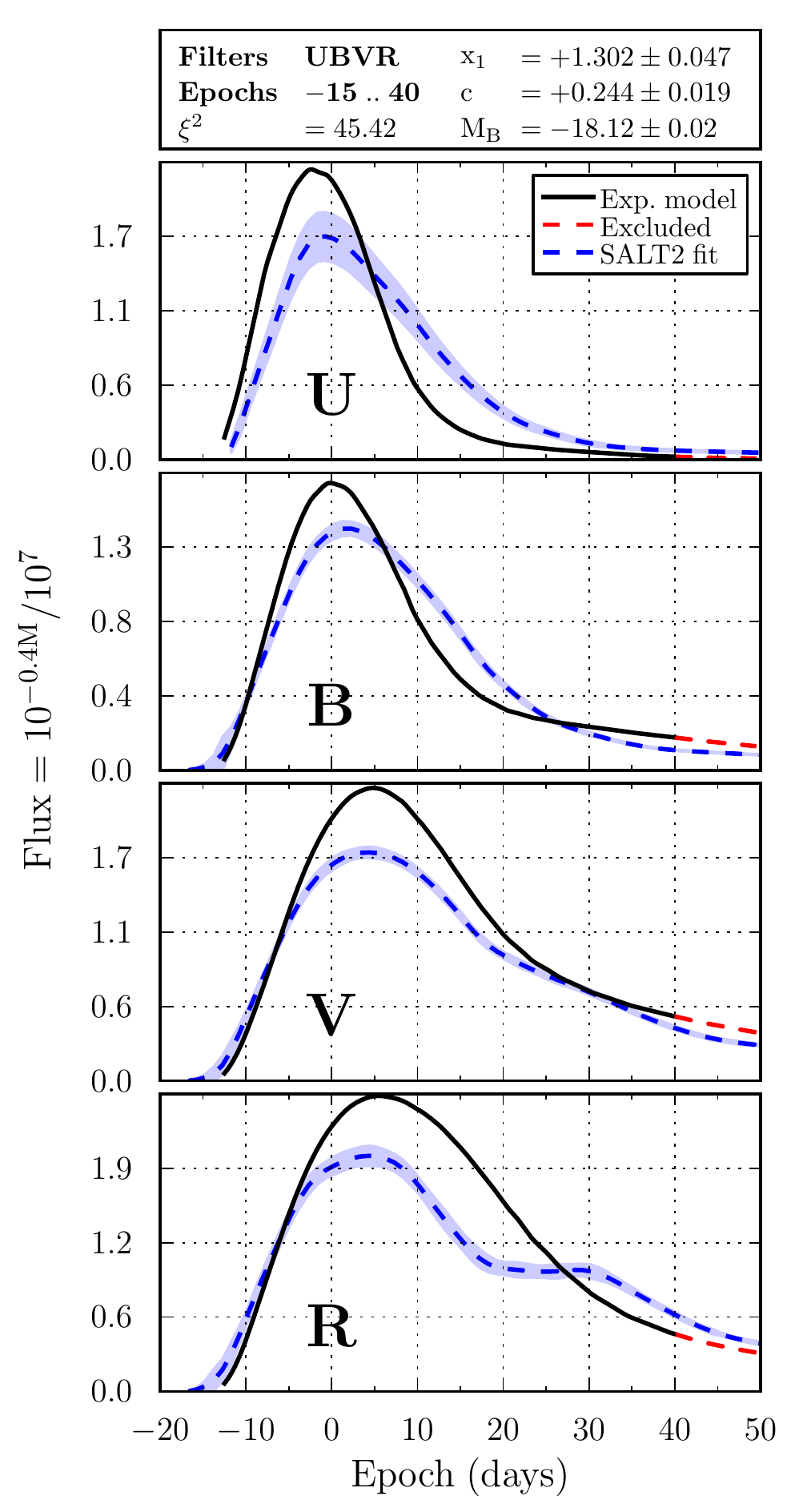}
\includegraphics[trim = 10mm 0mm 0mm 0mm, clip, scale=0.47]{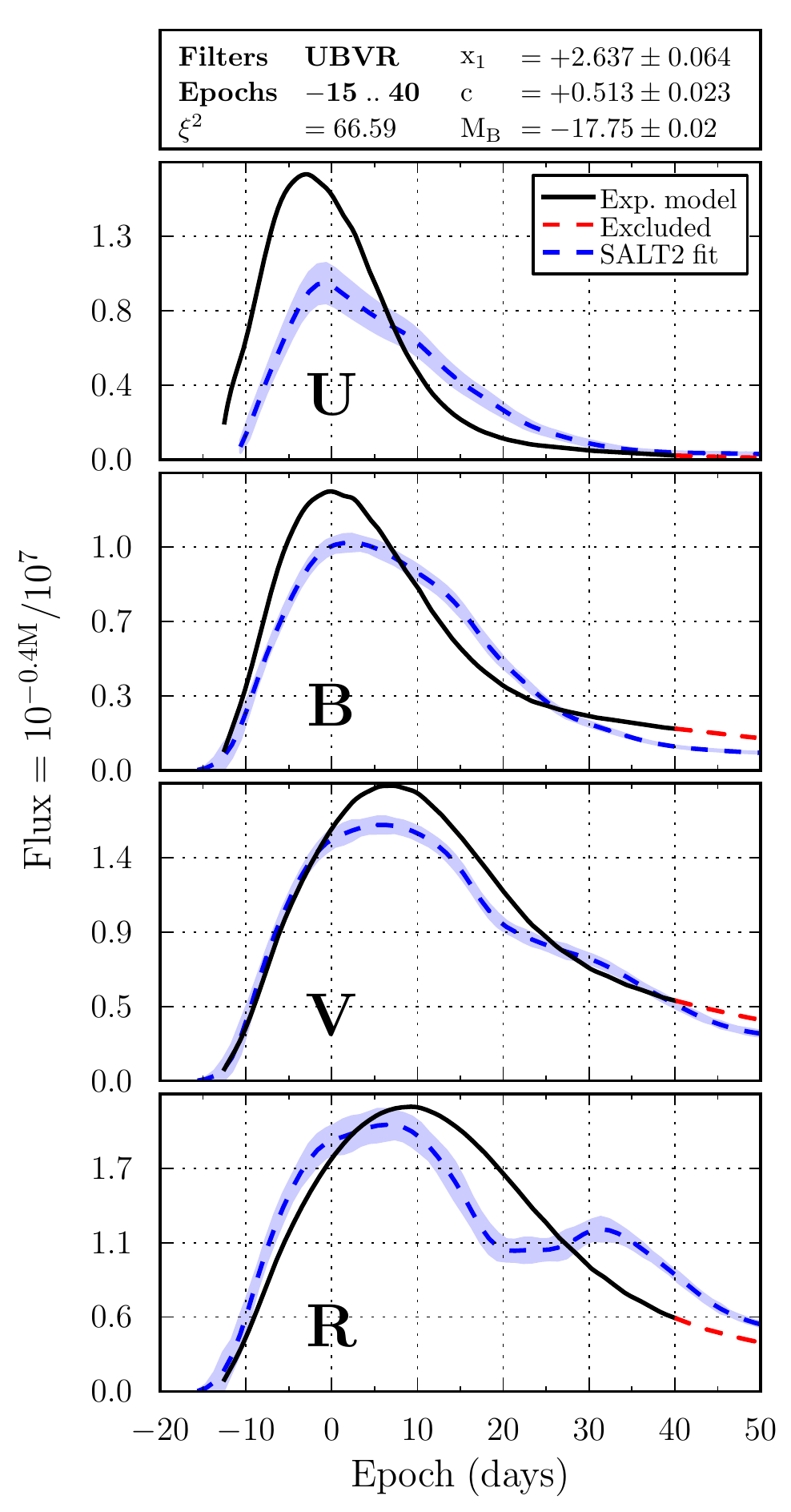}
\includegraphics[trim = 10mm 0mm 0mm 0mm, clip, scale=0.47]{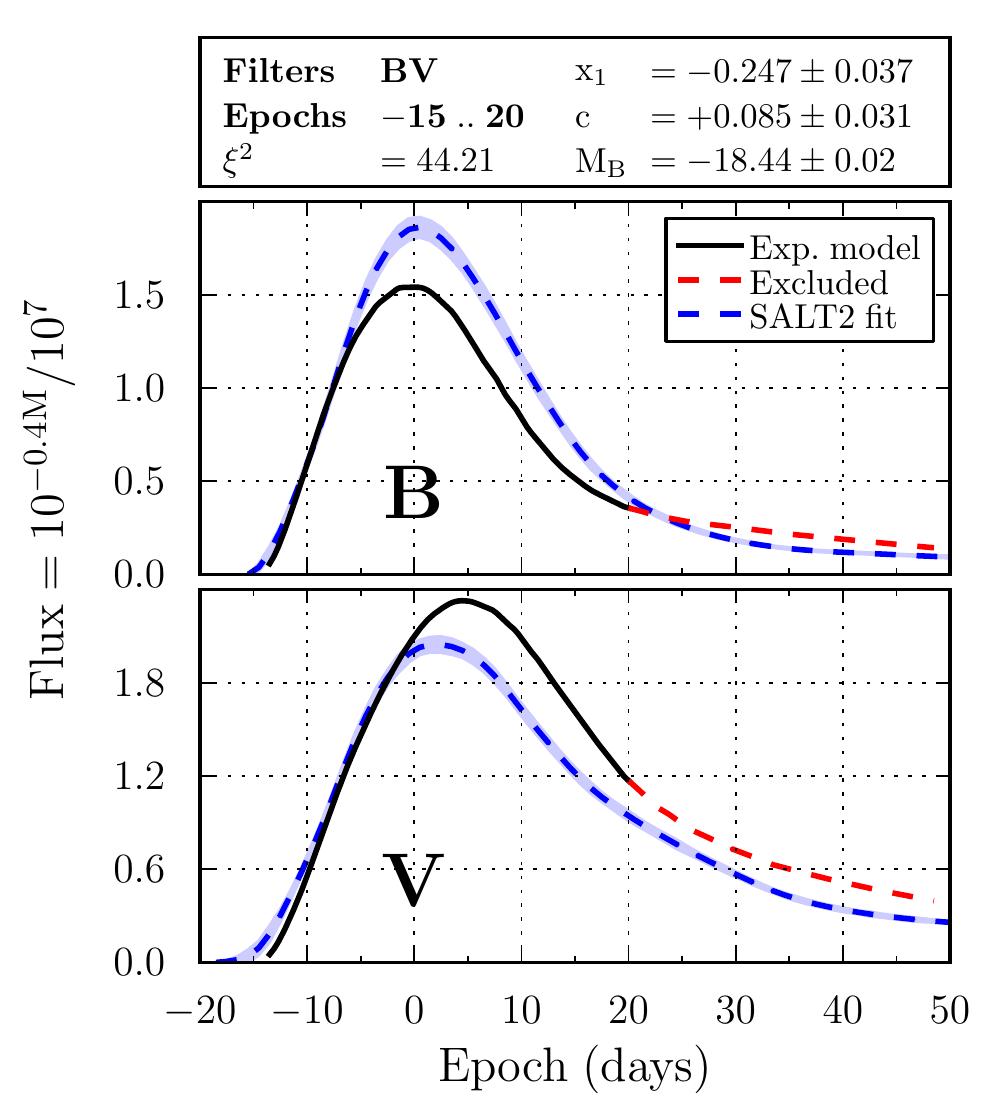}
\includegraphics[trim = 10mm 0mm 0mm 0mm, clip,scale=0.47]{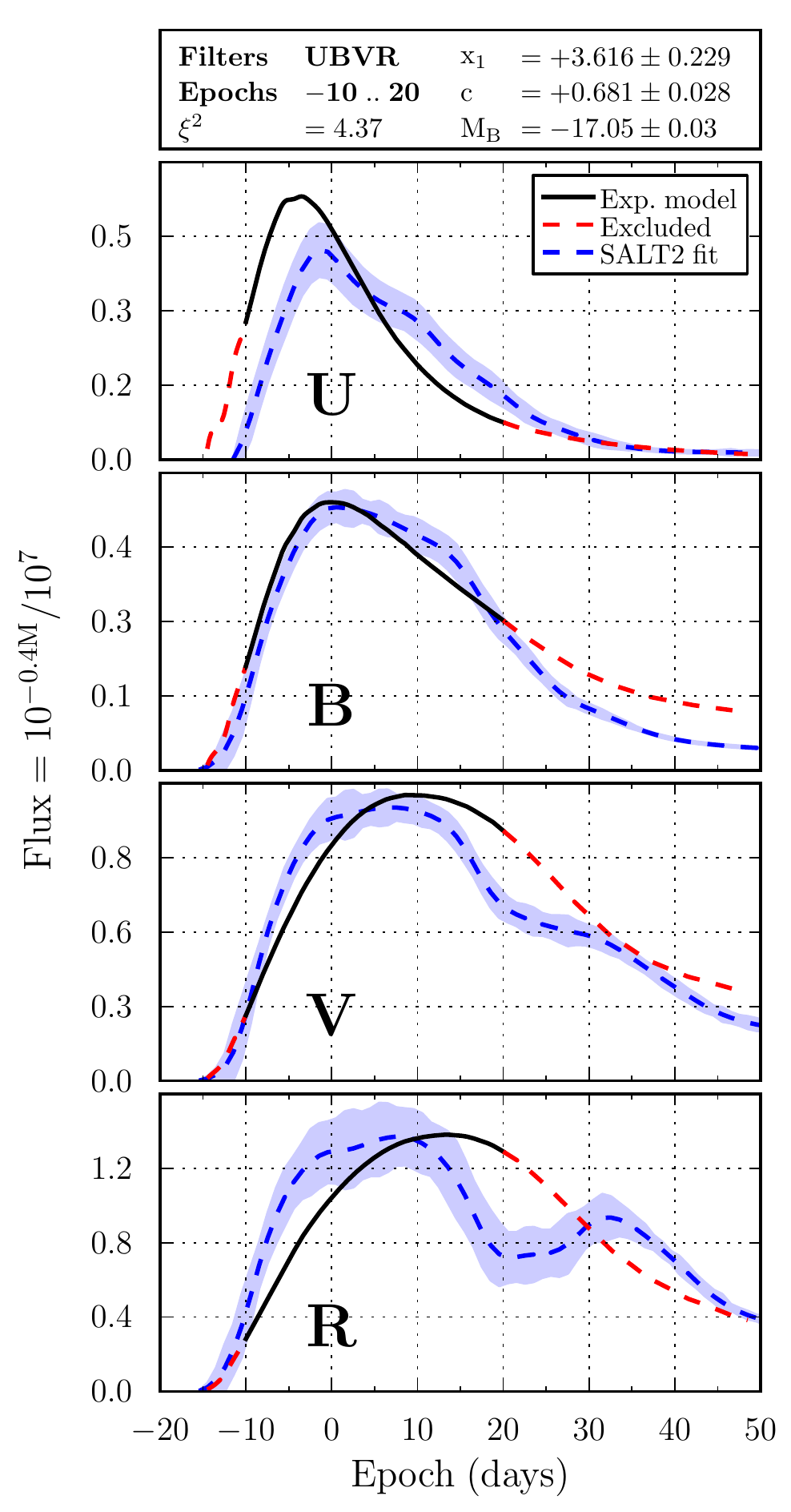}
\caption{
The light curves predicted by selected PD simulations (solid black lines), fit with the SALT2 model (dashed blue lines, with the blue shaded region indicating the 1$\sigma$ confidence interval). 
The dashed red parts of the model light curves were not included in the fit.
First panel: The {\pdone} model, fit in the $UBVR$ filter bands between -15 and +40 days from peak. 
Second panel: The same fit but for the {\pdtwo} model. 
Third panel: The \pdthree\ model, fit only in the $BV$ filter bands. 
Forth panel: The fits for the \Msims\ ({\pdfour}, {\pdfive} and {\pdsix}) are so poor that the fit does not converge when epochs from -15 to +40 days are included. 
The fit to the {\pdfour} model shown here includes epochs between -10 and +20 days.
Nevertheless, the SALT2 model has trouble accommodating the light curves at all, and is driven to a poorly constrained region of parameter space (very high stretch, dim magnitude), leading to large uncertainties on the SALT2 model. 
See \csec{sec:comparisons:lcevaluation} for a detailed discussion.
}
\label{fig:fits}
\end{figure*}

\begin{figure*}\centering
\includegraphics[trim = 1mm 5mm 4mm 3mm, clip,scale=0.7]{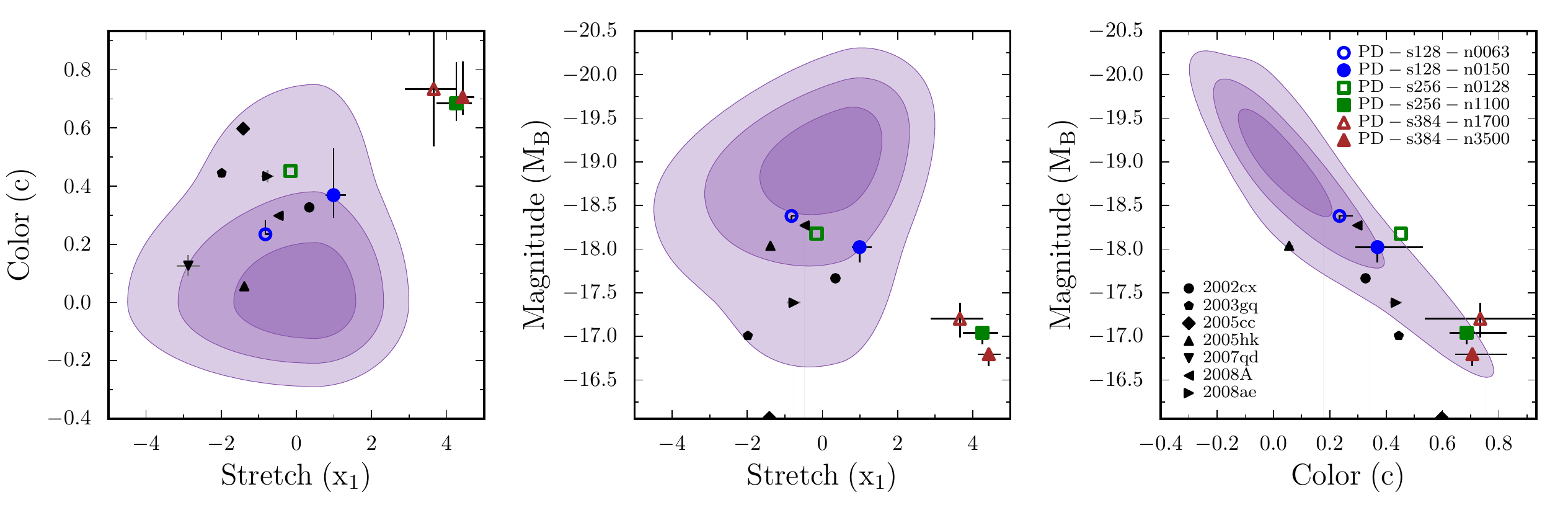}
\caption{
Fit results in stretch-color-magnitude space for the six simulations of the PD model (blue, red and green data points). 
The purple shaded contours represent the 68\%, 95\% and 99\% confidence contours of the population model of D13, marginalized over the variable which is missing from the respective panels.
The smaller black data points show the fitted SALT2 parameters of several observed under-luminous SNe. 
The magnitudes of 2005cc and 2007qd are outside the plotted range. 
The simulations roughly fall into two groups: the three simulations with few ignition points (blue points and empty green square) which lie within the 95\% contours of the population model, and the three simulations with many ignition points (filled green square and red data points) which are clearly incompatible with observed normal SNe Ia. 
See \csec{sec:comparisons:lcevaluation} for a detailed discussion.
}
\label{fig:pop}
\end{figure*}

We wish to derive quantitative statements about the degree to which the light curves predicted by our simulations agree with observations. For this purpose, we use the method proposed by \citet[][hereafter D13]{Diemer_2013}, which we briefly review here. The light curves presented in this paper were already analyzed in D13, and we expand on this analysis. Due to a minor change in the implementation of the SALT2 errors, the light curve fits and fitted parameters are slightly different from those shown in D13. 

Each set of simulated light curves is fit with the SALT2 data-driven model \citep[][]{Guy_2007, Guy_2010}.  
SALT2 is trained on a large sample of observed SNe, and thus summarizes information from a range of observed light curves. 
The model allows for three free parameters which are derived from the light curve fits, namely stretch, $x_1$, color, $c$, and magnitude, $m_{\rm B}$. 
Larger stretch corresponds to larger rise and decline times, color roughly corresponds to $B-V$ color at the epoch of peak brightness, and magnitude corresponds to an overall normalization. 
We convert $m_{\rm B}$ to an absolute B-Band peak magnitude, $M_{\rm B}$, but note that this magnitude does not necessarily correspond to the peak of the B-band light curve, since the best-fit parameters are derived from a simultaneous fit to all filter bands.  
Besides the best-fit parameters, we also derive a goodness-of-fit parameter, $\xi^2$, from each fit. Each set of explosion model light curves is fit 12 times: with three sets of filter bands ($UBVR$, $BV$ and $ugr$) and four sets of epoch ranges, allowing us to quantify the systematic uncertainty on the derived fit parameters (D13). 
While the quality of the light curve fit indicates whether the light curves predicted by a certain simulation resemble normal SN Ia light curves in general, a good fit does not guarantee that the best-fit parameters correspond to values which are observed in nature.  
In order to quantify the likelihood of a set of parameters being observed, D13 construct a population model in stretch-color-magnitude space, using data from large surveys and nearby SNe \citep{Kessler_2013}. 
We emphasize that all of the SNe used to construct the population model were well fit by the SALT2 model, meaning that the sample did not include peculiar events such as SNe Iax.

\fig{fig:fits} shows four sets of light curve fits. 
The top panels indicate the fitted filter bands and epoch ranges, as well as the best-fit parameters derived from the fit, and the goodness-of-fit parameter $\xi^2$. 
The left three panels show fits to the light curves of the three simulations with few ignition points. The fit quality is mediocre compared to other explosion models (D13), but the fits converge over a large range of epochs and filter bands. 
The right panel shows a fit to the light curves of one of the three simulations with many ignition points, {\pdfour}, which is representative of the {\pdfive} and {\pdsix} simulations as well. 
The fit barely converges as the SALT2 model is driven to an exotic region of parameter space, namely very high stretch and a dim magnitude.  
As there are essentially no observed SNe with comparable parameters, the error bars of the SALT2 model are inflated compared to the other fits.

While the light curve fits reveal interesting deviations of the six simulations of the PD model from standard SN Ia light curves, they cannot be used as a meaningful metric of quality for simulations of peculiar SN Ia, such as pure deflagration models. 
The SALT2 model is trained on normal rather than peculiar SNe Ia, and does not fit peculiar events well. However, the derived stretch, color and magnitude parameters still carry information about the character of the light curves, and can be used to compare them to observed events. 
Such a comparison is shown in \fig{fig:pop}. 
The best-fit stretch, color and magnitude for the six simulations of the PD model are shown with blue, green and red data points. 
The error bars on the parameters indicate the standard deviation of the values derived from the 12 different fits. 
The averaged value, indicated by the data points, represents the weighted average of the fit results, where fits with a lower $\xi^2$ receive a larger weight. 
The purple shaded contours show the population model of normal SNe Ia of D13. 
The smaller black data points in \fig{fig:pop} show the parameters of seven observed, peculiar SNe \citep[private communication with Mohan Ganeshalingam;][]{Foley_2013, Ganeshalingam_2013}. 

As expected, the PD models lie toward the red and dim end of the population model (right panel), as do the observed Iax SNe. 
However, the three simulations with few ignition points can be accommodated by the population model, meaning their parameters resemble those of normal, yet relatively red, SNe Ia. 
On the other hand, the three simulations with many ignition points are excluded by the population model, mainly due to their large stretch and red color. 
Their extreme parameter values are incompatible even with the observed SNe Iax.  The error bars on the best-fit parameters of the \Fsims\ are quite small, indicating that the fits to those models result in reliable estimates of the SALT2 parameters, despite the somewhat poor light curve fits.

Thus, our evaluation using the SALT2 model leads to the conclusion that the simulations with few ignition points are favored over those with many ignition points, and that the light curves predicted by the former are very roughly compatible with the observed light curves of under-luminous SNe Ia. 
In addition, our evaluation shows that the light curves predicted by the simulations with few ignition points are compatible with our population model. 
An important caveat to this statement is that the best-fit parameters derived from a light curve fit are only as reliable as the fit itself.  
Nevertheless, fitting the models with different epoch ranges and filter bands leads to a relatively robust determination of the best-fit parameters (relatively small error bars in \fig{fig:pop}).

\subsection{Locations in the \mdmplaneBB}
\label{sec:comparisons:phillips}

\begin{figure}
\centering
\includegraphics[width=3.25in]{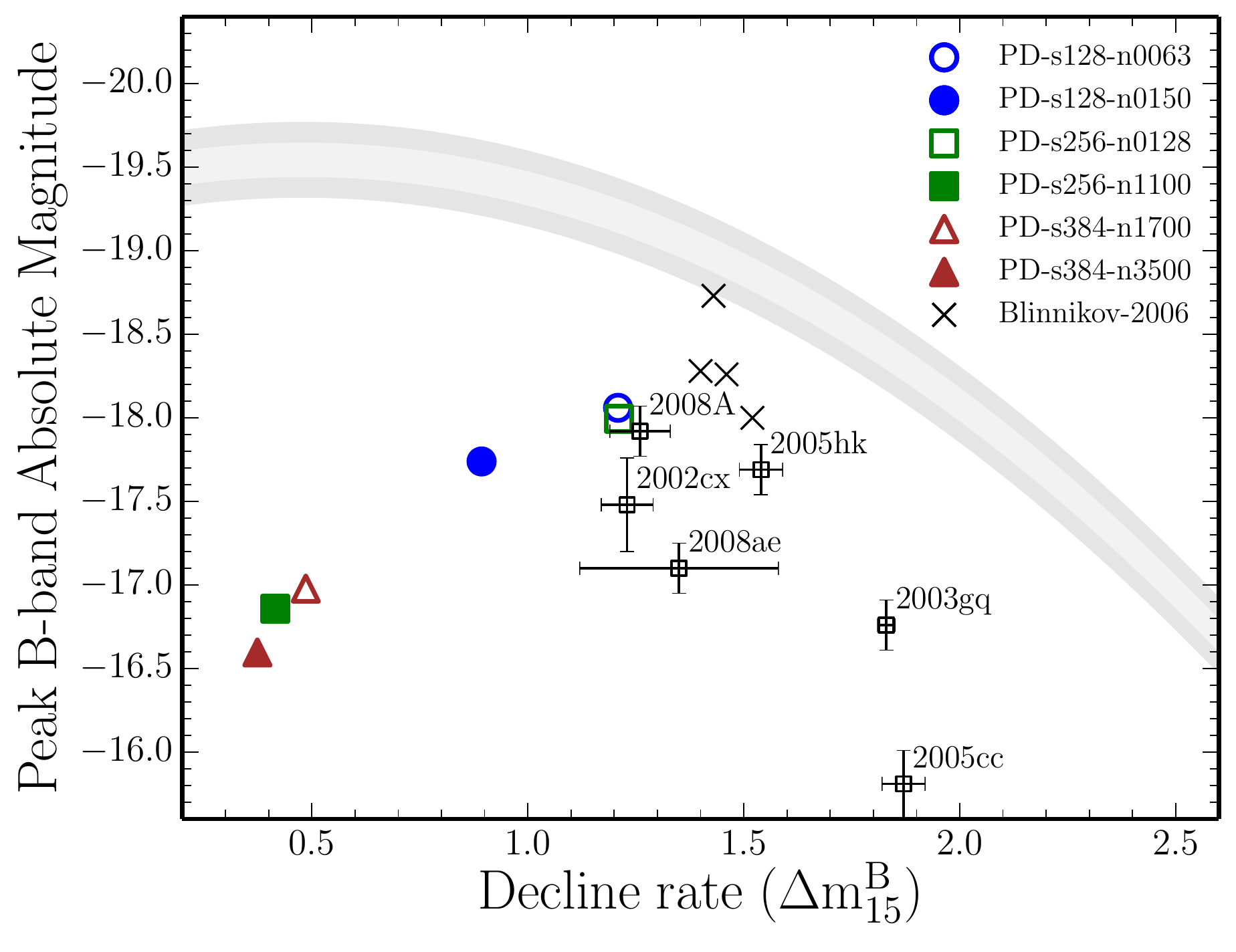}
\caption{
Locations of the six simulations of the PD model and six SNe Iax in the \mdmplaneBB.  
The gray shaded bands indicate the $1 \sigma$ and $2 \sigma$ contours of the Phillips relation with $\sigma = 0.11$ \citep{Phillips_1999}.
The filled and empty symbols represent the six simulations of the PD model.
The black crosses represent four models from Table 3 in \citet{Blinnikov_2006}.
The black squares with error bars show the locations of six SNe Iax \citep{Foley_2013, McClelland_2010}.}
\label{fig:phillips}
\end{figure}

\fig{fig:phillips} shows the locations of the six simulations of the PD model on the \mdmplaneBB.  
{For the six simulations we did, the values of these quantities are derived from the synthetic B-band light curves shown in \fig{fig:lc1}.}
The gray shaded curved bands indicate the region where normal, unreddened SNe Ia lie \citep{Phillips_1999}.
It is clear that all six simulations of the PD model lie well below this region.
The {\pdone} model lies closest to this region, but its \MB\ is still at least 1.0 magnitude fainter.  
The {\pdsix} model lies furthest away, and its \MB\ is nearly three magnitudes dimmer than normal SNe Ia for the same value of \dmfB.  
These results confirm that the PD model cannot account for normal SNe Ia \citep[see, e.g.,][]{Roepke_2007b}.

\fig{fig:phillips} also shows several observed SNe Iax in the \mdmplaneBB\ \citep{McClelland_2010, Foley_2013}. Their values are derived through low-order polynomial fitting to each light curve near its maximum brightness.  
One of the events in \citet{Foley_2013}, SNe 2008ha, has a very low peak luminosity (\MB\ $= -13.70$) and lies outside the plotted range of \MB.  
The \MB\ of SNe Iax are much fainter than those of normal SNe Ia, and their light curves decline more quickly than the light curves of normal SNe Ia. 
The distribution of SNe Iax in the \mdmplaneBB\ has not been established, i.e. it is uncertain whether SNe Iax lie on an extension of the Phillips relation, lie on a relation like the Phillips relation but different from it, or have some other distribution is. 
Thus, there is no particular distribution which we expect to be matched by our models.

It is clear, however, that the \Fsims\ lie close to, and even overlap, the region in the \mdmplaneBB\ where SNe Iax that have been observed, whereas the \Msims\ have far smaller values of \MB\ and \dmfB. 
Thus, \Fsims\ are favored over those \Msims\ as an explanation of the SNe Iax detected to date, a result that is consistent with the more global and rigorous analysis we carried out using the SALT2 model.

Finally, we note that the Phillips relation in \fig{fig:phillips} seems to be incongruent with the stretch-magnitude constraints of the population model of D13 (center panel of \fig{fig:pop}).  
While the simulations of the PD model lie many $\sigma$ off the Phillips relation, some of them can be accommodated by the population model. 
First, and most importantly, the population model contours are marginalized over color while the Phillips relation is not. 
Second, the decline rate \dmfB\ is derived from only two epochs in one particular filter band, whereas the SALT2 stretch parameter incorporates information from the entire set of light curves.  
The rising slope, for example, can have a significant influence on $x_1$ (D13).  
Finally, the Phillips relation is valid for normal SNe Ia, but does not accommodate under-luminous events \citep{Phillips_1999}, whereas the population model includes at least some peculiar events (D13). 
These observations demonstrate that it is dangerous to evaluate the light curves predicted by simulations solely based on how well they match the Phillips relation, particularly if the models represent unusual SNe Ia.

\subsection{Color evolution}
\label{sec:comparisons:colorevolution}

\begin{figure}
\centering
\includegraphics[width=3.25in]{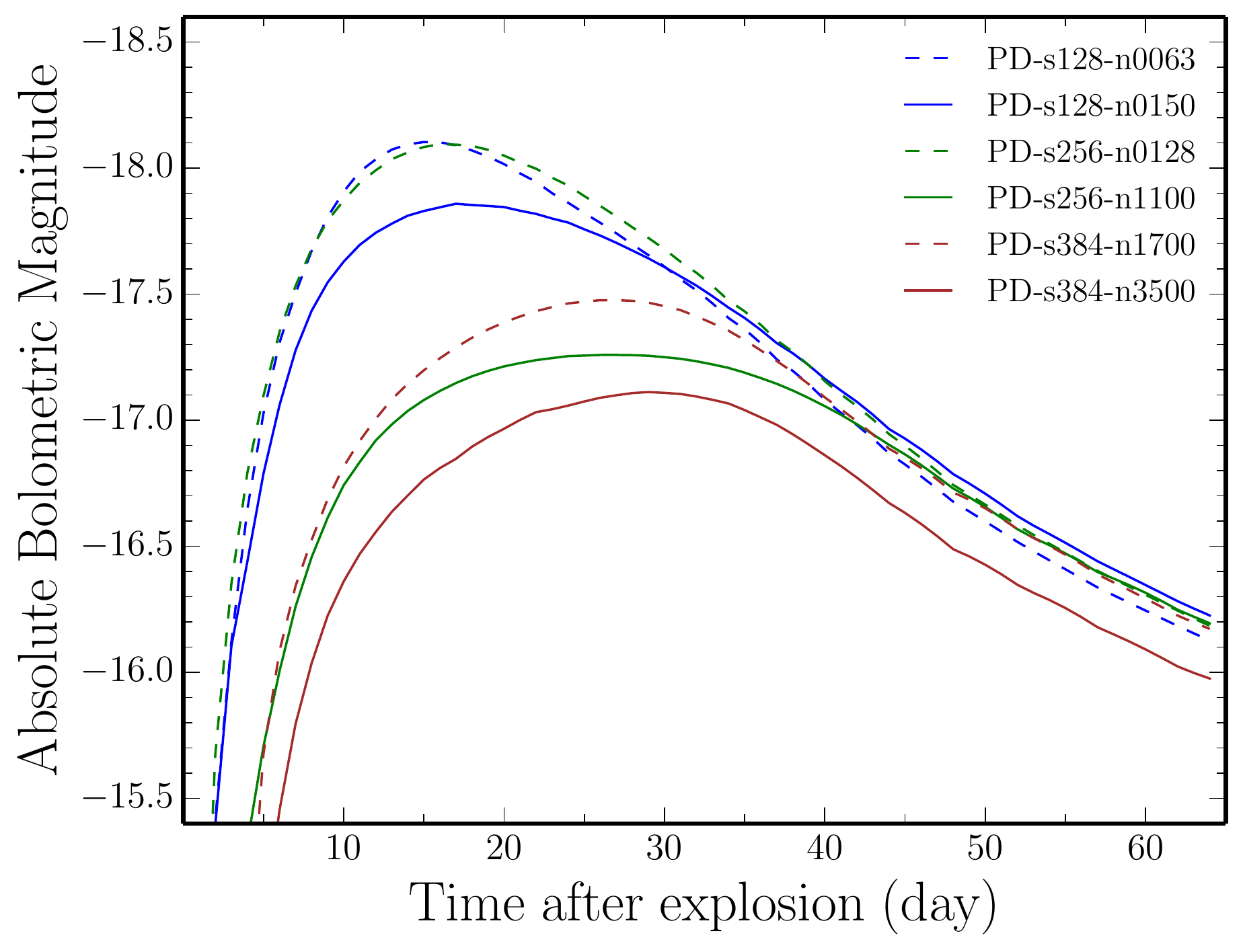}
\caption{
Absolute bolometric light curves for all six simulations of the PD model.  
The light curves for the \Fsims\ are much brighter and peak earlier than do those for the \Msims.
}
\label{fig:lc-bol}
\end{figure}

\begin{figure*}
\centering
\mbox{
\includegraphics[trim = 0mm 19mm 0mm 0mm, clip, scale=0.47]{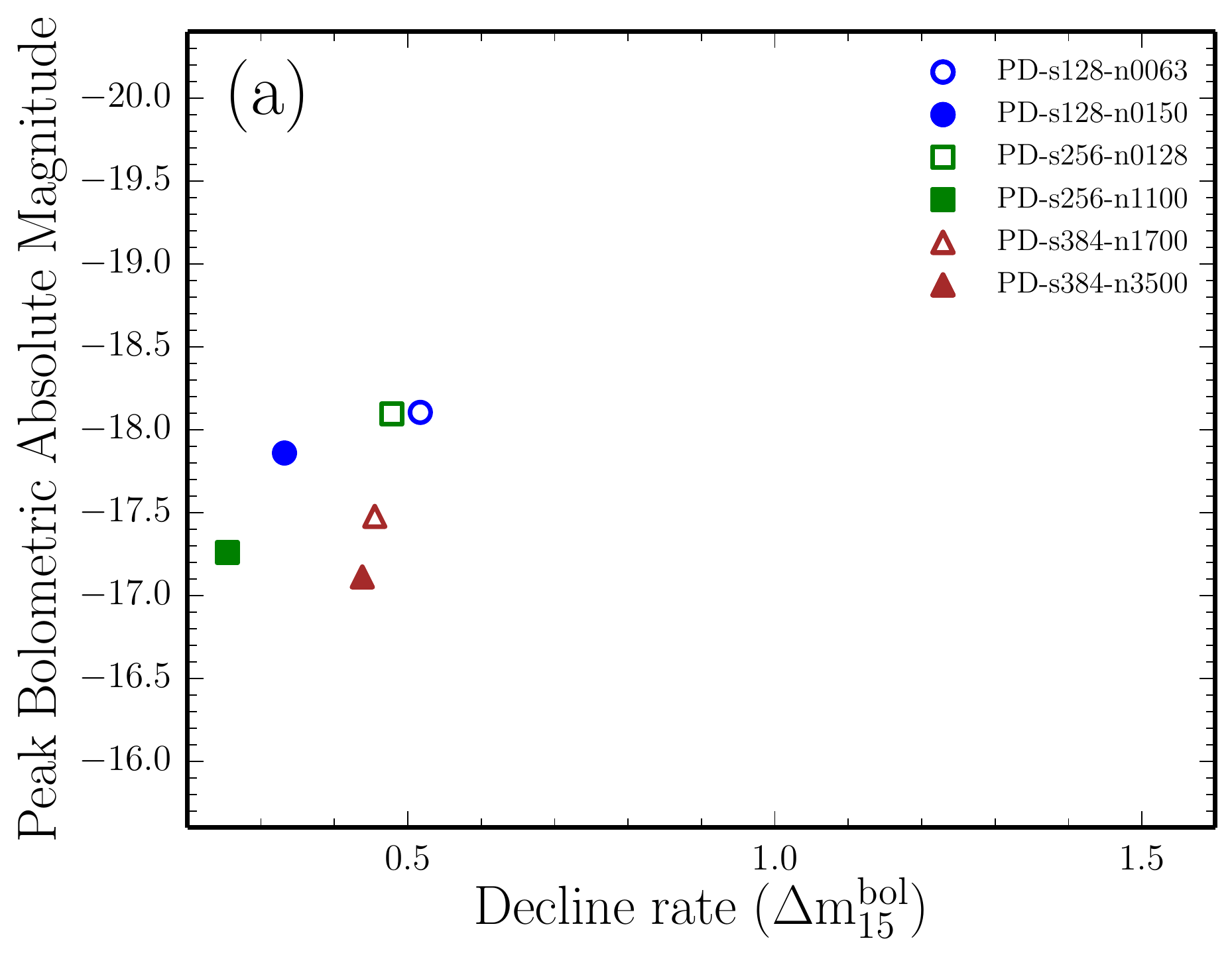}
\includegraphics[trim = 27mm 19mm 0mm 0mm, clip, scale=0.47]{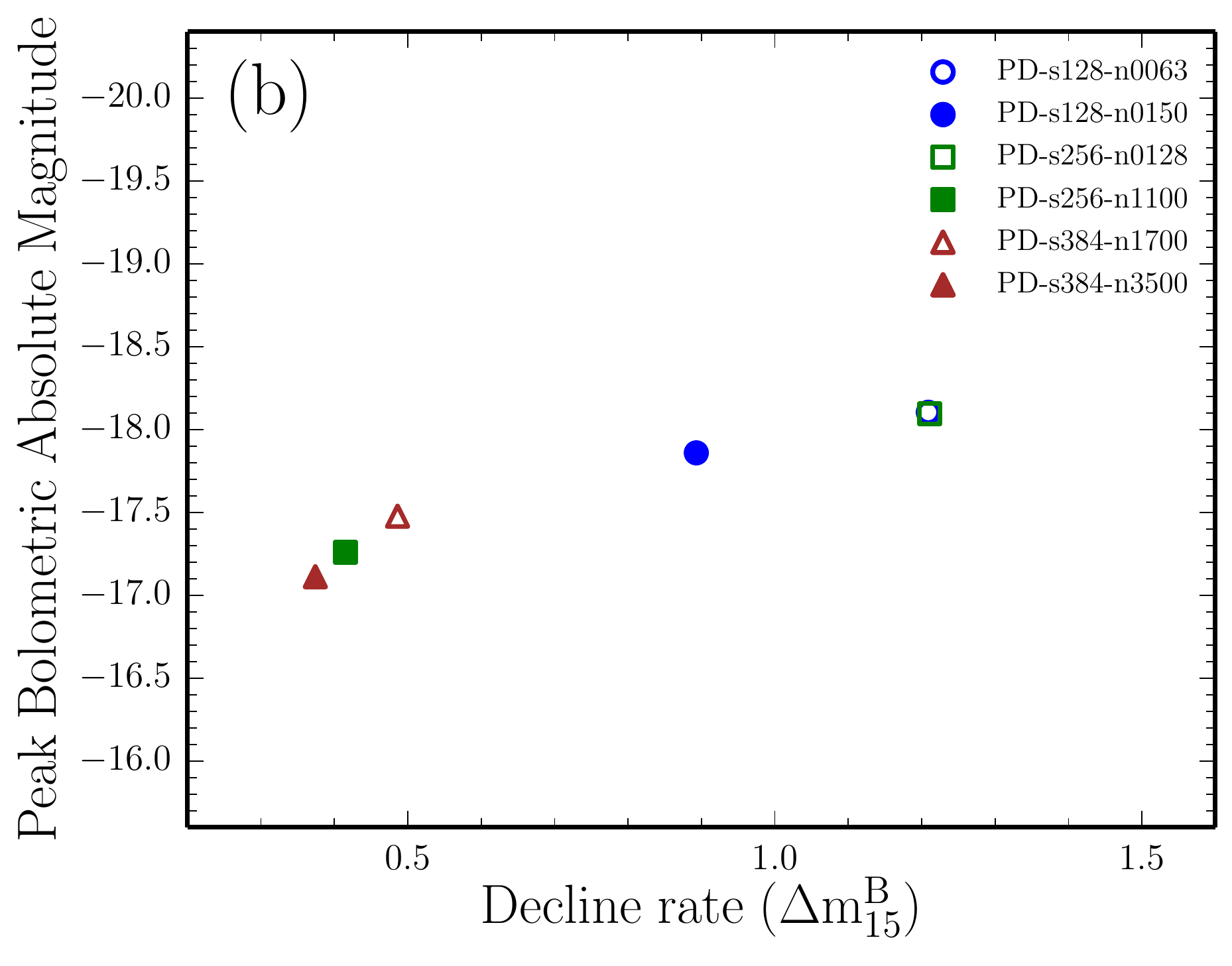}
} 
\mbox{
\includegraphics[trim = 0mm 0mm 0mm 0mm, clip, scale=0.47]{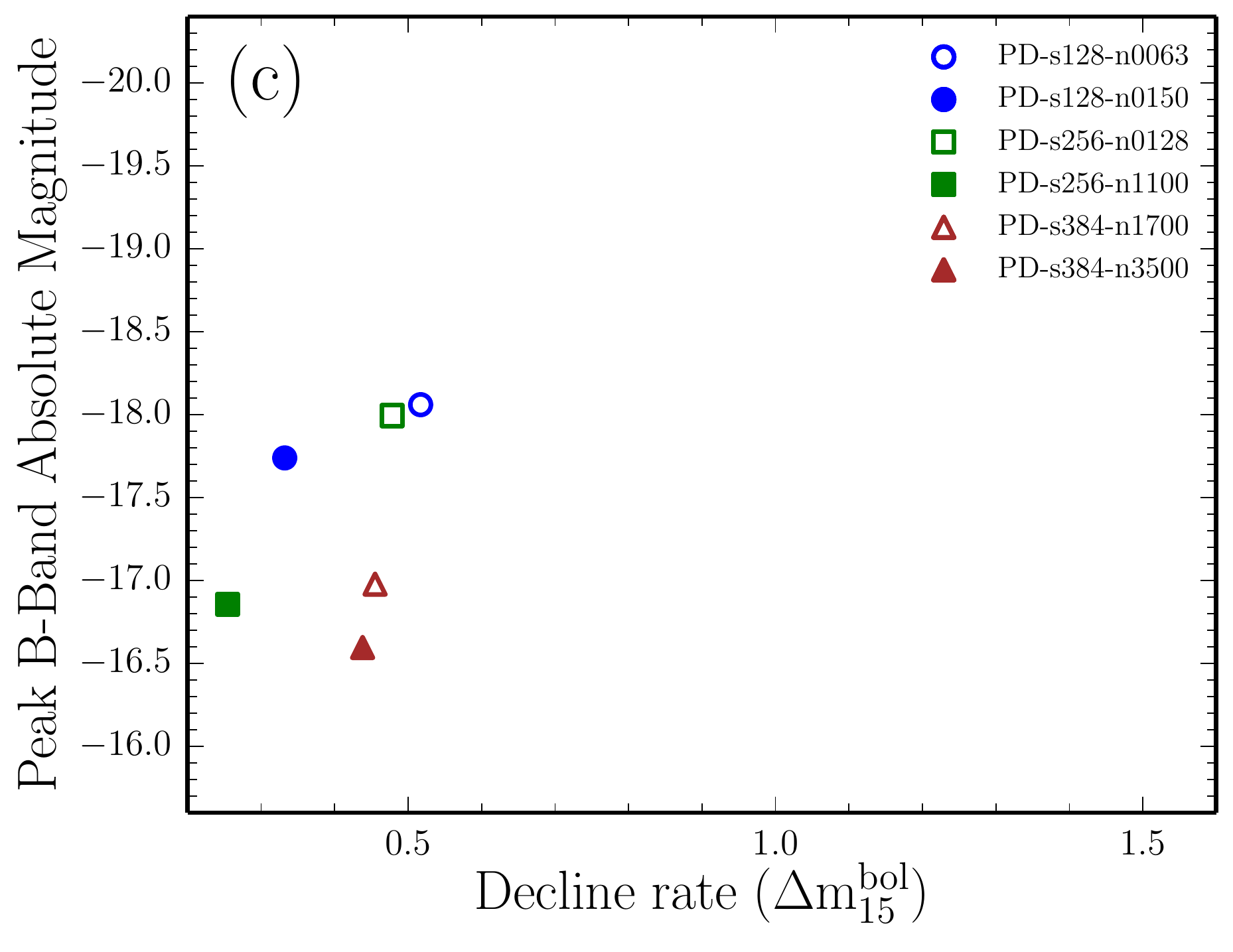}
\includegraphics[trim = 27mm 0mm 0mm 0mm, clip, scale=0.47]{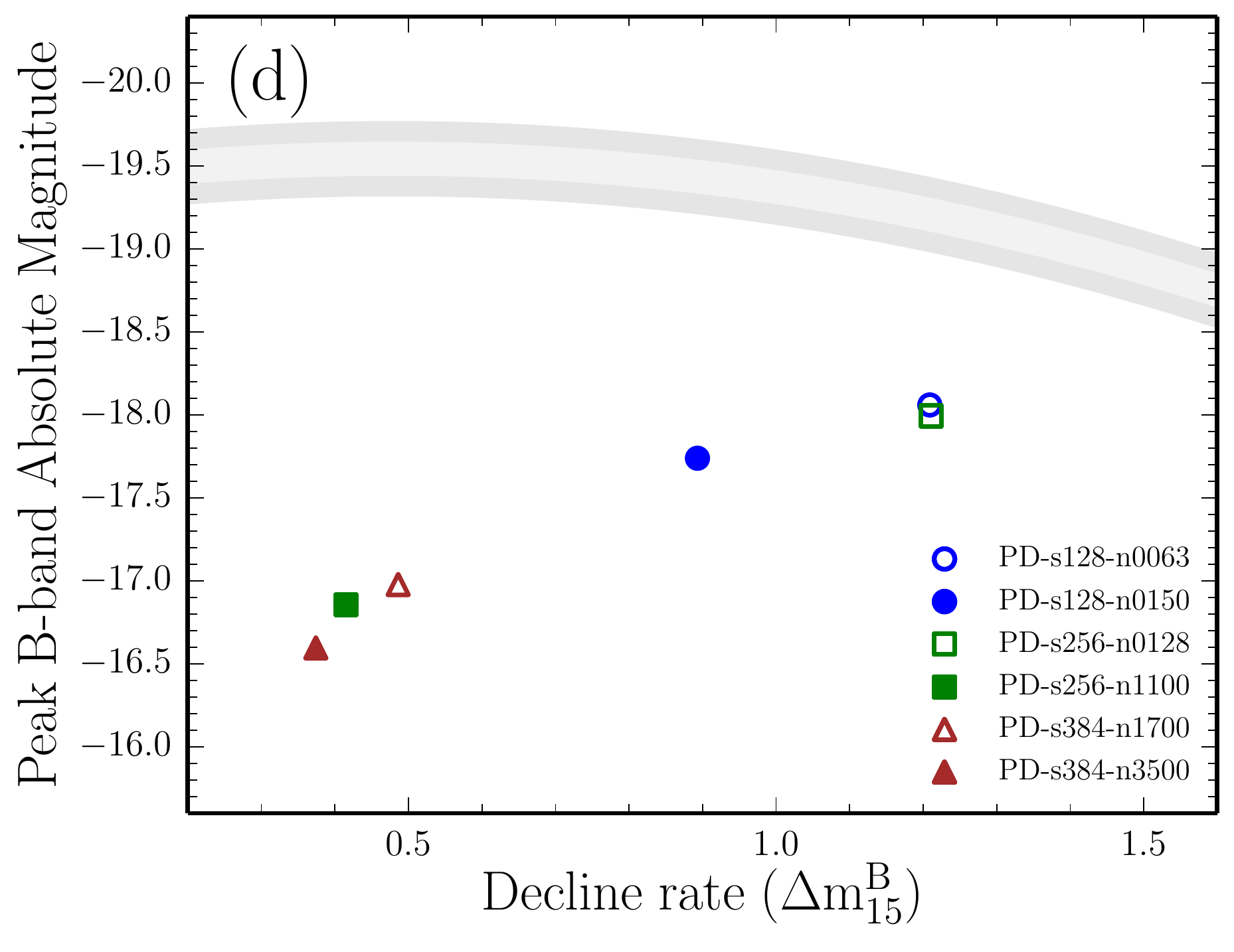}
}
\caption{
Locations of all six PD simulations in the four planes defined by the peak bolometric (\Mbol), and $B$-band (\MB) absolute magnitudes of their light curves, and the decline rates as measured by \dmfbol\ and \dmfB. 
The gray shaded bands in the panel (d) indicate the 1~$\sigma$ and 2~$\sigma$ contours of the Phillips relation with $\sigma = 0.11$ \citep{Phillips_1999}. 
The left panels demonstrate that the decline rate as defined by \dmfbol\ does not depend on \Mbol\ or \MB. 
In contrast, the right panels show that the decline rate is a strong function of the \Mbol\ and \MB, with light curves that are brighter declining more rapidly than those that are fainter, creating a trend opposite to the Phillips relation. 
Thus, the trend on the \mdmplaneBB\ is, at least to some extent, a result of the color evolution of the model light curves.
}
\label{fig:all-phillips-like}
\end{figure*}

\begin{figure}
\centering
\includegraphics[trim = 0mm 13mm 0mm 0mm, clip, width=3.25in]{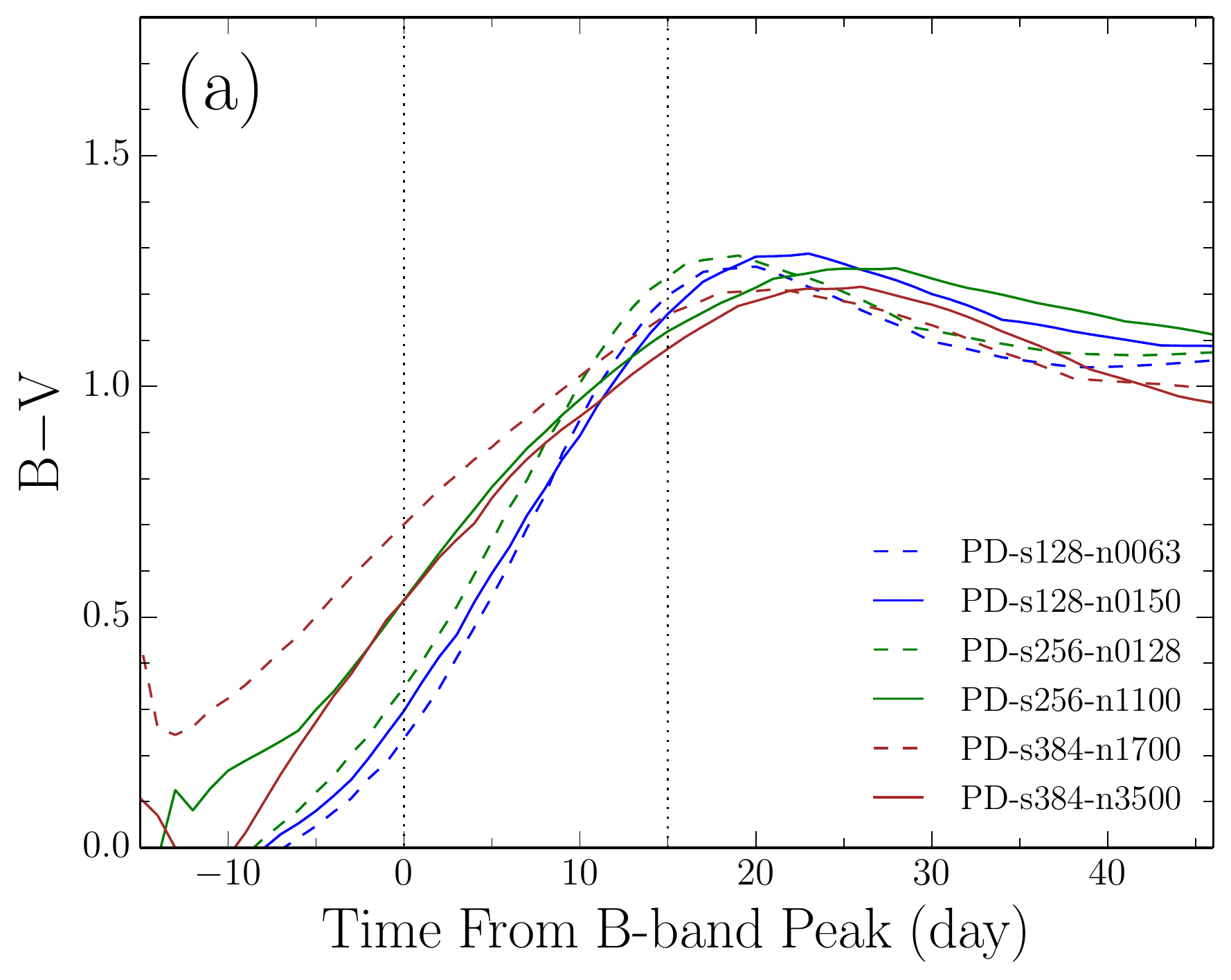}
\includegraphics[trim = 0mm 0mm 0mm 0mm, clip, width=3.25in]{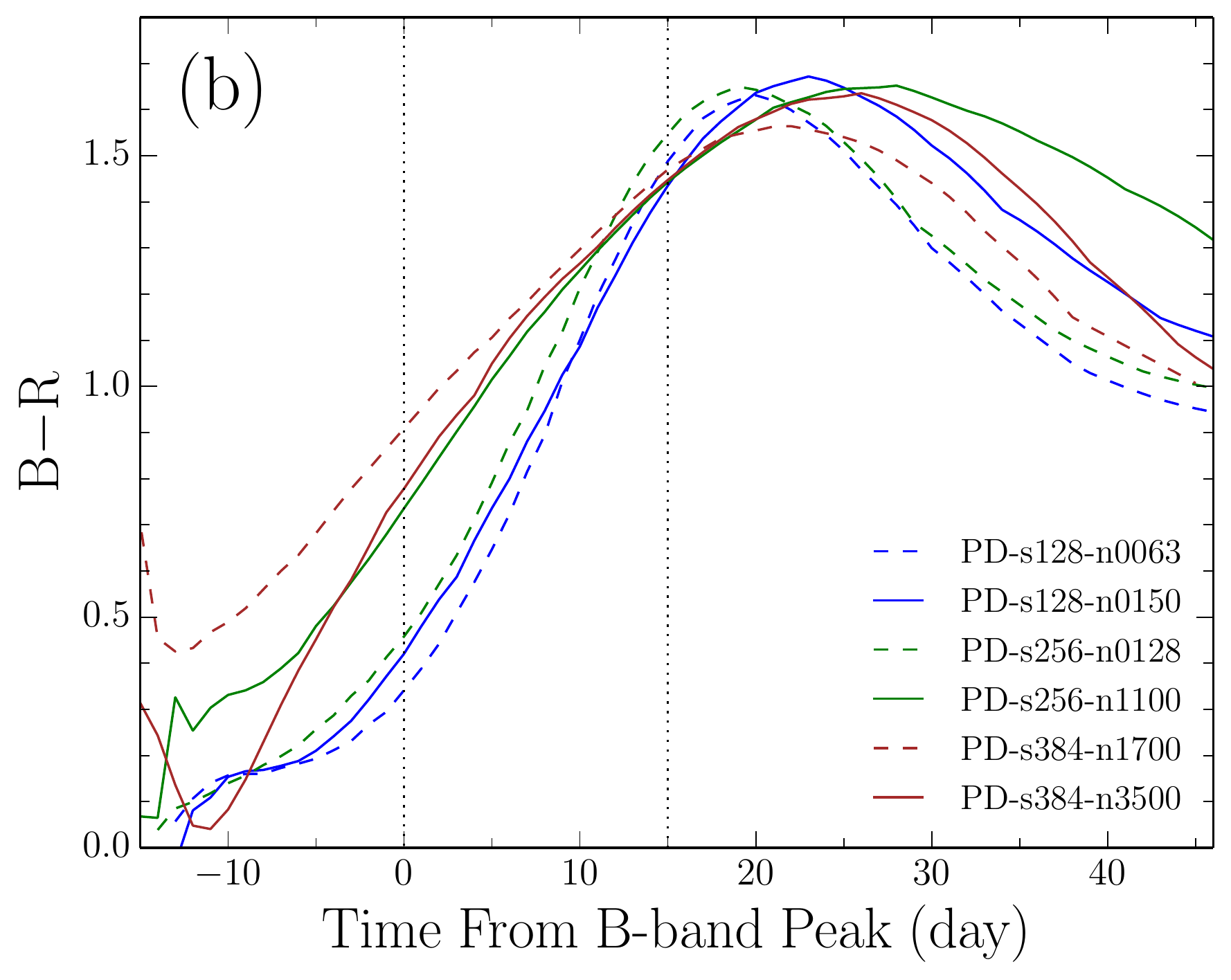}
\caption{
Color evolution of all six simulations.  
All of the simulations have relatively red colors.  
At maximum light, the \Fsims\ are much bluer than those \Msims.
Thereafter, the \Fsims\ (which are much brighter) become redder more rapidly; at fifteen days after maximum light in the B band, all of the simulations have similar colors.  
The differing rates of color evolution account for the differences in \dmfB.
}
\label{fig:color-evolution}
\end{figure}

Our simulations of the PD model lie along a diagonal line on the \mdmplaneBB, with a slope opposite to that of the Phillips relation. 
However, this distribution is a result of the particular ICs we chose to investigate, and simulations of a wider range of ICs can be expected to cover a larger region in the \mdmplaneBB. At the present time we do not know with confidence what ICs are realized in nature. 
Nevertheless, explaining the \MB--\dmfB\ trend of our models provides additional insight into the nature of the PD model.

\citet{Kasen_2007} used a 1D model to argue that the Phillips relation is due to the dependence of the color evolution of SNe Ia light curves on $M_\mathrm{Ni}$ (rather than the dependence of the diffusion time on $M_\mathrm{Ni}$, as had been emphasized in earlier studies). 
The color evolution is due to the onset of Fe II/Co II lines as the ejecta cools.  
These lines redistribute flux absorbed in the blue part of the spectrum to longer wavelengths.

In \figs{fig:lc-bol}, \ref{fig:all-phillips-like}, and \ref{fig:color-evolution}, we show that the  \MB--\dmfB\ locations of our simulations are also due to the dependence of the color evolution of their light curves on the amount of \nifs, but in addition, on its distribution in the ejecta. 
\fig{fig:lc-bol} shows the absolute bolometric light curves for all six simulations.  
The left panels of \fig{fig:all-phillips-like} show the locations of all six simulations in the \mdmplaneBbol\ and the \mdmplanebolbol. 
All six simulations have about the same \Mbol, and their distribution in the \mdmplanebolbol\ and in the \mdmplaneBbol\ is therefore roughly vertical.  
This suggests that the higher velocity of expansion of the ejecta in the \Fsims\ (which released more $E_{\mathrm nuc}$, producing a larger $E_K$) is compensated for by the higher opacities due to the larger amount of IGE in these same simulations; and that the converse is true for the \Msims.

In contrast, the right panels in \fig{fig:all-phillips-like} show that the six simulations lie along a diagonal line in the \mdmplanebolB\ and the \mdmplaneBB\ whose slope is opposite that of the Phillips relation.  
The difference between the distributions of the simulations in these two pairs of planes shows that the distributions in the latter pair are a result of the color evolution of the light curves. 
\fig{fig:color-evolution}, which shows the color evolution of the light curves as measured by $B$-$V$ and $B$-$R$, demonstrates this directly.

The origin of this color evolution appears to be the following.  
The simulations that produce more \nifs\ have higher peak luminosities than those that produce less \nifs, as expected, but the distribution of \nifs\ in the ejecta of the simulations with higher peak luminosities is very different than it is in those with lower peak luminosities.  
In the simulations with higher peak luminosities, the abundance of \nifs\ is unusually low in the core of the ejecta and unusually high throughout the rest of the ejecta.  
This can be seen from the second column in \fig{fig:slice}, which shows the distribution of the mass densities of \nifs\ in the ejecta, and \fig{fig:abund-vel}, which shows chemical abundances as a function of velocity, for each of the simulations.  
As a result, the temperature in the outer layers is initially high, but rapidly falls, as the heating of the outer layers due to the decay of the \nifs\ there declines and the energy deposited by this process diffuses out of the ejecta without being replaced by energy diffusing from the core.  
Fe II/Co II lines then form in the outer layers of the ejecta and redistribute the flux absorbed in the blue part of the spectrum to longer wavelengths.  
Thus these supernovae are blue around peak but then rapidly become red. 
In contrast, not only is there less \nifs\ in the ejecta of the simulations that have lower peak luminosities, the \nifs\ is also strongly concentrated in the core of the ejecta with virtually no \nifs\ present in the outer layers of the ejecta.  
The temperature in the outer layers of the ejecta is therefore low from the beginning.  
Consequently, these supernovae start relatively red and become redder with time.  

To summarize, the more luminous supernovae we simulated start blue but rapidly become red, whereas the less luminous supernovae we simulated start red and become redder. 
By 15 days after maximum light, all of the supernovae are equally red. 
This color evolution is the opposite of what is needed to produce the Phillips relation \citep{Kasen_2007}, and as a result our particular simulations of the PD model lie along a diagonal line in the \mdmplaneBB, with a slope opposite of that of the Phillips relation.  
However, we caution again that, because we performed a relatively small number of simulations of the PD model for a few ICs, and do not know what ICs occur in nature, we cannot say what the expected distribution of PD models in the \mdmplaneBB\ is.

\section{Discussion and Conclusions}
\label{sec:conclusions}

\subsection{Comparison with earlier work}

\citet{Roepke_2006} investigated the dependence of the properties of the hydrodynamic explosion phase of the PD model on three initial conditions:  the mass of the white dwarf progenitor star (1.367 and 1.403 $M_\sun$, the C/O ratio in the white dwarf progenitor star ($X_{^{12}{\rm C}}$ = 0.30, 0.46, and 0.62), and the abundance of  $^{22}$Ne (parametrized in terms of the metallicity: 0.5, 1.0, and 3.0 $Z_\sun$).  

The code \citet{Roepke_2006} used to simulate the reactive flow hydrodynamics is based on the Prometheus code, as is \texttt{FLASH}, and both use a solver based on the piece-wise parabolic method (PPM).  
However, there are many major differences between the simulations done by \citet{Roepke_2006} and those reported here.  
Among these are a 3D simulation of an octant of the star rather than a 3D simulation of the whole star, the use of an expanding grid rather than AMR, 17 million grid points vs. several billion grid points, three toroidal rings vs. a random distribution of ignition points within a confining sphere as the initial conditions, a very different sub-grid flame model that treats the interaction of the flame with the buoyancy-driven turbulence created by it in a very different way, and 20,000 Lagrangian tracer particles vs. 10 million tracer particles to determine the spatial distribution of the elements, as well as other differences in algorithms used and the treatment of the energy released by nuclear burning.  
Most of these differences reflect the dramatic difference in the computational resources available then vs. now.  
It is nevertheless interesting and useful to compare the results reported by \citet{Roepke_2006} and those we present here.
\footnote{
A single simulation of the PD model carried out only a year later by \citet{Roepke_2007b} made use of substantially greater computational resources, and was a 3D simulation of a whole star with 1 billion grid points and 150 thousand Lagrangian tracer particles, but provides many fewer quantitative results with which we can compare.  
Therefore, we focus on comparisons on the results of \citet{Roepke_2006}.
}

Despite these major differences, we find qualitative agreement between the quantities we can compare.  
\fig{fig:em}(a) shows the locations of the simulations of the PD model studied by \citet{Roepke_2006} in the ($E_\mathrm{nuc}$,$E_\mathrm{K}$)-plane.  
The three points corresponding to simulations that assumed a 1.367 $M_\sun$ white dwarf lie a bit to the left of the dashed line that shows the linear relationship that best fits the simulations we present here, while the three points corresponding to simulations that assumed a 1.403  $M_\sun$ white dwarf lie on it.  
It is not clear whether the values of $E_\mathrm{nuc}$ reported by \citet{Roepke_2006} include the contribution from the rather substantial mass that was assumed in the simulations to already have burned.  
If not, correcting for it would bring the former into better agreement with the dashed line and move the latter to the right of it (where they will be expected to lie, since they assumed a 1.403 $M_\sun$ white dwarf progenitor). 

\figs{fig:em}(c) and (d) show the locations of the simulations studied by \citet{Roepke_2006} in the ($E_\mathrm{nuc}$,$M_\mathrm{IGE}$)-plane and in the ($E_\mathrm{nuc}$,$M_{^{56}{\rm Ni}}$)-plane, and \figs{fig:Mek}(c) and (d) in the ($E_\mathrm{K}$,$M_\mathrm{IGE}$) and ($E_\mathrm{K}$,$M_{^{56}{\rm Ni}}$)-planes.  
The upper three points in each panel correspond to the simulations that assume a 1.367 $M_\sun$ white dwarf, while the lower three points correspond a 1.403 $M_\sun$ white dwarf.  
Here the agreement is not as good.  

In summary, we find qualitative agreement, and in some cases rough quantitative agreement, between the quantities reported by \citet{Roepke_2006} and presented here that we can compare, despite many major differences in the simulations.  
Therefore, the conclusion that the PD model produces under-luminous SNe Ia seems quite robust.

\subsection{Explosion-phase properties of the PD model}

We have conducted high-resolution, full-star 3D simulations of the PD model.  
We find that, in the particular set of six simulations we did, the properties of the hydrodynamic explosion phase, including the creation and the distribution of various elements, depend primarily on the number of ignition points \Nign.

In the \Fsims, the rate of nuclear burning is low at early times but increasing rapidly thereafter.  
The expansion of the white dwarf is therefore initially slow,  allowing more of the star to burn to IGE and IME before the nuclear flame quenches at low densities.  
In contrast, in the \Msims\ the rate of nuclear burning is high  at early times, but does not much increase.  
The expansion of the white dwarf is therefore fast, and less of the star can burn to IGE and IME before the flame quenches.

We also find that the spatial distributions of the overall mass density, as well as the densities of \nifs, Si, and C/O in the ejecta, depend sensitively on the number of ignition points.  
In the \Fsims, the burning bubbles accelerate away from the center of the star, leaving a relatively cold, high-density core composed of unburned C/O.  
Subsequently, the nuclear burning rate rapidly increases, creating an extended hot region consisting of IGE and IME surrounding the core; outside of this is a shell of unburned C/O.  
In contrast, in the \Msims, the burning bubbles in the core merge, producing a hot, less dense and more extended core composed primarily of IGE.  
The slow rate of nuclear burning at late times allows large plumes of hot ash to form, which rise toward the outer regions of the star.  
Around these plumes, cold,unburned C/O sinks toward the core but is unable to penetrate it fully.  
As a result, the distributions of {\nifs} and Si in the ejecta have a star-shaped pattern, while the distribution of unburned C/O has an inverse pattern.  
These differences in the amounts and the distributions of the various elements in the ejecta for simulations with few and many ignition points lead to clear observational signatures in their light curves, as we discuss below.

In the particular set of simulations of the PD model we performed, the number of ignition points is the most important factor governing the outcome of the simulations.  
However, it is clear that, had we been able to study a much larger range of possible ICs, the relative importance of these three parameters to the outcome of the simulations would vary greatly; and the outcome of the simulations would cover a much larger range in all of the properties, including the observed properties, of the resulting SNe Ia.

\subsection{Comparison with observations}

We calculated the light curves predicted by the simulations and compared them with observations.  
We find that the \Fsims\ produce light curves that rise more quickly, have a higher peak B-band absolute magnitude $M_\mathrm{B}$, and decline more rapidly than the \Msims.  
They are also redder.  
These properties are a result of the larger nuclear energy $E_\mathrm{nuc}$ released in these models, which results in larger kinetic  energies $E_\mathrm{K}$ and therefore light curves that rise and decline more rapidly, and the larger mass of {\nifs} produced in these models, which results in a larger $M_\mathrm{B}$.  
Finally, we find that, as a result, the $M_\mathrm{B}$ and the shapes of the light curves predicted by the simulations with few ignition points have properties similar to the under-luminous SNe Iax, while those with many ignition points do not.  
Thus the former are a more promising explanation of these events.

\subsection{Directions for future research}

We performed a relatively small number of simulations of the PD model because of the large computational cost of the 3D hydrodynamic simulations of the explosion phase, the post-processing of the Lagrangian tracer particles to determine the nucleosynthetic yields, and the radiation transfer calculations to determine the light curves.  
As mentioned above, simulations of the PD model for a wider range of ICs can be expected to cover a larger region in the \mdmplaneBB, on top of which we do not know with confidence which ICs are realized in nature. 
We are therefore unable to say what the distribution of simulations of the PD model in the \mdmplaneBB\ is, in general.

What we can say from the simulations we completed is that simulations with smaller numbers of ignition points and larger confining spheres (or spheres offset from the center of the star) would likely lead to models with low luminosities and faster decline rates (i.e., larger values of \dmfB), which would thus lie in the same region of the \mdmplaneBB\ as SNe Iax.  
However, we do not know with high confidence whether these ICs are realized in nature.

\figs{fig:burnedmass}, \ref{fig:slice},  and \ref{fig:abund-vel} suggest that such ICs might produce roughly the same amount of \nifs\ as our simulations that had many ignition points, but shift it outward in the ejecta.  
This would keep \MB\ roughly the same while producing faster decline rates. 
The results of two recent papers \citep{Jordan_2012b,Kromer_2013} support this conclusion.

\acknowledgements

We thank Klaus Weide, Norbert Flocke and Christopher Daley for the great help they provided during this work, and Brad Gallagher for creating the images in \fig{fig:volume}.  
We thank Katherine Riley, Mike Papka, and the staff at the Argonne Leadership Computing Facility at Argonne National Laboratory for help running our large simulations on Intrepid at ANL.  
We also thank the referee for valuable comments and suggestions that led to significant improvements in the paper. 
This work is supported in part by the U. S. Department of Energy under Contract No. B523820 and the National Science Foundation under Grant No. AST-0909132 to the Flash Center for Computational Sciences at the University of Chicago.  
This work used computational resources at ALCF at Argonne National Laboratory supported by the Office of Advanced Scientific Computing Research, Office of Science, U.S. Department of Energy under Contract No. DE-AC02-06CH11357.

\bibliographystyle{apj}
\bibliography{ms}

\end{document}